\documentclass[a4paper,fleqn]{cas-sc}

\usepackage[authoryear,longnamesfirst]{natbib}
\shortcites{Xu:2025,Li:2025,Sohrabi:2024,Elder:2007,Boettinger:2000,Asta:2009,Sosso:2016,Yao:2024,Righi:2023,Liu:2023,Barton:2022,Lazicki:2015,Pang2014SR,Smith:2013,Hawreliak:2011,Hawreliak:2008,Kalantar:2005,YANG2025104403,PhysRevResearch.6.013316,PhysRevB.93.214108} 
\usepackage{amssymb}
\usepackage{amsmath}
 \usepackage{color}
 \usepackage{graphicx}

\usepackage{arydshln} 
\newcommand{\vv}[1]{\boldsymbol{#1}} 
\newcommand{\mat}[1]{\boldsymbol{#1}} 

\makeatletter

\@addtoreset{equation}{section}
\makeatother
\graphicspath{{./figures/}}
\begin{document}
\shorttitle{PT Kinetics Model}
\shortauthors{D. N. Blaschke and A. Hunter and D. L. Preston}

\title[mode = title]{Phase Transformation Kinetics Model for Metals}

\author{Daniel N. Blaschke}[orcid =0000-0001-5138-1462]
\ead{dblaschke@lanl.gov}
\author{Abigail Hunter}[orcid =0000-0002-0443-4020]
\ead{ahunter@lanl.gov}
\author{Dean L. Preston}[orcid =0000-0002-8064-1523]
\ead{dean@lanl.gov}

\address{X Computational Physics, Los Alamos National Laboratory, Los Alamos, NM}


\nonumnote{LA-UR-24-32576}

\date{\today}
\begin{abstract}
We develop a new model for phase transformation kinetics in metals by generalizing the Levitas-Preston (LP) phase field model of martensite phase transformations (see  \cite{Levitas:2002,Levitas2:2002,Levitas:2003}) to arbitrary pressure.
Furthermore, we account for and track:
the interface speed of the pressure-driven phase transformation,
properties of critical nuclei,
as well as nucleation at grain sites and on dislocations and homogeneous nucleation.
The volume fraction evolution of each phase is described by employing KJMA \citep{Kolmogorov:1937,Johnson:1939,Avrami:1939, Avrami:1940,Avrami:1941} kinetic theory.
We then test our new model for iron under ramp loading conditions and compare our predictions for the $\alpha\to\epsilon$ iron phase transition to experimental data of \cite{Smith:2013}.
More than one combination of material and model parameters (such as dislocation density and interface speed) led to good agreement of our simulations to the experimental data,
thus highlighting the importance of having accurate microstructure data for the sample under consideration.
\end{abstract}



 
\maketitle
\setcounter{tocdepth}{2}
\tableofcontents

\section{Introduction}
\label{sec:introduction}

Phase transformations, particularly liquid-solid phase transformations, are ubiquitous in many manufacturing processes, and thus have received much focus from the community over the last several decades \citep{Boettinger:2000,Asta:2009,Sosso:2016,Gomez:2018,YANG2025104403}.
Less studied, however, are solid-solid phase transformations, in which there is a change in crystal structure that is accompanied by a sudden change in volume, that can occur in some metals under certain, typically extreme (i.e., high pressures or high rate) loading conditions (see e.g.  \cite{Sikka:1982,Gornostyrev:1999,Davis:2007,Rigg:2009,Smith:2013,Zong:2014,Lazicki:2015,Barton:2022,Liu:2023,Yao:2024,Li:2025}), but also due to a change in temperature \citep{Gomez:2018,Pei:2024}.
Under shock loading, for example, a metal can (partially) transform to the high pressure phase and back, which can have a strong influence on the extent of damage that can occur \citep{Resseguier:2008,Righi:2023}.
Such phase transformations have also been observed and studied in high entropy alloys, see e.g. \cite{Lahiri:2022,WU2023103730,Sohrabi:2024,JIANG2024103968}.

Often overlooked when considering solid-solid phase transformations is the time dependence of these transformations, or rather, the kinetics of the phase transformations.
In phase diagrams phase boundaries are typically shown as a sharp line.
However, many experimental studies have shown that depending on the loading conditions, the forward/reverse transition may happen above/below this equilibrium line \citep{Taylor:1991,Merkel:2020,PhysRevResearch.6.013316}.
The deviation from the equilibrium phase boundary is a result of the time it takes for the material to progress through a mixed phase composition and fully transform its crystal structure.

A prime example of this behavior can be seen in iron, which at pressures of around 13 GPa changes from body-centered-cubic (bcc) $\alpha$-iron to hexagonal-close-packed (hcp) $\epsilon$-iron (see e.g. \cite{Takahashi:1964,Barker:1974,Taylor:1991,Boettger:1997,Kalantar:2005,Kadau:2005,Jensen:2009,Bastea:2009,Hawreliak:2011,Smith:2013,PhysRevB.93.214108,Yao:2024}).
Under high-rate loading, such as shock compression, this phase transition starts at higher pressures than 13 GPa due to the time the phase transition takes.
Similarly, the reverse $\epsilon\to\alpha$ transition begins at much lower pressure, i.e. $\sim10$ GPa.
Many studies, particularly under quasi-static loading, have confirmed a pronounced pressure hysteresis, see e.g. \cite{Giles:1971,Taylor:1991,Merkel:2020,Righi:2023} and references therein.
Such a pressure hysteresis was observed also in other metals \citep{Sikka:1982,Rigg:2009,Zong:2014}.

Some atomistic studies have focused on the timescale of these types of phase transformations \citep{Bertrand:2013,Zong:2014,Pang2014SR,Gunkelmann:2015,Guo:2021,Ma:2022,PhysRevB.108.174109,Daphalapurkar:2024,Rawat:2025,Amadou:2025}.
A good theoretical understanding of the underlying dynamics, however, is lacking.
This is the gap we aim to fill with this work; i.e. we develop a new model enabling the prediction of the dynamics of this type of phase transition.
Specifically, given the initial material properties and loading conditions, our model can predict the time the phase transformation takes to complete as well as the ``overshoot'' pressure, i.e. the pressure at which the phase transition starts as a function of loading conditions (such as pressure rate if ramp loading is considered).

Phase field models are well suited to study phase transitions \citep{Elder:2007,Gomez:2018,Ansari:2021,Lahiri:2022,Tourret:2022,Yao:2024}.
For a review on the application of phase field models to the solid-solid phase transitions in shape memory alloys, see \cite{Xu:2025}, for example.

Our new model generalizes ideas of \cite{Levitas:2002,Levitas2:2002,Levitas:2003} as well as \cite{Kolmogorov:1937,Johnson:1939,Avrami:1939, Avrami:1940,Avrami:1941};
see also \cite{Levitas:2021}.
In particular, we generalize the Levitas-Preston (LP) phase field model of martensite phase transformations to arbitrary pressure
and then use it to calculate two-phase interface speeds and critical nuclei energies.
We derive expressions for the rates of homogeneous nucleation and nucleation at grain sites and dislocations, and employ KJMA kinetic theory to compute volume fraction evolution of the second phase; generalizations of this framework to more than two phases are left for future work.
We then apply our new model to iron under ramp loading conditions using an in-house research code written in Python and compare our predictions for the $\alpha\to\epsilon$ iron phase transition to experimental data of \cite{Smith:2013}.

In Section \ref{sec:meth} we develop our general theory and introduce some simplifying approximations.
We emphasize that our model describes only the phase transition kinetics, and within a larger simulation it must be coupled with an appropriate flow stress model (such as e.g. \cite{PTW:2003}).
In contrast to other (typically phenomenological) phase transition kinetics models, our present work explicitly calculates the nucleation rates of the new phase on various types of microstructure, as detailed in Section \ref{sec:nucleation}.
Furthermore, we focus on the simpler case of only two phases for simplicity; a generalization to multiple phases is left for future work.
In Section \ref{sec:ironexample} we proceed to successfully test our model on the $\alpha\to\epsilon$ iron phase transition under ramp loading conditions, since our equations simplify considerably when pressure becomes a linear function of time.
We leave the reverse $\epsilon\to\alpha$ transformation as well as shock loading simulations to future work.

\section{Methodology}
\label{sec:meth}

We construct a phase field model of solid-solid phase transformations based on a Landau (Gibbs) free energy that is a quartic polynomial in a single order parameter $\eta$.
In particular, we generalize the Levitas-Preston (LP) phase field model of martensite phase transformations \citep{Levitas:2002,Levitas2:2002,Levitas:2003}, which  is only valid in a neighborhood of zero pressure, to arbitrary pressure.
A number of approximations are made to keep the model tractable, as elucidated in the respective subsections below and subsequently summarized in Section \ref{sec:summarizeapprox}.

We designate contractions of tensors $\mat{A}=\{A_{\alpha\beta}\}$ and $\mat{B}=\{B_{\alpha\beta}\}$ over one index as $\mat{A}\cdot\mat{B}=\{A_{\alpha\beta}B_{\beta\gamma}\}$.
Contractions of a tensor and a vector, $\vv{v}=\{v_\alpha\}$, will also be denoted by a single raised dot:
$\mat{A}\cdot\vv{v}=\{A_{\alpha\beta}v_\beta\}$.
Contractions over two indices will be denoted as $(\mat{A}:\mat{B})= A_{ij} B_{ij}$.
A $\otimes$ denotes a tensor (dyadic) product.

\subsection{Gibbs Free Energy}
\label{sec:Gibbs}

In the LP model for a single martensitic variant, the specific (per unit volume) Gibbs free energy depends on the stress tensor, temperature, and an order parameter, $\eta$, that vanishes in austenite (A) and equals unity in the martensite (M)
\begin{align}
G(\sigma, T, \eta) &= -\mat{\sigma}:\mat{S}:\mat{\sigma}/2 - \mat{\sigma}:\mat{\epsilon}^t \, \phi(\eta) 
  - \mat{\sigma}:\mat{\epsilon}^T + f(T, \eta)    \, .
\label{eq:gibbs1}
\end{align}
Here $\mat{S}$ is the second-order (rank four) elastic compliance tensor, $T$ is the temperature, $\mat{\epsilon}^t$ is the A$\rightarrow$M transformation strain tensor, $\mat{\epsilon}^T$ is the thermal strain tensor, and $f(T,\eta)$ is the thermal (stress independent) part of the free energy.  In contrast to the LP model, the third- and fourth-order elastic compliances have been dropped from Equation \eqref{eq:gibbs1}. Additional relations include
\begin{align}
\mat{S}(\eta) &= \mat{S} + (\mat{S}_1 - \mat{S}_0) \phi_{S}(\eta)\,, \nonumber \\
\mat\epsilon^T &= \mat\epsilon^{T0} + (\mat\epsilon^{T1} - \mat\epsilon^{T0}) \phi_T(\eta) \, , \nonumber \\
\phi(\eta) &= a\eta^2 + (4-2a)\eta^3 + (a-3)\eta^4\, ,
\qquad 0<a<6\, .
\label{eq:gibbs2}
\end{align}
The functions $\phi_{S}(\eta)$ and $\phi_T(\eta)$ are of the same form as $\phi(\eta)$ but with parameters $a_{S}$ and $a_T$. The thermal part of the free energy is
\begin{equation}
f(T,\eta) = A(T)\eta^2 + [4\Delta G(T) - 2A(T)]\eta^3 + [A(T)-3\Delta G(T)]\eta^4 + f(T,0)\, ,
\label{eq:gibbs3}
\end{equation}
where
\begin{equation}
\Delta G(T) = f(T,1)-f(T,0) 
\label{eq:gibbs4}
\end{equation}
and $A(T)$ is a temperature dependent coefficient.
$\eta=0$ and $\eta=1$ indicate phases A and M, respectively.

In the model constructed herein we reduce the number of parameters by taking $\phi_{S}(\eta)=\phi_T(\eta)=\phi(\eta)$, that is, we assume $a_{S}=a_T=a$.
Our goal is to derive a model with the smallest set of model parameters necessary to capture the essential physics.
As mentioned above, we generalize the LP model from a neighborhood of $P=0$ to arbitrary $P$. This entails, in particular, generalizing $f(T,\eta)$ to $f(P, T, \eta)$, and therefore replacing Equation \eqref{eq:gibbs4} by
\begin{equation}
\Delta G(P,T) = f(P,T,1)-f(P,T,0) 
\label{eq:gibbs5}
\end{equation}
and $A(T)$ by $A(P,T)$.
Throughout this paper both $G$ and $\sigma$ are in units of GPa.

\subsubsection{Transformation strain and work}
\label{sec:trans_strain}

The transformation strain tensor $\epsilon^t$ may be decomposed into deviatoric and diagonal terms
\begin{equation}
\epsilon^t_{ij} = e^t_{ij} + \frac13 \epsilon^t_{kk}\delta_{ij}\, .
\label{eq:tstrain1}
\end{equation}
The $e^t_{ij}$ are the transformation strain deviators which are traceless: $e^t_{ii} = 0$. From the general differential relation $d \epsilon_{kk} = d\ln V$ it follows that 
\begin{equation}
\epsilon^t_{kk} = \ln \left(\frac{V_f}{V_i}\right)\, ,
\label{eq:tstrain2}
\end{equation}
where $V_i$ ($V_f$) is the initial (final) specific volume of the transforming material. Equations \eqref{eq:tstrain1} and \eqref{eq:tstrain2} imply that the transformation strain deviator quantifies the structural change at constant volume. As in the LP model, $\mat\epsilon^t$ is taken to be independent of $T$ --- it is defined at a reference temperature, here the A-M equilibrium temperature, $T_e(P)$, and the $T$ dependence is taken into account by thermal expansion. The transformation strain tensor for the A$\rightarrow$M phase transformation may be written
\begin{equation}
\epsilon^t_{ij}(P) = e^t_{ij}(P) + \frac{1}{3} \ln \left[\frac{\rho_0(P,T_e)}{\rho_1(P,T_e)}\right] \delta_{ij}\, ,
\label{eq:tstrain3}
\end{equation}
where $\rho_{0,1}(P,T)$ are the equations of state of A and M.

As an example, consider the transformation strain for the cubic-tetragonal phase transformation (PT) in NiAl. For one of its three martensitic variants (\cite{Levitas2:2002}; Equation (48))
\begin{equation}
\mat\epsilon^{t}(P=0) = \text{diag}\!\left\{0.215;-0.078;-0.078\right\} \, .
\label{eq:tstrain4}
\end{equation}
One readily obtains
$e^t_{11} = 0.195$, $e^t_{22}=e^t_{33}=-0.0977$, $\epsilon^t_{kk} = 0.059$.
Equation \eqref{eq:tstrain3} gives $\rho_0 = 1.061\rho_1$.

As is well known, the stress tensor may be decomposed into deviatoric and isotropic terms
\begin{equation}
\sigma_{ij} = s_{ij} -P\delta_{ij} \, ,
\label{eq:tstrain6}
\end{equation}
where the $s_{ij}$ are the traceless stress deviators.
Equations \eqref{eq:tstrain3} and \eqref{eq:tstrain6} yield the $P$-dependent transformation work
\begin{equation}
\mat\sigma:\mat\epsilon^t = s_{ij}e^t_{ji}(P) - P \ln \left\{ \frac{\rho_0\left[P,T_e(P)\right]}{\rho_1\left[P,T_e(P)\right]}\right\} \, .
\label{eq:twork1}
\end{equation}

\subsubsection{Thermal expansion}
\label{sec:texp}

The thermal strain tensor is given by
\begin{equation}
\epsilon^T_{ij} = \alpha_{ij}\left[T-T_e(P)\right]\, 
\label{eq:texp1}
\end{equation}
where $T_e$ is the chosen reference temperature. The $\alpha_{ij}$ are the thermal expansion tensors, which are of order the linear thermal expansion coefficient. In general, $\alpha_{ij} \sim 10^{-5}$K$^{-1}$, thus for $T-T_e = 10^3$K we have $\epsilon^T_{ij}\sim0.01$. The components of the transformation strain tensor are of order 0.1; see Equation \eqref{eq:tstrain4}. Since the thermal expansion work is roughly an order of magnitude smaller than the transformation work, we henceforth neglect it in the Gibbs free energy.

\subsubsection{Elastic energy}
\label{sec:Eener}

For the elastic energy term in the Gibbs free energy, we need the compliance tensor which in the isotropic limit is \citep{Wallace:1972}
\begin{equation}
S_{ijkl} = \left(\frac{1}{9B} -
\frac{1}{6\mu}\right) \delta_{ij}\delta_{kl} + \frac{1}{4\mu}\left(\delta_{ik}\delta_{jl} + \delta_{il}\delta_{jk}\right)\, 
\label{eq:Eener1}
\end{equation}
where $B$ and $\mu$ are the pressure- and temperature-dependent bulk and shear moduli, respectively. 

Equations \eqref{eq:tstrain6} and \eqref{eq:Eener1} result in
\begin{align}
\mat\sigma:\mat{S}:\mat\sigma/2 &= \frac{1}{2}\left[s_{ij}S_{ijkl}s_{kl} - 2P s_{ij}S_{ijkk} + P^2S_{iijj}\right] \, \nonumber \\
&= \frac{P^2}{2B} + \frac{s_{ij}s_{ji}}{4\mu} \, .
\label{eq:Eener2}
\end{align}
The maximum value of $s_{ij}s_{ji}$ is limited by plastic flow: $s_{ij}s_{ji} \leq 2\sigma^2/3$ where $\sigma$ is the flow stress, hence $|s_{ij}| \lesssim \sqrt{2/3}\sigma$. We now compare $|e^t_{ij}|$ to $|s_{ij}|/4\mu$. In general, $|e^t_{ij}| \sim 0.1$. For copper at $P=0, T\sim300$K, and plastic strain rates of order 10$^4$s$^{-1}$, $\sigma\approx500$MPa and $\mu\approx50$GPa, therefore $|s_{ij}|/4\mu \lesssim \sigma/2\sqrt{6}\mu \approx 0.002$. We conclude that the $s_{ij}s_{ij}$ term in Equation \eqref{eq:Eener2} is negligible in comparison to the $s_{ij}e^t_{ij}$ term in Equation \eqref{eq:twork1}. Metals with higher flow stresses, e.g., V and Ta, also have larger shear moduli, so the ratio $\sigma/\mu$ is still of order 0.01. 

Equation \eqref{eq:gibbs2} and $\mat\sigma:\mat{S}_{0,1}:\mat\sigma \approx P^2/2B_{0,1}$, where $B_{0,1} \equiv B_{0,1}(P,T)$, gives our elastic energy term
\begin{equation}
-\frac{P^2}{2}\left[\frac{1}{B_0} + \left(\frac{1}{B_1} - \frac{1}{B_0} \right)\phi(\eta)\right]\, .
\label{eq:Eener3}
\end{equation}

\subsubsection[\texorpdfstring{$G(s_{ij}, P, T, \eta)$}{G(sij,P,T,eta)}]{$\boldsymbol{G(s_{ij}, P, T, \eta)}$}
\label{sec:G}

Combining Equations \eqref{eq:gibbs1}, \eqref{eq:twork1} and \eqref{eq:Eener3}, and neglecting the thermal expansion term, we obtain
\begin{equation}
G(s_{ij}, P, T, \eta) = -\frac{P^2}{2}\left[\frac{1}{B_0}+\left(\frac{1}{B_1}-\frac{1}{B_0}\right)\phi(\eta)\right]-s_{ij}e^t_{ji}(P)\phi(\eta) + P\ln\left(\frac{\rho_0}{\rho_1}\right)\phi(\eta)+f(P,T,\eta)\, .
\label{eq:G1}
\end{equation}
The Gibbs free energies of A (where $\eta=0$) and M (where $\eta=1$) are given by
\begin{align}
\mbox{A:}\qquad G_0(P,T) &\equiv G(s_{ij}, P,T,0) = -\frac{P^2}{2B_0} + f(P,T,0) \, , \nonumber \\
\mbox{M:}\qquad G_1(P,T) &\equiv G(s_{ij}, P,T,1) = -\frac{P^2}{2B_1} - s_{ij}e^t_{ij}(P) + P\ln\left(\frac{\rho_0}{\rho_1}\right)+ f(P,T,1) \, .
\label{eq:G2}
\end{align}
Combining Equations \eqref{eq:G1}, \eqref{eq:G2}, and \eqref{eq:gibbs3} yields
\begin{equation}
G(s_{ij}, P, T, \eta) = -\left[\mat\sigma:\mat\epsilon^t + \left(B^{-1}_1 - B^{-1}_0\right)P^2/2\right]\phi(\eta) + G_0(P,T) + A\eta^2 + \left(4\Delta G-2A\right)\eta^3 + \left(A-3\Delta G\right)\eta^4 \, ,
\label{eq:G3}
\end{equation}
where $B_{0,1}\equiv B_{0,1}(P,T)$, $A\equiv A(P,T)$, and $\Delta G$ and $\mat\sigma:\mat\epsilon^t$ are given by Equations \eqref{eq:gibbs5} and \eqref{eq:twork1}. The Gibbs energy, Equation \eqref{eq:G3}, is the basis of the present model. To render subsequent expressions in simpler form we define
\begin{equation}
W_{t\lambda} = \mat\sigma:\mat\epsilon^t + \left(B^{-1}_1 - B^{-1}_0 \right)P^2/2 \, ,
\label{eq:G4}
\end{equation}
which is the sum of the transformation and elastic work terms for A$\rightarrow$M.

Next we obtain the conditions for the A$\rightarrow$M and M$\rightarrow$A phase transformations from $\partial^2G/\partial\eta^2 \leq 0$ at $\eta=0$ and $\eta=1$, respectively. From Equation \eqref{eq:G3} we get
\begin{equation}
\frac{\partial^2 G}{\partial \eta^2} = -W_{t\lambda}\left\{2a + 12\eta\left[2-a+(a-3)\eta\right]\right\} + 2A + 12\eta\left[2\Delta G -A +\left(A-3\Delta G\right)\eta\right]\, .
\label{eq:G5}
\end{equation}
The instability conditions are
\begin{align}
\mbox{A}\rightarrow \mbox{M}:\qquad W_{t\lambda} &\geq \frac{A(P,T)}{a} \, , \nonumber \\
\mbox{M}\rightarrow \mbox{A}:\qquad W_{t\lambda} &\leq \frac{6\Delta G(P,T) - A(P,T)}{6-a} \, .
\label{eq:G6}
\end{align}
(These are the same as in \citep{Levitas:2002}, Equation \eqref{eq:Eener1} for $\sigma:\epsilon^t \rightarrow W_{t\lambda}$, $\Delta G^{\theta} \rightarrow \Delta G (P,T)$, and $A\rightarrow A(P,T)$.)

Equality of the left- and right-hand sides of Equation \eqref{eq:G6} yields equations for the A$\rightarrow$M and M$\rightarrow$A spinodals, that is, the limits of metastability
\begin{align}
\mbox{A}\rightarrow \mbox{M}:\qquad& A(P,T) -a W_{t\lambda} = 0 \, , \nonumber \\
\mbox{M}\rightarrow \mbox{A}:\qquad& 6\Delta G(P,T) - A(P,T) -(6-a) W_{t\lambda} = 0 \, .
\label{eq:G7}
\end{align}
The coexistence curve (equilibrium phase boundary) is the curve in the $P-T$ plane satisfying
\begin{equation}
G(s_{ij}, P, T, 1) = G(s_{ij}, P, T, 0) \, ,
\label{eq:G8}
\end{equation}
which reduces to
\begin{equation}
\Delta G(P, T) = W_{t\lambda} \, .
\label{eq:G9}
\end{equation}
Thus, the elastic work and the transformation work, in particular the stress deviators, shift the A-M equilibrium boundary. We shall denote the solutions of Equations \eqref{eq:G7} and \eqref{eq:G9} for $T$ as a function of $P$ as follows:
\begin{align}
\mbox{A}\rightarrow \mbox{M spinodal}:\qquad& T_c(P) \, , \nonumber \\
\mbox{M}\rightarrow \mbox{A spinodal}:\qquad& \bar{T}_c(P) \, , \nonumber \\
\mbox{coexistence curve}:\qquad& T_e(P) \, .
\label{eq:G10}
\end{align}

\begin{figure}[pos=!htb]
\centering
\includegraphics[width=0.65\textwidth]{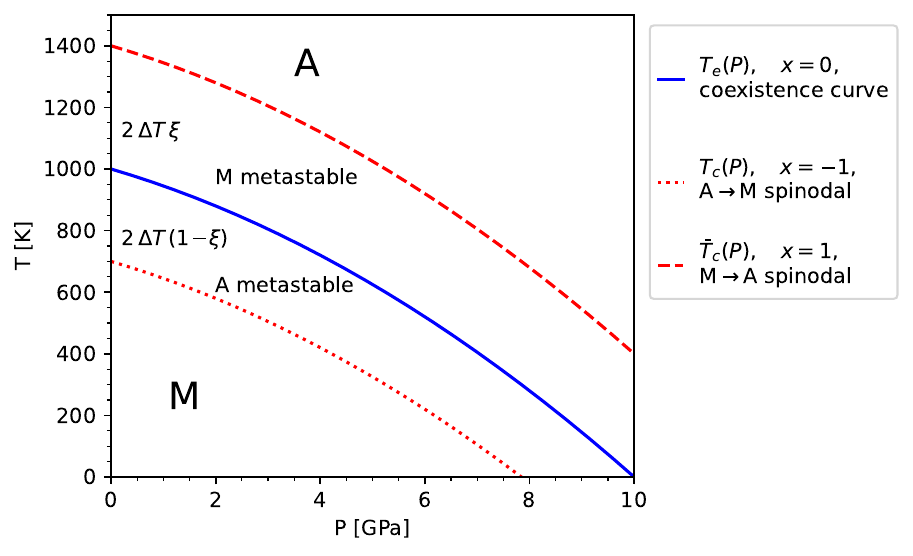}
\caption{Coexistence curve and spinodals.
Parameters $\Delta T$ and $\xi$ can be estimated using Eq. \eqref{eq:G12} and $x$ is defined in Eq. \eqref{eq:G18}.}
\label{fig:coexistspinodal}
\end{figure}

We now assume that the spinodals are well approximated as translations of the coexistence curve along the $T$ axis in the $T-P$ plane. Two parameters, $\Delta T$ and $\xi$, control the separation of the spinodals and their shifts relative to the coexistence curve as illustrated in Figure \ref{fig:coexistspinodal}.
Specifically,
\begin{align}
\bar{T}_c (P) &= T_e (P) + 2\xi \Delta T \, , \nonumber \\
T_c(P) &= T_e(P) - 2(1-\xi)\Delta T \, ,
\label{eq:G11}
\end{align}
where $0 < \xi < 1$.
A generalization to include pressure dependence in $\Delta T$ and/or $\xi$ could be considered in future work if new experimental data becomes available that requires such a generalization.
If the spinodals have been determined experimentally or calculated on $\left[P_1, P_2\right]$ and their deviations from $T_e(P) \,\pm$ constants are modest, then our model can be used to estimate the PT kinetics by using the approximations
\begin{align}
\Delta T &\approx \frac{1}{2} (P_2 - P_1)^{-1} \int^{P_2}_{P_1} \left[\bar{T}_c(P) - T_c(P)\right] dP\, , \nonumber \\
\xi &\approx \frac{\int^{P_2}_{P_1} \left[\bar{T}_c(P) - T_e(P)\right] dP}{\int^{P_2}_{P_1} \left[\bar{T}_c(P) - T_c(P)\right] dP} \, ,
\label{eq:G12}
\end{align}
i.e. experimental (or simulated) data within this pressure interval can be used to determine approximations to the model parameters $\Delta T$ and $\xi$.

We now expand $A(P,T)$ around the A$\rightarrow$M spinodal $T_c(P)$
\begin{equation}
A(P,T) = A(P,T_c(P)) + A'_T(P) \left[T-T_c(P)\right]+\ldots \, ,
\label{eq:G13}
\end{equation}
where
\begin{align}
A(P,T_c(P)) &= aW_{t\lambda} \,, &
A'_T(P) &= \frac{\partial A(P,T)}{\partial T}\bigg|_{T_c(P)} \, .
\label{eq:G14}
\end{align}
Similarly, $\Delta G(P, T)$ is expanded around the coexistence curve
\begin{equation}
\Delta G(P,T) = \Delta G(P,T_e(P)) + \Delta G'_T(P)\left[T-T_e(P)\right]+\ldots \, ,
\label{eq:G15}
\end{equation}
where
\begin{align}
\Delta G(P,T_e(P)) &= W_{t\lambda} \,, &
\Delta G'_T(P) &= \frac{\partial G(P,T)}{\partial T}\bigg|_{T_e(P)} \, .
\label{eq:G16}
\end{align}
$\Delta G = G_1 - G_0$ is positive and increasing (negative and decreasing) for $T-T_e$ positive and increasing (negative and decreasing), hence $\Delta G'_T > 0$. Substituting Equation \eqref{eq:G13} and Equation \eqref{eq:G15} with $T=\bar{T}_c(P)$ into Equation \eqref{eq:G7} for the M$\rightarrow$A spinodal yields
\begin{equation}
\frac{A'_T(P)}{6 \Delta G'_T(P)} = \frac{\bar{T}_c - T_e}{\bar{T}_c - T_c} = \xi \, ,
\label{eq:G17}
\end{equation}
a relation that will be used below.
In general $\xi$ is a function of $P$, but we take it to be a constant.

We now introduce a variable, $x$, that increases from 0 to 1 as $T$ increases from $T_e(P)$ to the M$\rightarrow$A spinodal, $\bar{T}_c(P)$, and decreases from 0 to -1 as $T$ is decreased from $T_e(P)$ to the A$\rightarrow$M spinodal, $T_c(P)$. Its definition is 
\begin{align}
x &= \frac{T-T_e}{2\Delta T \left[ \xi \theta(T-T_e) + (1-\xi) \theta(T_e-T)\right]} 
= \begin{cases}
 \frac{T-T_e}{2\xi\Delta T} = \frac{T-T_e}{\bar{T}_c - T_e} \, , \qquad &T\geq T_e \, , \\[1.5ex]
 \frac{T-T_e}{2(1-\xi)\Delta T} = \frac{T-T_e}{T_e-T_c} \, , \qquad &T \leq T_e \, .
 \end{cases}
\label{eq:G18}
\end{align}

In the A phase region (includes stable A and metastable M) we have
\begin{align}
A(P,T) &= aW_{t\lambda} + 12\Delta G'_T\Delta T \xi \left[1-\xi(1-x)\right] \, , \nonumber \\
\Delta G(P,T) &= W_{t\lambda} + 2 \Delta G'_T \Delta T \xi x \, , \qquad 0\leq x \leq 1 \, ,
\label{eq:G19}
\end{align}
where we used Equation \eqref{eq:G17}.
For the M phase region (includes stable M and metastable A) we find
\begin{align}
A(P,T) &= aW_{t\lambda} + 12\Delta G'_T\Delta T \xi (1-\xi)(1+x) \, , \nonumber \\
\Delta G(P,T) &= W_{t\lambda} + 2 \Delta G'_T \Delta T (1-\xi) x \,, \qquad -1\leq x \leq 0 \,.
\label{eq:G20}
\end{align}

\subsection{Speeds of A-M interfaces}
\label{sec:struct}

In this section we obtain an analytic solution for the profiles and speeds of diffuse A-M interfaces as a function of temperature and stress tensor. We incorporate an athermal threshold that accounts for the stress fields due to crystal defects, the Peierls barrier, and tilt and twin boundaries. 

We begin by expressing the Gibbs potential in terms of
\begin{align}
s_1 &= A-aW_{t\lambda} \,, &
s_2 &= 12 \left(\Delta G - W_{t\lambda}\right) \, .
\label{eq:struct1}
\end{align}
The result is
\begin{equation}
G(s_{ij}, P, T, \eta) = G_0(P,T) + s_1 \eta^2 \left[ 1-\left(6-\mathcal{P}\right)\eta/3 + \left(4-\mathcal{P}\right)\eta^2/4\right]\, ,
\label{eq:struct2}
\end{equation}
where $\mathcal{P} = s_2/s_1$. This result coincides with Equation (41) in \cite{Levitas:2003}.

The speeds of the A--M interfaces, as well as the structures (profiles in $\eta$) of those interfaces (next section) are obtained from the time-dependent Ginzburg-Landau (TDGL) equation
\begin{equation}
\frac{\partial \eta}{\partial t} = - \kappa \left(\frac{\partial G}{\partial \eta} - 2 \beta \frac{\partial^2 \eta}{\partial x^2} \right)\, ,
\label{eq:TDGL}
\end{equation}
where $\kappa > 0$ and $\beta > 0$ are the kinetic and gradient energy coefficients with dimensions of volume/energy-time and energy/length, respectively. Note that $G_0(P,T)$ makes no contribution, hence the interface speed is given by Equation (13) in \cite{Levitas:2010}
\begin{equation}
c = \frac{\kappa s_2 \beta^{1/2}}{\sqrt{4s_1 - s_2}}\, ,
\label{eq:struct3}
\end{equation}
where $s_1$ and $s_2$ are defined in Equation \eqref{eq:struct1} and inertial effects have been neglected.
In the austenite phase region $s_2 >0$, thus $c>0$. In contrast, in the martensite phase region we have $s_2<0$, hence $c<0$. In summary, the interface speed is positive (negative) for interface motion directed from A toward M (from M toward A).

Following \cite{Levitas:2010}, we incorporate an athermal threshold for interface propagation. Experimental results and sharp interface theory indicate that the interface speed depends on the excess of the thermodynamic driving force over the athermal threshold, $K$. For M$\rightarrow$A, i.e., growing A nuclei, the driving force is $s_2/12$ and $c>0$ is a function of $s_2/12 - K$. The driving force for A$\rightarrow$M is $-s_2/12$, thus the interface propagates provided $-s_2/12 >K$, or equivalently $s_2/12 + K <0; \, c<0$ is a function of $s_2/12 + K$. For $s_2$ positive (negative) the threshold $K$ is subtracted from (added to) $s_2/12: \, s_2/12 \rightarrow s_2/12 - \text{sign}(s_2)K = \Delta G - W_{t\lambda} - \text{sign}(s_2)K$ which is equivalent to 
\begin{equation}
W_{t\lambda} \rightarrow W_{t\lambda} + \text{sign}(s_2)K \, .
\label{eq:struct4}
\end{equation}
The athermal threshold is taken into account by making the replacement Equation \eqref{eq:struct4} in $s_1$ and $s_2$
\begin{align}
s_1 &\rightarrow s_1 - a \,\text{sign}(s_2) K \,, &
s_2 &\rightarrow s_2 \left(1-\frac{K}{|s_2|/12}\right) \, ,
\label{eq:struct5}
\end{align}
and then making the replacements Equation \eqref{eq:struct5} in Equation \eqref{eq:struct3} to obtain
\begin{equation}
c=\frac{\kappa \beta^{1/2} s_2 \,\mathcal{R}\!\left(1-12 K/|s_2|\right)}{\left[4s_1 - s_2 + 4 \text{sign}(s_2) \left(3-a\right)K\right]^{1/2}} \, ,
\label{eq:struct6}
\end{equation}
where $\mathcal{R}(z)=z\theta(z)$ is the ramp function that gives $c=0$ if $|s_2|/12<K$.

Next we derive an expression for the interface speed in terms of the variable $x$ defined in Equation \eqref{eq:G18}. We use the expansions, Equations \eqref{eq:G13} and \eqref{eq:G15}, for $A(P,T)$ and $\Delta G(P,T)$. We first obtain two results for $c$, one for $x\geq0$ and the other for $x\leq0$, and then merge them into a single expression for $c$.

Equation \eqref{eq:G19}, \eqref{eq:G20}, and \eqref{eq:struct1} yield the following expressions for $s_1$ and $s_2$ in the A $(x>0)$ and M $(x<0)$ phase regions
\begin{align}
s_1 &= 12\Delta G'_T \Delta T \cdot 
\begin{cases} 
\xi \left[1-\xi(1-x)\right] \, , \qquad &x\geq0 \, , \\
\xi (1-\xi)(1+x) \, , \qquad &x\leq0 \, ,
\end{cases} \nonumber \\
 s_2 &= 24 \Delta G'_T \Delta T \cdot
 \begin{cases}
\xi x \, , \qquad &x\geq0\, ,\\
(1-\xi)x\,, \qquad &x\leq0 \, .
 \end{cases}
\label{eq:struct7}
\end{align}
Since $\Delta G'_T > 0$ and $0<\xi<1$ it follows that $s_1 > 0$ for $-1\leq x \leq 1$ and $\text{sign}(s_2) = \text{sign}(x)$.

Equations \eqref{eq:struct6} and \eqref{eq:struct7} give
\begin{equation}
c = 6\kappa \beta^{1/2} g \cdot 
\begin{cases} 
\frac{\xi \mathcal{R}\left(1-K/g\xi|x|\right)x}{\left\{3g\xi\left[1+(1-2\xi)(1-|x|)\right]+ \text{sign}(x)(3-a)K\right\}^{1/2}} \, , \qquad &x>0 \, , \\[1.5ex]
\frac{(1-\xi)\mathcal{R}\left(1-K/g (1-\xi)|x|\right)x}{\left\{3g(1-\xi) \left[1+(2\xi-1)(1-|x|)\right]+ \text{sign}(x)(3-a)K\right\}^{1/2}} \, , \qquad &x<0 \, ,
\end{cases} 
\label{eq:struct8}
\end{equation}
where $g=2\Delta G'_T \Delta T$. This result can be written as the following single expression
\begin{equation}
c = 2\kappa  \frac{\left[3\beta g \Psi(\xi,x)\right]^{1/2} \mathcal{R}\left(1 - K/g \Psi(\xi,x) |x|\right) x}{\left\{1+ \text{sign}(x) \left[(1-2\xi)(1-|x|) + (1-a/3) K/g \Psi(\xi,x)\right]\right\}^{1/2}} \, ,
\label{eq:struct9}
\end{equation}
where $\Psi(\xi,x) = \xi \theta(x)+(1-\xi)\theta(-x)$.

For $K \neq 0$ the interface speed is zero for $|x|\leq K/g \Psi$, that is, for
\begin{equation}
 -\frac{K}{g(1-\xi)} \leq x \leq \frac{K}{g\xi}\, .
\label{eq:struct10}
\end{equation}
Thus, an athermal threshold eliminates phase transitions in a band around the coexistence curve. If $K\geq g\Psi$ then $c=0$ throughout the regions of metastability, i.e., there is no growth of nuclei. We assume $K < g\Psi$. The denominator in Equation \eqref{eq:struct9} is real and non-zero for $K/g\Psi <|x| <1$ provided $a<6$.

In the absence of an athermal threshold
\begin{equation}
c = 2\kappa \sqrt{3\beta g} \cdot 
\begin{cases} 
\left[\frac{\xi}{1+(1-2\xi)(1-x)}\right]^{1/2} x \, , \qquad x\geq0 \, , \\
\left[\frac{1-\xi}{1+(2\xi-1)(1+x)}\right]^{1/2} x \, , \qquad x\leq0 \, .
\end{cases} 
\label{eq:struct11}
\end{equation}
The interface speed is clearly a continuous function of $x$, but its derivative with respect to $x$ is not continuous unless $\xi = 1/2$
\begin{equation}
\frac{dc}{dx}\bigg|_0 = 2\kappa \sqrt{3\beta \Delta G'_T \Delta T} \cdot 
\begin{cases} 
\sqrt{\frac{\xi}{1-\xi}} \, , \qquad x\rightarrow0^+ \, , \\
\sqrt{\frac{1-\xi}{\xi}} \, , \qquad x\rightarrow0^- \, .
\end{cases} 
\label{eq:struct12}
\end{equation}

As the surface of a product-phase nucleus expands it intersects other growing nuclei. As the volume fraction, $\lambda_V$ of the product phase increases the mean or effective interface speed decreases to zero as $\lambda_V \rightarrow 1$. Also, as $\lambda_V \rightarrow 1$ the volume available for product-phase nucleation decreases hence the effective nucleation rate, $\dot{N}$, decreases. These volume-fraction-induced decreases in $c$ and $\dot{N}$ are accounted for in KJMA kinetic theory \citep{Kolmogorov:1937,Johnson:1939,Avrami:1939, Avrami:1940,Avrami:1941,Bruna:2006}, which we employ in this model; see Section \ref{sec:kjma}.

\subsection{Propagating Interface Profiles}
\label{sec:interfaceprofiles}

For $G$ given in Equation \eqref{eq:struct2} we obtain
\begin{align}
\frac{\partial G}{\partial \eta} &= s_1\left[2\eta - (6-\mathcal{P})\eta^2 + (4-\mathcal{P})\eta^3\right]
\nonumber\\
&= (4s_1-s_2)\eta(\eta-1)\left(\eta-\frac{2s_1}{4s_1-s_2}\right)
\,.\label{eq:dGdeta}
\end{align}
Using Equation \eqref{eq:dGdeta} the TDGL equation \eqref{eq:TDGL} can be written
\begin{align}
\frac1{2\kappa\beta}\frac{\partial\eta}{\partial t} &= \frac{\partial^2\eta}{\partial x^2} - \frac{s_1}{\alpha\beta}\eta(\eta-1)(\eta-\alpha)
\,, \label{eq:TDGLrewrite}
\end{align}
where
\begin{equation}
\alpha = 2s_1 / (4s_1-s_2)\,.
\end{equation}
In order to solve \eqref{eq:TDGLrewrite} we approximate the stresses, hence $s_1$ and $s_2$, as constants.
This is tantamount to the neglect of inertial (mass density) effects.
However, as discussed in \cite{Levitas:2010}, inertial effects can be approximately taken into account by replacing the constant stress tensor by the average of the stress tensors in the A and M.

For interfaces propagating at a constant speed, $c$, the order parameter is a function of $x-ct\equiv z$.
The TDGL equation \eqref{eq:TDGLrewrite} then assumes the form
\begin{equation}
\frac{d^2\eta}{dz^2}+\frac{c}{2\kappa\beta}\frac{d\eta}{dz} - \frac{s_1}{\alpha\beta}\eta(\eta-1)(\eta-\alpha)=0
\,. \label{eq:detadz2}
\end{equation}
This is a nonlinear autonomous differential equation, that is, the independent variable does not explicitly appear in the equation.
An autonomous equation of order $n$ can always be replaced by a non-autonomous equation of order $n-1$.
For Equation \eqref{eq:detadz2} the procedure is to express $d\eta/dz$ as a function of $\eta$.
\begin{align}
\frac{d\eta}{dz} &= f(\eta)
\,,\nonumber\\
\frac{d^2\eta}{dz^2} &= \frac{d}{dz}f(\eta) = \frac{d\eta}{dz}\frac{df}{d\eta} = f'(\eta)f(\eta)
\,; \label{eq:nonautonomousdiffeq}
\end{align}
see \cite{Bender:1978}.
Equations \eqref{eq:detadz2} and \eqref{eq:nonautonomousdiffeq} result in
\begin{equation}
f'(\eta)f(\eta) + \frac{c}{2\kappa\beta}f(\eta) = \frac{s_1}{\alpha\beta}\eta(\eta-1)(\eta-\alpha)
\,. \label{eq:diffeqforfofeta}
\end{equation}
Since the right hand side is a cubic polynomial with no constant term it is obvious that
\begin{equation}
f(\eta) = f_1\eta + f_2\eta^2
\,, \label{eq:fofetaansatz}
\end{equation}
where $f_{1,2}$ are constants to be determined.
Substituting Equation \eqref{eq:fofetaansatz} in Equation \eqref{eq:diffeqforfofeta} gives us
\begin{align}
\left[f_1^2+\frac{c}{2\kappa\beta}f_1-\frac{s_1}{\beta}\right]\eta + \left[3f_1f_2+\frac{c}{2\kappa\beta}f_2+\frac1\beta\left(3s_1-\frac{s_2}{2}\right)\right]\eta^2
+\left(2f_2^2-\frac{s_1}{\alpha\beta}\right)\eta^3 = 0
\,.
\end{align}
Equating the three coefficients to zero yields three equations for the three unknowns, namely $f_1$, $f_2$, and the interface speed, $c$.
The solution is
\begin{align}
f_1&=\frac12\sqrt{\frac{4s_1-s_2}{\beta}}\,, &
f_2&=-f_1\,, &
c&=\kappa s_2\sqrt{\frac{\beta}{4s_1-s_2}}
\,. \label{eq:coeffsfic}
\end{align}
Equations \eqref{eq:nonautonomousdiffeq}, \eqref{eq:fofetaansatz}, and \eqref{eq:coeffsfic} give
\begin{equation}
\frac{d\eta}{dz}=f(\eta)=\eta(1-\eta)f_1\,,
\end{equation}
hence
\begin{equation}
f_1 z = f_1\cdot (x-ct) = \int\frac{d\eta}{\eta(1-\eta)} = \ln\left(\frac{\eta}{1-\eta}\right) \, .
\end{equation}
Solving for $\eta$
\begin{equation}
\eta(x,t) = \left\{1 + \exp\left[-\frac12\sqrt{\frac{4s_1-s_2}{\beta}}(x-ct)\right]\right\}^{-1}
\,. \label{eq:etasolution}
\end{equation}
Since $\eta\sim0$ as $x\to-\infty$, $\eta\sim1$ as $x\to+\infty$, and $\eta(0,0)=1/2$, we define the interface width, $\Delta$, by
\begin{equation}
\frac{d\eta(x,0)}{dx}\Bigg|_0 = \frac{1}{\Delta}
\,,
\end{equation}
which gives
\begin{equation}
\Delta = 8\sqrt{\frac{\beta}{4s_1-s_2}}
\,. \label{eq:Deltasol}
\end{equation}
The effect of an athermal threshold can be incorporated by making the replacements \eqref{eq:struct5} in \eqref{eq:etasolution} and \eqref{eq:Deltasol}.

\subsection{Three-dimensional critical nuclei energies}
\label{sec:criticalnuclei3D}

In this section we obtain approximations for the three-dimensional M and A critical nuclei energies based on the one-dimensional nuclei energies $E^\text{M,A}$ given in Equations \eqref{eq:critnucenergyM} and \eqref{eq:critnucenergyA} in Appendix \ref{sec:criticalnuclei1D}.

\subsubsection{Widths of martensitic critical nuclei}

The width of a M nucleus is defined as
\begin{equation}
W_M = \frac{1}{\eta^M(0)}\int^\infty_{-\infty}\eta^M(z) dz
\,,
\end{equation}
where $\eta^M(z)$ is given in Equation \eqref{eq:etaM}.
We obtain
\begin{equation}
W_M(x,\xi) = 2\sqrt{\frac{\beta}{s_1}} C_M(x,\xi) \ln\left[\frac{C_M(x,\xi)+1}{C_M(x,\xi)-1}\right]
\,, \label{eq:widthM}
\end{equation}
where
\begin{align}
C_M(x,\xi) &= \frac{6 - \mathcal{P}(x,\xi) + \sqrt{\mathcal{P}^2(x,\xi) - 3\mathcal{P}(x,\xi)}}{3\sqrt{4 - \mathcal{P}(x,\xi)}}
\nonumber\\
&= \frac{2\left[-x + 3\xi(1+x)\right] + \sqrt{2x\left[2x - 3\xi(1+x)\right]}}{3\sqrt{2\xi(1+x)\left[2\xi(1+x) - x\right]}}
\,,
\end{align}
\begin{equation}
\mathcal{P}(x,\xi) = \frac{s_2}{s_1} = \frac{2x}{\xi(1+x)}\leq0
\,, \qquad\qquad -1\leq x\leq0 \,.
\end{equation}
The variable $x$ is defined in Equation \eqref{eq:G18} and $s_{1,2}$ were determined in Equation \eqref{eq:struct7}.
On the coexistence curve we have $\mathcal{P}=0$ and $\mathcal{P}\sim-\infty$ as $x\to -1$, which is the A$\longrightarrow$M spinodal (metastability limit).
Since $C_M(0,\xi)=1$ it follows that $W_M$ is logarithmically divergent on the coexistence curve.
As $x\to-1$, $C_M(x,\xi)$ diverges as $(1+x)^{-1/2}$.
However,
\begin{equation}
C_M(-1+\epsilon,\xi)\ln\left[\frac{C_M(-1+\epsilon,\xi)+1}{C_M(-1+\epsilon,\xi)-1}\right] = 2 + \frac34\xi\epsilon + \ldots
\,, \label{eq:critnucenergyM3D}
\end{equation}
therefore $W_M=4\sqrt{\beta/s_1}$ on the A$\longrightarrow$M spinodal $T_c(P)$.

\subsubsection{Widths of austenitic critical nuclei}

Since $\eta\sim1$ as $z\to\pm\infty$ the definition of the width of an A nucleus is
\begin{equation}
W_A = \frac{1}{1-\eta^A(0)}\int^\infty_{-\infty}\left[1-\eta^A(z)\right] dz
\end{equation}
where $\eta^A(z)$ is presented in Equation \eqref{eq:etaA}.
We find
\begin{equation}
W_A(x,\xi) = 2\sqrt{\frac{\beta}{s_1}}\sqrt{\frac{2}{2-\mathcal{P}(x,\xi)}}C_A(x,\xi)\ln\left[\frac{C_A(x,\xi)+1}{C_A(x,\xi)-1}\right]
\,, \label{eq:widthWA3D}
\end{equation}
where
\begin{align}
C_A(x,\xi) &= \frac{4\left[3-\mathcal{P}(x,\xi)\right]+\sqrt{2\mathcal{P}(x,\xi)\left[6 - \mathcal{P}(x,\xi)\right]}}{3\sqrt{2\left[2-\mathcal{P}(x,\xi)\right]\left[4-\mathcal{P}(x,\xi)\right]}}
\nonumber\\
&= \frac{2\left[3-2x-3\xi(1-x)\right] + \sqrt{2x\left[3-x-3\xi(1-x)\right]}}{3\sqrt{2(1-\xi)(1-x)\left[2-x-2\xi(1-x)\right]}}
\,,
\end{align}
\begin{equation}
\mathcal{P}(x,\xi) = \frac{2x}{1-\xi(1-x)}
\,, \qquad\qquad 0\leq x\leq1 \,.
\end{equation}
$\mathcal{P}=0$ on the coexistence curve and $\mathcal{P}=2$ on the M$\longrightarrow$A spinodal.
Like M nuclei, $C_A(0,\xi)=1$ hence $W_A$ is logarithmically divergent on the coexistence curve.
The series expansion of $C_A\ln\left[(C_A+1)/(C_A-1)\right]$ around the M$\longrightarrow$A spinodal is given by Equation \eqref{eq:critnucenergyM3D} with $\xi\to1-\xi$.
Also
\begin{equation}
\sqrt{\frac{2}{2-\mathcal{P}(x,\xi)}} = \frac{1}{\sqrt{1-\xi}}\left[\frac{1}{\sqrt{1-x}} - \frac{\xi}{2}\sqrt{1-x} + \mathcal{O}(1-x)^{3/2}\right]
\,.
\end{equation}
Using Equation \eqref{eq:widthWA3D} we get
\begin{equation}
W_A(x,\xi) = 4\sqrt{\frac{\beta}{(1-\xi)s_1}}\left[\frac{1}{\sqrt{1-x}} + \mathcal{O}(1-x)^{1/2}\right]
\,,
\end{equation}
so $W_A$ diverges on the M$\longrightarrow$A spinodal $\bar{T}_c(P)$.

\begin{figure}[pos=!htb]
\centering
\includegraphics[width=0.5\textwidth,trim=7.pt 10.pt 7.pt 5.pt, clip]{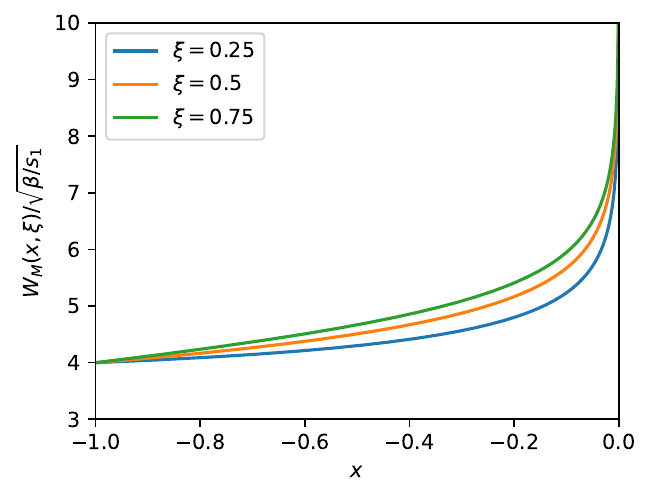}%
\includegraphics[width=0.5\textwidth,trim=7.pt 10.pt 7.pt 5.pt, clip]{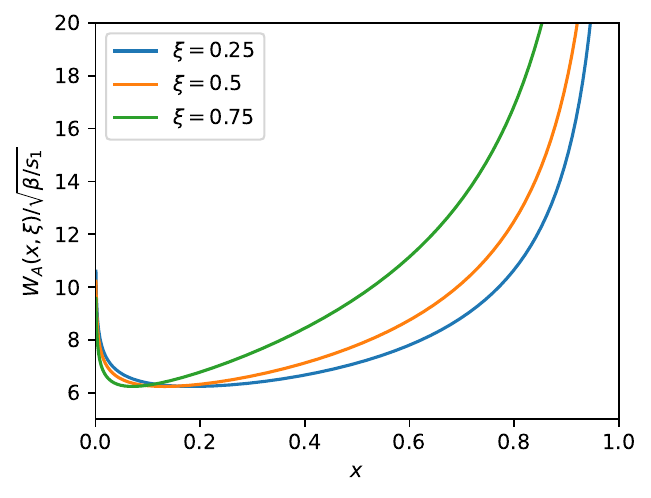}
\caption{We show the widths of martensitic (left) and austenitic (right) critical nuclei as functions of $x$ for different values of parameter $\xi$.
Values $x=-1$ (left) and $x=1$ (right) correspond to the A$\to$M and M$\to$A spinodals, respectively.}
\label{fig:WMA1D}
\end{figure}

The left panel of Figure \ref{fig:WMA1D} is a plot of $W_M(x,\xi)/\sqrt{\beta/s_1}$ for $x<0$ and $\xi=0.25$, 0.5, and 0.75.
The right panel of Figure \ref{fig:WMA1D} shows $W_A/\sqrt{\beta/s_1}$ for $0<x<1$ and $\xi=0.25$, 0.5, and 0.75.

We mention in passing that $C_A(x,\xi)=C_M(-x,1-\xi)$, hence $C_M$ and $C_A$ can be replaced by a single function, namely $C(x,\xi)\equiv C_M(x,\xi)$.
Then $C_M(x,\xi)\to C(x,\xi)$ in Equation \eqref{eq:widthM} for $W_M(x,\xi)$, and $C_A(x,\xi)\to C(-x,1-\xi)$ in Equation \eqref{eq:widthWA3D} for $W_A(x,\xi)$.

\subsubsection{One-dimensional nuclei energies revisited}

Remarkably the arguments of the logarithms in Equations \eqref{eq:critnucenergyM} and \eqref{eq:critnucenergyA} for the one-dimensional nuclei energies $E^M$ and $E^A$ are identical to the arguments of the logarithms in the corresponding expressions for the widths of the nuclei, i.e. Equations \eqref{eq:widthM} and \eqref{eq:widthWA3D}.
Consequently the one-dimensional nuclei energies may be written as follows:
\begin{align}
E^M(x,\xi) &= \frac89\sqrt{\beta s_1}\frac{\mathcal{P}^2-3\mathcal{P}}{(4-\mathcal{P})^2}
\left[1 + \frac{3(4-\mathcal{P})}{\mathcal{P}^2-3\mathcal{P}} - \frac{6-\mathcal{P}}{3\sqrt{4-\mathcal{P}}}\ln\left(\frac{C_M+1}{C_M-1}\right)\right]
\,,\nonumber\\
E^A(x,\xi) &= \frac89\sqrt{\beta s_1}\frac{3\mathcal{P}-\mathcal{P}^2}{(4-\mathcal{P})^2}
\left\{\sqrt{\frac{2-\mathcal{P}}{2}}\left[\frac{3(4-\mathcal{P})}{3\mathcal{P}-\mathcal{P}^2} - 1\right] - \frac{6-\mathcal{P}}{3\sqrt{4-\mathcal{P}}}\ln\left(\frac{C_A+1}{C_A-1}\right)\right\}
\,,
\end{align}
where $\mathcal{P}\equiv\mathcal{P}(x,\xi)$ and $C_{M,A}\equiv C_{M,A}(x,\xi)$.

\subsubsection{Three-dimensional critical nuclei energies}

The one-dimensional (single-coordinate) nuclei energies $E^{M,A}$ are proportional to $\sqrt{\beta s_1}$.
The gradient energy coefficient $\beta$ has dimensions of energy / length and $s_1$ has the dimensions of energy / volume, hence $E^{M,A}\sim\sqrt{\beta s_1}\sim$ energy / area.
In a three-dimensional $x$-$y$-$z$ space the single-coordinate ($z$) M and A nuclei may be envisioned as infinite-planar sheets parallel to the $x$-$y$ plane with thicknesses $W_{M,A}$ along the $z$ axis.
In view of the complexity of solving the time-independent GL equation for the critical nuclei profiles and energies in three dimensions, we approximate the three-dimensional critical nuclei energies $E_c^M$ and $E_c^A$ as equal to the energies of the M and A sheets within squares with sides equal to the thicknesses, which are the effective widths of the corresponding three-dimensional critical nuclei:
\begin{align}
E_c^M(x,\xi) &= E^M(x,\xi)W_M^2(x,\xi)
\,,\nonumber\\
E_c^A(x,\xi) &= E^A(x,\xi)W_A^2(x,\xi)
\,.
\end{align}
The widths $W_M$ and $W_A$ are given in Equations \eqref{eq:widthM} and \eqref{eq:widthWA3D}.
It follows that $E_c^{M,A}\sim \beta^{3/2}s_1^{-1/2}$ with dimensions of energy.

We now obtain estimates of $\beta^{3/2}s_1^{-1/2}$.
In \cite{Levitas:2003}, the parameter $\beta$ is estimated for NiAl for M-M interface widths of 0.3 nm and 1 nm;
the corresponding values of $\beta$ are $2.3\times10^{-11}$ N and $2.6\times10^{-10}$ N, or equivalently, 0.14 eV nm$^{-1}$ and 1.6 eV nm$^{-1}$.
Data on NiAl also indicate that a representative value of $s_1$ is 500 MPa $\approx 3$ eV nm$^{-3}$ \citep{Levitas:2003}.
The corresponding values of $\beta^{3/2}s_1^{-1/2}$ are 0.03 eV and 1.17 eV.


\begin{figure}[pos=!htb]
\centering
\includegraphics[width=0.5\textwidth,trim=7.pt 10.pt 7.pt 5.pt, clip]{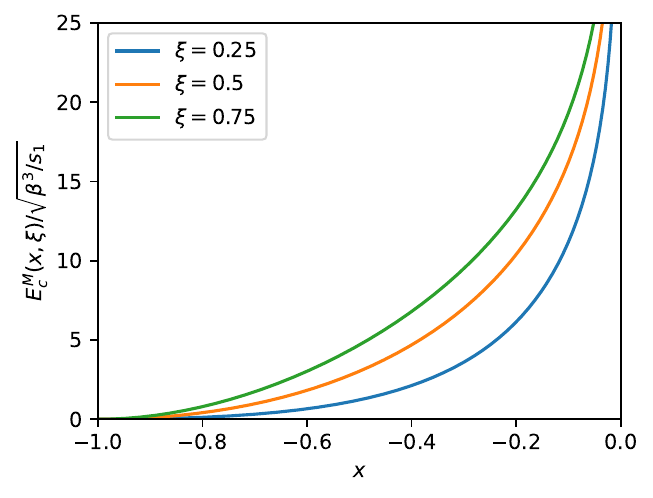}%
\includegraphics[width=0.5\textwidth,trim=7.pt 10.pt 7.pt 5.pt, clip]{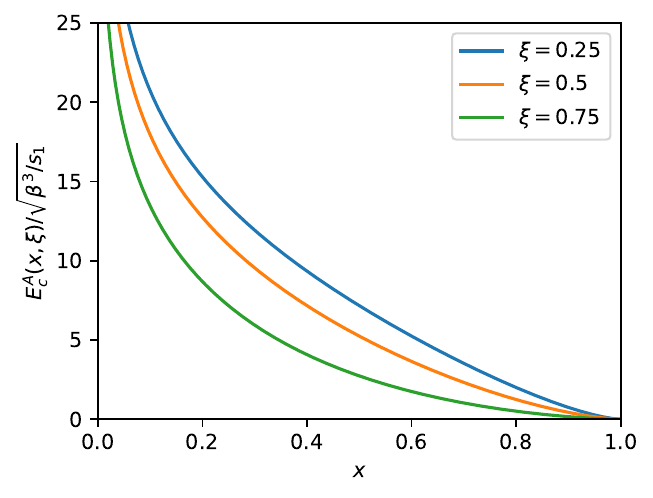}
\caption{We show the martensitic (left) and austenitic (right) critical nuclei energies as functions of $x$ for different values of parameter $\xi$.}
\label{fig:EcMA}
\end{figure}

The left panel of Figure \ref{fig:EcMA} is a plot of $E_c^M(x,\xi) / \sqrt{\beta^3 / s_1}$ for $-1\leq x\leq 0$ and $\xi=0.25$, 0.5, and 0.75.
The right panel of Figure \ref{fig:EcMA} shows $E_c^A(x,\xi) / \sqrt{\beta^3 / s_1}$ for $0\leq x\leq 1$ and $\xi=0.25$, 0.5, and 0.75.

\subsection{Nucleation and rates}
\label{sec:nucleation}

We proceed to determine the nucleation rates due to various types of microstructure after setting the stage and deriving some basic equations for the nucleation of the new phase on each type of microstructure.

\subsubsection{Nucleation at Grain Sites}
\label{sec:grainsites}

We begin by deriving the critical nucleus energy for homogeneous nucleation in the approximation of sharp A-M interfaces;
the energy per unit area of the interface is $\gamma_\text{AM}$.
The energy of a second-phase droplet of radius $r$ is
\begin{equation}
W(r) = 4\pi r^2 \gamma_\text{AM} - \frac43 \pi r^3\Delta G \, ,
\end{equation}
where $\Delta G=G_A-G_M$.
$W(r)$ is an increasing function at small $r$ but decreasing at sufficiently large $r$.
The radius $r_c$ of the critical nucleus is the solution of $dW/dr=0$
\begin{equation}
r_c=\frac{2\gamma_\text{AM}}{\Delta G}
\,. \label{eq:defrc}
\end{equation}
The critical nucleus energy for homogeneous nucleation is
\begin{equation}
\varepsilon_c^\text{hom} = W(r_c) = \frac{16\pi}{3}\frac{\gamma^3_\text{AM}}{\Delta G^2}
\,. \label{eq:epsilonchom}
\end{equation}
\cite{Clemm:1955} calculated the critical energies for nucleation at two-, three-, and four-grain junctions, that is, at grain boundaries, edges, and  corners (or vertices).
In each case they computed the volume and surface area of the nucleus plus the matrix (austenite) grain boundary area that is eliminated in forming the M nucleus.
In all three cases the critical energy is proportional to $\varepsilon_c^\text{hom}$
\begin{align}
\varepsilon_{cd}^\text{grain}(k) &= \varepsilon_c^\text{hom} f_d^\text{grain}(k)
\,, & 
f_d^\text{grain}(k) &= \frac{\left[b_d(k)-2 k a_d(k)\right]^3}{36\pi c_d^2(k)}
\,, & 
k &= \frac{\gamma_\text{AA}}{2\gamma_\text{AM}}
\,,
\end{align}
where $\gamma_\text{AA}$ is the grain boundary energy between two A grains and $d$ is the grain site dimension:
$d=0$ for grain corners, $d=1$ for grain edges, and $d=2$ for grain boundaries.
Clemm and Fisher emphasize that nucleation occurs only for $k$ less than a critical value that we denote $k^c_d$.
The critical values are
\begin{align}
k^c_2 &= 1\,, && \text{grain boundaries (2-grain junctions)}
,\nonumber\\
k^c_1 &= \sqrt{3}/2\,, && \text{edges (3-grain junctions)}
,\nonumber\\
k^c_0 &= \sqrt{2/3}\,, && \text{corners (4-grain junctions)}
\,,
\end{align}
i.e. nucleation at grain boundaries occurs only if $\gamma_\text{AA}<2\gamma_\text{AM}$, at grain edges if $\gamma_\text{AA}\lesssim1.73\gamma_\text{AM}$, and at grain corners only if $\gamma_\text{AA}\lesssim1.63\gamma_\text{AM}$.

\paragraph{Grain boundary nucleation:}\ 
\\
For nucleation at grain boundaries ($d=2$)
\begin{align}
a_2(k) &= \pi(1-k^2)
\,, &
b_2(k) &= 4\pi(1-k)
\,, &
c_2(k) &= \frac{2\pi}{3}\left(2-3k+k^3\right)
\,, \label{eq:GBnuc}
\end{align}
hence
\begin{equation}
f_2^\text{grain}(k) 
=\frac12\left(2 - 3 k + k^3\right)
\,.
\end{equation}
In the interest of model simplicity we approximate this by $(1-k/k^c_2)^n$ and then choose a rational value for $n$ that very nearly minimizes the root mean square (rms) error.
We find that the minimum rms error is 0.009841 for $n=1.6646$;
we use $n=5/3$ for which the rms error is 0.009846.
Thus we have
\begin{equation}
f^\text{grain}_2(k) \approx (1-k/k^c_2)^{5/3} = (1-k)^{5/3}
\qquad\qquad \forall k\in[0,1]
\,.
\end{equation}

\paragraph{Grain edge nucleation:}

\begin{align}
a_1(k) &= 3\beta (1-k^2) - k\sqrt{3-4k^2}
\,, \nonumber\\
b_1(k) &= 12\left(\pi/2 - \alpha - k\beta\right)
\,, \nonumber\\
c_1(k) &= 2\left[\pi - 2\alpha + \frac{k^2}{3}\sqrt{3-4k^2} - \beta k(3-k^2)\right]
\,, \nonumber\\
\alpha &= \sin^{-1}\!\left(\frac{1}{2\sqrt{1-k^2}}\right)
\,, \qquad\qquad
\beta = \cos^{-1}\!\left(\frac{k}{\sqrt{3(1-k^2)}}\right)
\,,
\end{align}
hence
\begin{align}
f_1^\text{grain}(k) 
&= \frac{1}{2 \pi }\left[k^2\sqrt{3-4 k^2} +3k\left( k^2 -3\right) \beta +3 \pi -6 \alpha\right]
\,.
\end{align}
In this case we obtain an approximate expression for $f_1^\text{grain}$ of the form $(1-k/k^c_1)^n$.
The minimum rms error is 0.005939 for $n=2.073$;
we employ $n=2$ for which the rms error is 0.01039.
Therefore
\begin{equation}
f_1^\text{grain}(k) \approx \left(1-k/k_1^c\right)^2 = \left(1-2k/\sqrt{3}\right)^2
\qquad\qquad \forall k\in[0,k^c_1]
\,.
\end{equation}

\paragraph{Grain corner nucleation:}

\begin{align}
a_0(k) &= 3\left\{2\phi(1-k^2) - K\left[(1-k^2-K^2/4)^{1/2} - K/\sqrt{8}\right]\right\}
\,, \nonumber\\
b_0(k) &= 24\left(\pi/3 - k\phi - \delta\right)
\,, \nonumber\\
c_0(k) &= 2\left\{4(\pi/3 - \delta) + k K\left[\left(1-k^2-K^2/4\right)^{1/2} - K/\sqrt{8}\right] - 2k\phi\left(3-k^2\right)\right\}
\,, \nonumber\\
K &= \frac43 \left(\frac32 -2k^2\right)^{1/2} - \frac23 k
\,, \qquad\qquad
\phi = \sin^{-1}\!\left(\frac{K}{2\sqrt{1-k^2}}\right)
\,, \nonumber\\
\delta &= \cos^{-1}\!\left[\frac{\sqrt{2} - k\left(3-K^2\right)^{1/2}}{K\sqrt{1-k^2}}\right]
\,.
\end{align}
For $f_0^\text{grain}(k)$ of the form $(1-k/k^c_0)^n$ the minimum rms error is 0.004820 for $n=2.346$;
we use $n=5/2$ with an rms error of 0.01480.
Hence

\begin{equation}
f_0^\text{grain}(k) \approx \left(1-k/k_0^c\right)^{5/2} = \left(1-k/\sqrt{2/3}\right)^{5/2}
\qquad\qquad \forall k\in[0,k^c_0]
\,.
\end{equation}
To summarize, the $d$-dependent exponents in our approximations for $f_d^\text{grain}$ are $5/3$, 2, and $5/2$ for $d=2$, 1, and 0, respectively. Thus, the $d$ dependence of the exponent is $5/(d/2+2)$.
In this model we employ the approximate expression
\begin{equation}
f_d^\text{grain}(k) = \left(1-k/k_d^c\right)^{\frac{5}{d/2+2}}
\qquad\qquad \forall k\in[0,k^c_d]
\,. \label{eq:fdgrainofk}
\end{equation}
The rms error is $\sim1\%$ for $d=1$ or 2 and $\sim1.5\%$ for $d=0$.

\subsubsection{Nucleation on Dislocations}

Nucleation on dislocations was studied in the $\alpha\to\epsilon$ iron phase transition using molecular dynamics simulations in \cite{Luu:2019}.
\cite{Cahn:1957} calculated the activation energy for second-phase nucleation on dislocations.
A nucleus lies along the core of the dislocation and has a circular cross-section of radius $r(z)$ where $z$ is the distance along the dislocation line.
At large distances from the maximum radius of the nucleus the radius approaches a constant $r_0$.
The corresponding free energy per unit length is comprised of three terms
\begin{equation}
G(r) = -\frac{\mu b_B^2}{4\pi}\tilde\kappa \ln r + 2\pi \gamma_\text{AM}r - \pi \Delta G r^2 + \text{const.}
\label{eq:freeenergy_disloc}
\end{equation}
Here $\mu$ is the shear modulus, $b_B$ is the magnitude of the Burgers vector, $\tilde\kappa=1$ for screw dislocations and $\tilde\kappa=(1-\nu)^{-1}$ for edge dislocations, and $\nu$ is Poisson's ratio.
We shall average over edge and screw components hence $\tilde\kappa\to (1-\nu/2)/(1-\nu)$.
The minimum nucleus radius, $r_0$, minimizes the free energy \eqref{eq:freeenergy_disloc}, i.e. it is the solution of $G'(r_0)=0$
\begin{align}
r_0 &= \frac{\gamma_\text{AM}}{2\Delta G}\left(1-\sqrt{1-\alpha}\right)
\,, &
\alpha &= \frac{\mu b_B^2 \tilde\kappa \Delta G}{2\pi^2\gamma^2_\text{AM}}
\,, \label{eq:r0andalpha}
\end{align}
where $0\leq\alpha\leq1$.
Cahn interprets the material at $r\leq r_0$ as a ``sub-critical metastable cylinder of the second phase''.
The free energy of formation of a nucleus, i.e. the energy of the critical nucleus, $\varepsilon_c^\text{dis}$, follows from \eqref{eq:freeenergy_disloc} by accounting for surface slope
\begin{equation}
2\pi \gamma_\text{AM} r \to 2 \pi \gamma_\text{AM} r \sqrt{1+\left(dr/dz\right)^2}
\,,
\end{equation}
and subtracting $G(r_0)$, therefore
\begin{align}
\varepsilon_c^\text{dis} &= \int_{-\infty}^\infty\left[-\frac{\mu b_B^2}{4\pi}\tilde\kappa\ln\left(\frac{r}{r_0}\right)
+ 2\pi\gamma_\text{AM}\left(r\sqrt{1+r'^2}-r_0\right) - \pi\Delta G\left(r^2-r^2_0\right)\right] dz
\nonumber\\
&\equiv \int_{-\infty}^\infty\mathcal{L}(r,r')dz
\,, \label{eq:integral17}
\end{align}
where $r'=dr/dz$.
The shape and size of the nucleus, $r(z)$, is an extremum of the integral \eqref{eq:integral17}.
Consequently $r(z)$ is a solution of the Euler-Lagrange equation
\begin{equation}
\frac{d}{dz}\frac{\partial \mathcal{L}(r,r')}{\partial r'} - \frac{\partial\mathcal{L}(r,r')}{\partial r} = 0
\label{eq:EulerLagrange}
\end{equation}
subject to the boundary conditions $r'(0)=0$ and $r'\sim0$ as $z\to\pm\infty$.
Since there is no explicit dependence on $z$ in $\mathcal{L}$ --- Equation \eqref{eq:EulerLagrange} is an autonomous differential equation --- we have the identity \cite[pp.24,25]{Weinstock:1952}
\begin{equation}
\frac{d}{dz}\left(r'\frac{\partial \mathcal{L}}{\partial r'} - \mathcal{L}\right) = r'\left[\frac{d}{dz}\left(\frac{\partial \mathcal{L}}{\partial r'}\right) - \frac{\partial \mathcal{L}}{\partial r}\right] \, ,
\label{eq:weinstockidentity}
\end{equation}
which is trivially verified utilizing $\left(\partial \mathcal{L}/\partial z =0\right)$
\begin{equation}
\frac{d \mathcal{L}}{dz} = r' \frac{\partial \mathcal{L}}{\partial r} + r''\frac{\partial \mathcal{L}}{\partial r'}
\,.
\end{equation}
Since the right hand side of \eqref{eq:weinstockidentity} vanishes,
\begin{equation}
r' \frac{\partial \mathcal{L}}{\partial r} - \mathcal{L} = c
\end{equation}
where $c$ is a constant of integration.
Substituting $\mathcal{L}$ from Eq. \eqref{eq:integral17} yields
\begin{equation}
r' \frac{\partial \mathcal{L}}{\partial r'} - \mathcal{L} = \frac{\mu b_B^2}{4\pi} \tilde\kappa \ln\left(\frac{r}{r_0}\right)
+2\pi \gamma_\text{AM} r_0\left(1-\frac{r/r_0}{\sqrt{1+r'^2}}\right)
+\pi \Delta G r_0^2 \left(\frac{r^2}{r_0^2} -1\right) = c
\,.
\end{equation}
The boundary condition $r'=0$ for $r=r_0$ implies $c=0$.
Solving for $r'$ we obtain
\begin{equation}
\frac{dr}{dz}=\sqrt{q^2(r/r_0)-1}
\,,
\end{equation}
\begin{equation}
q(r/r_0) = \frac{r}{r_0}\left[1 + \left(1-2\frac{r_0}{r_c}\right)\ln\left(\frac{r}{r_0}\right) + \frac{r_0}{r_c}\left(\frac{r^2}{r_0^2}-1\right)\right]^{-1}
\,,
\end{equation}
where $r_c=2\gamma_\text{AM}/\Delta G$ (see Equation \eqref{eq:defrc}).
It is convenient to express $\mathcal{L}$ in terms of $q(r/r_0)$
\begin{equation}
\mathcal{L}(r) = 2\pi \gamma_\text{AM} r\left[q\left(\frac{r}{r_0}\right) - q^{-1}\left(\frac{r}{r_0}\right)\right]
\,,
\end{equation}
where $r_0$ is defined in \eqref{eq:r0andalpha}.
Substitution in Equation \eqref{eq:integral17} results in
\begin{align}
\varepsilon_c^\text{dis} &= 4\pi \gamma_\text{AM} r_0^2 \int_1^{r_\text{max}/r_0} dx\, x\sqrt{1-q^{-2}(x)}
\nonumber\\
&= \varepsilon_c^\text{hom} \cdot\frac{3}{16}\left(1-\sqrt{1-\alpha}\right)^2\int_1^{r_\text{max}/r_0}dx \, x\sqrt{1-q^{-2}(x)}
\end{align}
which is equivalent to Equation (6) in \cite{Cahn:1957};
$\varepsilon_c^\text{hom}$ is given in \eqref{eq:epsilonchom}.
For $x\gtrsim1$, $q(x)=1+\sqrt{1-\alpha}\,(x-1)^2/2 + \mathcal{O}(x-1)^3$,
hence it is increasing quadratically.
At sufficiently large $x$ it reaches a maximum and then decreases to unity at $r_\text{max}/r_0$, i.e. the upper limit of integration is the solution of $q(r_\text{max}/r_0)=1$.

\begin{figure}[pos=!htb]
\centering
\includegraphics[width=0.473\textwidth]{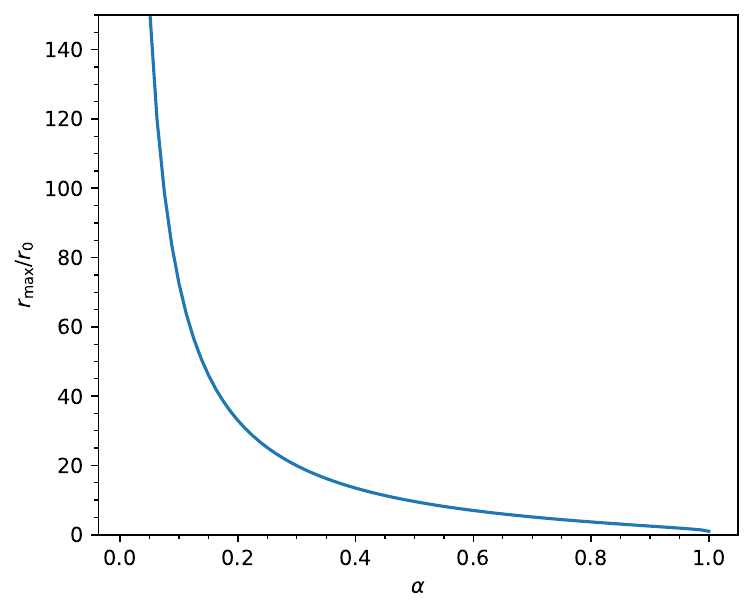}%
\includegraphics[width=0.527\textwidth]{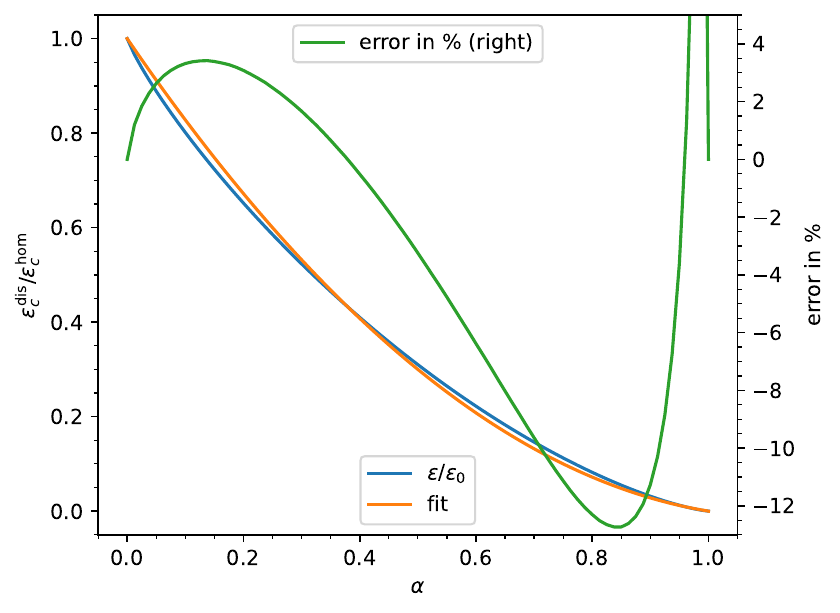}
\caption{We present $r_\text{max}/r_0$ and $\varepsilon_c^\text{dis}/\varepsilon_c^\text{hom}$ versus $\alpha$, as well as values of our fit $f^\text{dis}(\alpha)$ and the corresponding errors.}
\label{fig:nucldisloc}
\end{figure}

In Figure \ref{fig:nucldisloc}
we present $r_\text{max}/r_0$ and $\varepsilon_c^\text{dis}/\varepsilon_c^\text{hom}$ versus $\alpha$, as well as values of our fit
\begin{equation}
\varepsilon_c^\text{dis}/\varepsilon_c^\text{hom}=f^\text{dis}(\alpha) \approx (1-\alpha)\left(1-\frac45\alpha\right)\theta(1-\alpha)
\label{eq:fitfdisofalpha}
\end{equation}
and the corresponding errors.

\subsubsection{Nucleation Rates}

In this model we account for homogeneous nucleation, nucleation at $d$-dimensional grain sites ($d=0,1,2$), and nucleation by dislocations.
The critical nuclei energies for grain site and dislocation nucleation are $\varepsilon_c^\text{hom} f_d^\text{grain}$ and $\varepsilon_c^\text{hom}f^\text{dis}$, respectively;
$f_d^\text{grain}$ and $f^\text{dis}$ are given in Equations \eqref{eq:fdgrainofk} and \eqref{eq:fitfdisofalpha}.

We now obtain expressions for the rates of homogeneous nucleation and nucleation at grain sites and dislocations.

\paragraph{Homogeneous nucleation:}\ 
\\
It is generally assumed that the rate of homogeneous nucleation is proportional to an Arrhenius exponential, i.e. $\exp\left(-\varepsilon_c^\text{hom}/k_\text{B}T\right)$, where
$\varepsilon_c^\text{hom}$ is given in \eqref{eq:epsilonchom}.
The prefactor is taken to be the product of $\nu\sim\nu_D\sim10^{13}$s$^{-1}$ (Debye frequency) and the atomic number density, $n$
\begin{equation}
\dot{N}^\text{hom} = \nu_D n \exp\left(-\varepsilon_c^\text{hom}/k_\text{B}T\right)
\,. \label{eq:Ndothom}
\end{equation}

\paragraph{Grain site nucleation rates:}\ 
\\
In these cases we have
\begin{equation}
\dot{N}_d^\text{grain} = S_d I_d^\text{grain} \, ,
\end{equation}
where $I_d^\text{grain}$ is the nucleation frequency per unit $d$-volume ($d=0,1,2$) of a $d$-dimensional grain site,
and $S_d$ is the total grain site $d$-volume per unit polycrystal volume.

The nucleation frequency is again expected to be proportional to $\nu\sim\nu_D$.
We approximate the number of atoms per unit $d$-volume of a $d$-dimensional site as
\begin{equation}
n\delta^{3-d} \, ,
\end{equation}
where $\delta$ is an effective grain boundary thickness and $n$ is the number density of atoms.
The nucleation frequencies per unit $d$-volume are
\begin{equation}
I_d^\text{grain} = \nu_D n \delta^{3-d} \exp\left(-\varepsilon_c^\text{hom}f_d^\text{grain}(k)/k_\text{B}T\right)
\,. \label{eq:cornerfreq}
\end{equation}
We now consider $S_d$, the grain site $d$-volume per unit polycrystal volume.
Clearly $S_d$ must be proportional to $D^{d-3}$ where $D$ is the mean grain diameter:
$S_d=s_dD^{d-3}$, where $s_d$ is a constant.
\cite{Cahn:1956} chose to approximate the grains as truncated octahedra (tetrakaidecahedra) which are the Voronoi (Wigner-Seitz) cells of the bcc lattice.
These polyhedra have 14 faces, 36 edges, and 24 vertices.
If $D$ is the distance between the square faces then the edge length is $a=D/2\sqrt{2}$.
The grain volume is $V=8\sqrt{2}a^3=D^3/2$ and the area is $A=(6+12\sqrt{3})a^2=3(1+2\sqrt{3})D^2/4\approx 3.348D^2$.
The grain boundary area per unit volume is
\begin{equation}
S_2 = \frac{A/2}{V} = \frac34\left(1+2\sqrt{3}\right)D^{-1}\approx 3.348 D^{-1}
\,, \label{eq:Stwo}
\end{equation}
where the factor of $1/2$ accounts for the sharing of a grain boundary by adjacent grains.
Edges are shared by three grains and corners / vertices by four grains, hence the edge length per unit volume is
\begin{equation}
S_1 = \frac13\frac{36a}{V}=6\sqrt{2}D^{-2}\approx 8.485D^{-2}
\,,
\end{equation}
and the number of corners per unit volume is
\begin{equation}
S_0 = \frac14 \frac{24}{V} = 12D^{-3}
\,. \label{eq:cornerspervol1}
\end{equation}
As an alternative to the Voronoi cells of the bcc lattice we consider the Voronoi cells of the fcc lattice, namely rhombic  dodecahedra with 12 faces, 24 edges, and 14 vertices.
The $S_d$ are then given by
\begin{align}
S_2 &= 3D^{-1}
\,, \nonumber\\
S_1 &= 4\sqrt{3}D^{-2} \approx 6.928 D^{-2}
\,, \nonumber\\
S_0 &= (7/2)\sqrt{2}D^{-3} \approx 4.950 D^{-3}
\,. \label{eq:cornerspervol2}
\end{align}
Note that only $S_0$ differs significantly between the two choices for grain structure.

\paragraph{Nucleation rates on dislocations:}\ 
\\
In this case the nucleation rate is proportional to the product of $\nu\sim\nu_D$ and the number of atoms per unit volume in the dislocation cores.
Consider a volume $V$ of crystal containing a length $l$ of dislocation.
We approximate the volume of the dislocation core as $lb_B^2$ thus there are $nlb_B^2$ atoms in the core, and the number of core atoms per unit volume is $nlb_B^2/V=n\varrho_\text{dis}b_B^2$.
It follows that
\begin{equation}
\dot{N}^\text{dis} = \nu_D n \varrho_\text{dis} b_B^2 \exp\left(-\varepsilon_c^\text{hom}f^\text{dis}(\alpha)/k_\text{B}T\right)
\,. \label{eq:Ndotdis}
\end{equation}

\subsection{Volume Fraction Evolution: KJMA Kinetic Theory}
\label{sec:kjma}

In this section we employ KJMA (Kolmogorov, Johnson, Mehl, Avrami) kinetic theory to calculate the time dependence of the product-phase volume fraction; see \cite{Kolmogorov:1937,Johnson:1939,Avrami:1939, Avrami:1940,Avrami:1941,Bruna:2006}.
Homogeneous as well as heterogeneous nucleation at dislocations and grain sites---boundaries, edges, and corners---are taken into account.

\subsubsection{Homogeneous and dislocation nucleation}

At the heart of KJMA kinetic theory is the introduction of an ``extended'' volume fraction, $\lambda_E$, which is the sum of the volumes of all growing nuclei without accounting for nucleus-nucleus impingement or for the elimination of nucleation in transformed material.
Typically, growth of the new phase will stop on the area of contact with another nucleus and ignoring this effect allows $\lambda_E$ to grow larger than 1, which must be corrected for by relating it to a ``physical'' volume fraction, $\lambda$.
If the nucleation sites are randomly distributed, as is the case for homogeneous nucleation and nucleation on dislocations (assuming no dislocation patterning), then the physical volume fraction, $\lambda$, is simply related to $\lambda_E$:
\begin{equation}
\lambda = 1-\exp{(-\lambda_E)}
\,.
\end{equation}
For a nucleation rate $\dot N$ and interphase interface speed $c$ the extended volume fraction at time $t$ is given by
\begin{align}
\lambda_E(t) &= \frac43 \pi \int_0^t \dot{N}(t') r^3(t,t') dt'
\,, &
r(t,t') &= \int_{t'}^t c(t'') dt''
\,, \label{eq:extendedlambda}
\end{align}
which accounts for $t'$ the growth of all critical nuclei formed during $[0,t]$.
In the simplest case, i.e. constant $\dot{N}$ and $c$
\begin{equation}
\lambda(t) = 1-\exp{\left(-\frac{\pi}3 \dot{N}c t^4\right)}
\,. 
\end{equation}

\subsubsection{Nucleation at grain sites}

When critical nuclei are formed on a single grain boundary or edge then impingement between those growing nuclei must be taken into account to calculate the time-dependence of the product phase physical volume per unit area or length.
As nucleation and growth continue, impingement results in late-time growth occurring primarily perpendicular to the boundary or edge, hence in two planes parallel to the boundary or on a  cylinder coaxial with the edge.
Multiplying the boundary or edge physical product-phase volume per unit area or length by $S_2$ or $S_1$ yields the corresponding extended volume fraction.

\paragraph{Grain boundary nucleation}
\ \\
Consider a critical nucleus formed at time $t'$ on a grain boundary $B$.
Its interaction with a plane $A$ parallel to $B$ at a distance $z\leq r(t,t')$ is a circle of radius $\left[r^2(t,t') - z^2\right]^{1/2}$.
During $(t',t'+dt')$ there are $I_2^\text{grain}(t') dt'$ critical nuclei formed per unit area of $B$.
At a later time $t$ the extended area of the intersections of these growing $t'$-nuclei with $A$ is
\begin{equation}
d\lambda_{2,E} (t,t',z) = \pi\left[r^2(t,t') - z^2\right]I_2^\text{grain}(t') dt'
\end{equation}
per unit area of $B$.
The extended area fraction on $A$ due to all nuclei formed up to time $t$ is
\begin{equation}
\lambda_{2,E} (t,z) = \pi\int_0^t dt' I_2^\text{grain}(t') \left[r^2(t,t') - z^2\right] \theta\left(r^2(t,t') - z^2\right)
\,,
\end{equation}
and the physical area fraction is given by
\begin{equation}
\lambda_2(t,z) = 1-\exp{\left[-\lambda_{2,E}(t,z)\right]}
\,.
\end{equation}
The physical volume at time $t$ of all nuclei growing from unit area of the fiducial grain boundary $B$ is 
\begin{equation}
2\int_0^{r(t,0)} \lambda_2(t,z) dz
\,,
\end{equation}
where the factor of two accounts for growth into both grains bounded by $B$ and $r(t,0)\geq r(t,t')\geq z$.
The extended volume fraction for a polycrystal with grain boundary area per unit volume $S_2$ is
\begin{equation}
\lambda_{2,E}^\text{grain}(t) = 2 S_2 \int_0^{r(t,0)} \lambda_2(t,z) dz
\,. \label{eq:boundarypervol}
\end{equation}

\paragraph{Grain edge nucleation}
\ \\
We now consider a critical nucleus formed on an edge $E$ at time $t'$.
Its radius at time $t$ is $r(t,t')$ given in Equation \eqref{eq:extendedlambda}.
Its intersection with a line $F$ parallel to $E$ at a distance $z\leq r(t,t')$ is a segment of length $2\sqrt{r^2(t,t')-z^2}$.
There are $I_1^\text{grain}(t') dt'$ nuclei formed per unit length of $E$ during $(t',t'+dt')$.
At time $t\geq t'$ the extended length of the intersections with $F$ is
\begin{equation}
d\lambda_{1,E}(t,t',z) = 2\sqrt{r^2(t,t')-z^2} I_1^\text{grain} (t') \theta\left(r(t,t')-z\right) dt'
\end{equation}
per unit length, the extended length fraction on $F$ due to nuclei formed at $t'\leq t$ is
\begin{equation}
\lambda_{1,E}(t,z) = 2\int_0^t dt' \sqrt{r^2(t,t')-z^2} I_1^\text{grain} (t') \theta\left(r(t,t')-z\right)
\,,
\end{equation}
and the physical length fraction is
\begin{equation}
\lambda_1(t,z) = 1-\exp{\left[-\lambda_{1,E}(t,z)\right]}
\,.
\end{equation}
Rotating $F$ around $E$ gives a physical area per unit length of $E$ of $2\pi z\lambda_1(t,z)$, and then integrating over all $z$ gives the physical volume of the growing nuclei per unit length of $E$
\begin{equation}
2\pi \int_0^{r(t,0)} dz z \lambda_1(t,z)
\,. \label{eq:volumelambda1}
\end{equation}
The product of $S_1$, i.e. the total edge length per unit polycrystal volume, and Equation \eqref{eq:volumelambda1} gives the extended volume fraction of the growing nuclei originating on edges since impingement is not taken into account
\begin{equation}
\lambda_{1,E}^\text{grain}(t) = 2\pi S_1 \int_0^{r(t,0)} dz z \lambda_1(t,z)
\,. \label{eq:volfracgrowingnuc}
\end{equation}

\paragraph{Grain corner nucleation}
\ \\
The grain corner nucleation frequency, $I_0^\text{grain}$, is given in Equation \eqref{eq:cornerfreq}.
If at time $t'$ there are $N(t')$ corners per unit volume without nuclei then the number per unit volume remaining at time $t'+dt'$ is $N(t'+dt')=N(t') - N(t')I_0^\text{grain}(t')dt'$, hence $\dot{N}(t')=-N(t')I_0^\text{grain}(t')$ with solution
\begin{equation}
N(t') = S_0 \exp\left[-\int_0^{t'} I_0^\text{grain}(t'')dt''\right]
\,,
\end{equation}
where $S_0\sim D^{-3}$, the number of corners per unit volume, is given in Equations \eqref{eq:cornerspervol1} (bcc Voronoi grains) and \eqref{eq:cornerspervol2} (fcc Voronoi grains).
Since the number of corners nucleating per unit volume between $t'$ and $t'+dt'$ is $N(t')I_0^\text{grain}(t')dt'$, and the radius at time $t$ of a $t'$-nucleus is $r(t,t')$, then the extended volume fraction at time $t$ is
\begin{equation}
\lambda_{0,E}^\text{grain}(t) = \frac43 \pi S_0\int_0^t dt' r^3(t,t') I_0^\text{grain}(t')\exp\left[-\int_0^{t'}I_0^\text{grain}(t'')dt''\right]
\,.
\end{equation}
For $I_0^\text{grain}$ constant we recover Equation (6) in \cite{Cahn:1956}.
(Note: A factor of $I_0^\text{grain}$ is missing from the first line of his Equation (6) and from an expression in the preceding text.)

\subsubsection{Extended volume fraction: approximation and summary}

Since $r(t,t')/r(t,0)$ equals unity for $t'=0$ and drops to zero at $t'=t$ we employ the linear approximation
\begin{equation}
\frac{r(t,t')}{r(t,0)}\approx 1-\frac{t'}{t}
\end{equation}
in our sub-models for the extended volume fractions.
Note that this approximation eliminates an integration over the interphase interface space.
Also, in Equations \eqref{eq:boundarypervol} and \eqref{eq:volfracgrowingnuc} we change the integration variable from $z$ to $x=z/r(t,0)$.
We obtain
\begin{align}
\left[\begin{array}{c}
\lambda_E^\text{hom}(t) \\
\lambda_E^\text{dis}(t)
\end{array}\right] &= \frac43 \pi r^3(t,0)\int_0^t dt' \left(1-\frac{t'}{t}\right)^3
\left[\begin{array}{c}
\dot{N}^\text{hom}(t') \\
\dot{N}^\text{dis}(t')
\end{array}\right]
\,,\nonumber\\
\lambda_{2,E}^\text{grain}(t) &= 2s_2\frac{r(t,0)}{D}\left[1-\int_0^1dx\exp\left\{-\pi r^2(t,0)\int_0^{t(1-x)} dt' I_2^\text{grain}(t')\left[\left(1-\frac{t'}{t}\right)^2-x^2\right]\right\}\right]
\,,\nonumber\\
\lambda_{1,E}^\text{grain}(t) &= \pi s_1\frac{r^2(t,0)}{D^2}\left[1-2\int_0^1dx x \exp\left\{-2r(t,0)\int_0^{t(1-x)} dt' I_1^\text{grain}(t')\sqrt{\left(1-\frac{t'}{t}\right)^2-x^2}\right\}\right]
\,, \nonumber\\
\lambda_{0,E}^\text{grain}(t) &= \frac43\pi s_0\frac{r^3(t,0)}{D^3}\int_0^t dt' \left(1-\frac{t'}{t}\right)^3 I_0^\text{grain}(t') \exp\left\{-\int_0^{t'} I_0^\text{grain}(t'') dt''\right\}
\,, \label{eq:volfracevolution}
\end{align}
where $\dot{N}^\text{hom}$, $\dot{N}^\text{dis}$, and $I_d^\text{grain}$ are given in Equations \eqref{eq:Ndothom}, \eqref{eq:Ndotdis}, and \eqref{eq:cornerfreq}, respectively, and the constants $s_d=S_dD^{3-d}$ are given in Equations \eqref{eq:Stwo}--\eqref{eq:cornerspervol2}.

Using Mathematica we can analytically integrate over $t'$ in the expressions for $\lambda^\text{grain}_{2,E}$ and $\lambda^\text{grain}_{0,E}$ in Equation \eqref{eq:volfracevolution} to get (after substituting \eqref{eq:cornerfreq} for $I_2^\text{grain}$ and $I_0^\text{grain}$)
\begin{align}
\lambda^\text{grain}_{2,E} &= 2\frac{s_2}D r(t,0)\left[1-\int_0^1dx\exp\left(-\pi r(t,0)^2 \nu_DN_A \frac{\delta}{\bar{V}}\exp\left[-\pi r(t,0)^2\delta\frac{\nu_DN_A}{\bar{V}}J_2(t,x)\right]\right)\right]
\,,\nonumber\\
J_2(t,x) &= \int_0^{t(1-x)} dt' \left[\left(1-\frac{t'}t\right)^2-x^2\right]e^{-A_2/t'^2} 
\nonumber\\
&= \frac{e^{\frac{-A_2}{B^2}}}{3t}(x-1)\left(2A_2+(x-1)(2x+1)t^2\right) + \frac{A_2}{t}E_1\!\!\left(\!\frac{A_2}{B^2}\!\right)
+\sqrt{A_2\pi}\left(2A_2+3(x^2-1)t^2\right)\frac{\text{erfc}\Big(\frac{\sqrt{A_2}}{B}\Big)}{3t^2}
\,,\nonumber\\
B &= t(1-x)
\,,\nonumber\\
\lambda_{0,E}^\text{grain}(t) &= \frac43\pi s_0\frac{r^3(t,0)}{D^3}\int_0^t dt' \left(1-\frac{t'}{t}\right)^3 \left(\frac{\nu_D N_A}{\bar{V}} \delta^{3} e^{-{A_2}/{t'^2}}\right)\exp\left[{-\frac{\nu_D N_A}{\bar{V}} \delta^{3}J_0(t')}\right]
\,,\nonumber\\
J_0(t) &= \int_0^t dt' e^{-{A_0}/{t'^2}}=
te^{-{A_0}/{t^2}} - \sqrt{A_0\pi}\,\text{erfc}\left(\sqrt{A_0}/t\right)
\,,\nonumber\\
A_d &= \left(\frac{16\pi}3\right)\frac{\gamma_{AM}^3}{\left(\Delta G_{P}'\right)^2k_\text{B}T}\frac{f_d^\text{grain}(k) }{\dot{P}^2}
\,, \label{eq:volfracevolution_grains}
\end{align}
where the exponential integral $E_1(z)=\int_z^\infty \frac{e^{-t}}{t} dt$  and the complementary error function erfc$(x)=1-\text{erf}(x)$.

\subsection{Summary of approximations made}
\label{sec:summarizeapprox}

Our phase field model of solid-solid phase transformation kinetics is based on a Landau (Gibbs) free energy that is a quartic polynomial in a single order parameter $\eta$.
The Gibbs energy incorporates a quadratic coupling to small-strain elasticity via the second-order elastic compliance tensor $\mat{S}$, a transformation strain tensor $\epsilon^t$, a thermal strain tensor $\epsilon^T$, and a stress-independent term;
all the above are functions of $\eta$.
We simplify the model by approximating the $\eta$ dependencies of the $\mat{S}$, $\epsilon^T$, and $\epsilon^t$ terms by a single function of $\eta$.

Classical nucleation is approximately evaluated for homogeneous nucleation and nucleation on grain sites (faces, edges, vertices) and dislocations.
The total nucleation rate is a sum over these five types of nucleation sites where each term is of the form $C \exp(- \epsilon^{hom} f / k_B T)$; $\epsilon^{hom}$ is the Arrhenius barrier for homogeneous nucleation.
Our expressions for the prefactors $C$ and barrier reduction factors $f<1$ for non-homogeneous nucleation are computed without accounting for differences in the elastic properties of the phases, deviations of the nuclei shapes from spherical or cylindrical, or the stress fields around nucleating sites (Eshelby-like terms). 
Neither autocatalytic nor athermal nucleation are included.
However, other types of nucleation, e.g. from shear banding \citep{Javanbakht:2016,EhsanEsfahani:2020,Clayton:2024}, point defects, impurities, etc., as well as twin dynamics could be incorporated in future work.

As mentioned in the Introduction we employ classical KJMA theory to compute volume fraction evolution. 
Expressions for the interface speeds and nucleation rates are required for the KJMA computations.
The interface speeds are proportional to the kinetic energy coefficient $\kappa$ in the time-dependent Ginzburg-Landau equation, and the nucleation rates are functions of the interface energy $\gamma$.
We approximate $\kappa$ as a constant, though it may depend on both $P$ and $T$, and $\gamma$ is taken to be isotropic;
we do not consider the dependence of $\gamma$ on pressure or the decrease in $\gamma$ with temperature \citep{Hunter:2022}.
We note that classical KJMA theory does not account for heterogeneity of nucleation sites, non-convex nuclei, anisotropic growth, elastic shielding, or the possibility that growing nuclei can grow over other growing nuclei, a shortcoming in the KJMA treatment of impingement.
We do not account for the Gibbs-Thomson effect, i.e. the pressure increase, hence increase in free energy, due to interface curvature, an effect that may be important in the early stages of a phase transformation when the product phase is comprised of small regions in the parent phase.
The model does not account for heat flow (dissipation) or the latent heat release due to the phase transformation.

Although our model is based on mesoscale physics, we briefly mention two atomic-scale mechanisms relevant to the bcc $\to$ hcp transition.
The Burgers mechanism involves compression of the bcc unit cell along one of its three fourfold axes and displacements of adjacent atomic planes in opposite directions \citep{Burgers:1934,Toledano:1996}.
The Pitsch-Schrader mechanism involves atomic shuffling, i.e. a collective displacement of atomic planes resulting in a specific relative orientation of the bcc and hcp phases \citep{PitschSchrader}.

Generalizations of the current framework to account for the current limitations listed above are envisioned for future follow up work.
These will also include the generalization to describing multiple phases, as well as partial phase transformations (observed e.g. under quasistatic loading in \cite{Merkel:2020}).



\section{Model Application: The \texorpdfstring{$\boldsymbol{\alpha}$(bcc)$\to\boldsymbol{\epsilon}$(hcp)}{alpha(bcc)->epsilon(hcp)} transition in iron under ramp loading}
\label{sec:ironexample}

\begin{figure}[pos=!ht]
\centering
\includegraphics[width=0.5\textwidth]{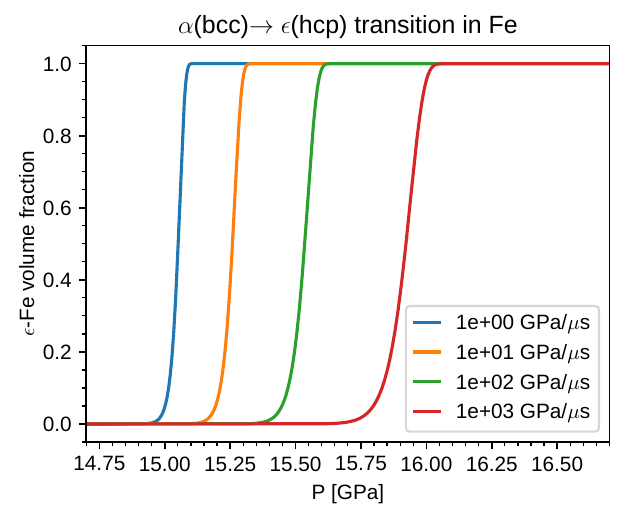}%
\includegraphics[width=0.5\textwidth]{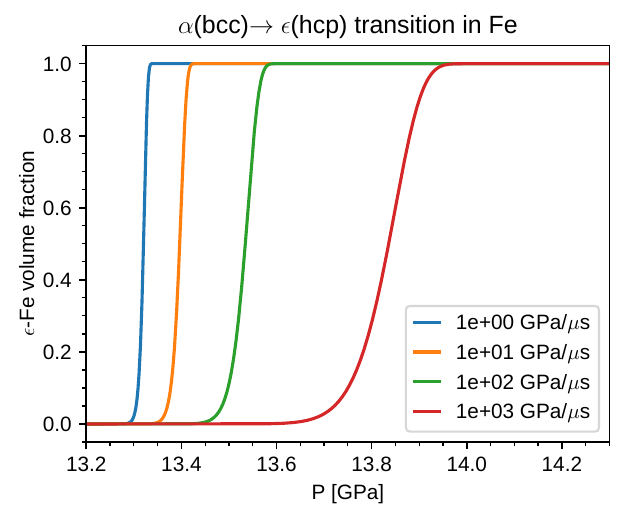}
\caption{We show the $\epsilon$ Fe volume fraction versus ramp pressure for constant loading rates of 1, 10, 100, and 1000 GPa/$\mu$s at 300K.
The left pane includes only homogeneous nucleation 
whereas the right pane includes the additional effects of nucleation on dislocations (with dislocation density 10$^{12}$m$^{-2}$) and nucleation on grain boundaries with average grain diameters of $D=100\mu$m and grain boundary thicknesses $\delta$ of 0.1nm.
Clearly, the epsilon volume fraction rapidly increases to unity at pressures $\sim$13--16 GPa for the chosen loading rates.}
\label{fig:iron_fig1}
\end{figure}

\begin{figure}[pos=!ht]
\centering
\includegraphics[width=0.5\textwidth]{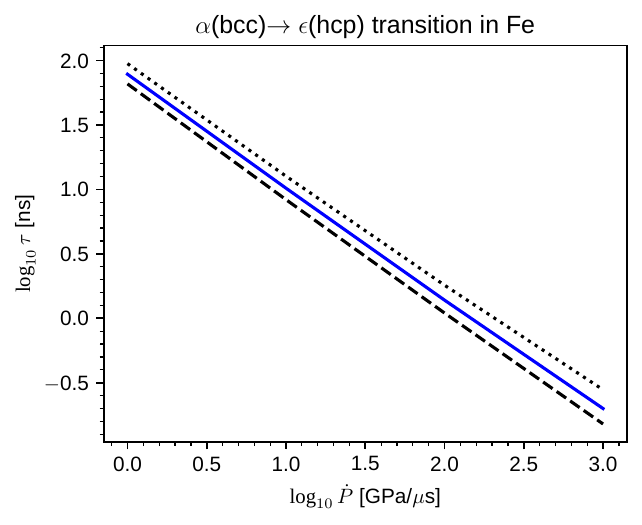}%
\includegraphics[width=0.5\textwidth]{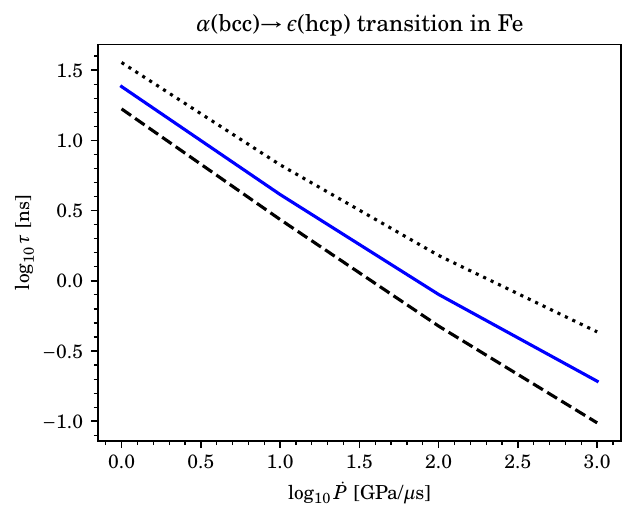}
\put(-350,150){$\tau\approx71\dot{P}^{-0.86}\textrm{ns}$}
\caption{This plot shows that the log of $\tau$, which is the time scale (relaxation time)
of the PT, is a decreasing linear function of the log of the loading rate.
The left pane includes only homogeneous nucleation 
whereas the right pane includes the additional effects of nucleation on dislocations (with dislocation density 10$^{12}$m$^{-2}$) and nucleation on grain boundaries with average grain diameters of $D=100\mu$m and grain boundary thicknesses $\delta$ of 0.1nm.
The lower (upper) dashed (dotted) line shows the effect of increasing (decreasing) the interface speed by a factor of ten.}
\label{fig:iron_fig2}
\end{figure}

At temperatures below approximately 800K iron is in the bcc $\alpha$ phase at low pressure and in the hcp $\epsilon$ phase at elevated pressure \citep{Takahashi:1964}.
The $\alpha\to\epsilon$ transition begins at 13 GPa on the shock Hugoniot at a temperature of approximately 350K.
A well-known series of plate impact experiments were carried out by \cite{Barker:1974} to investigate the shock-induced $\alpha\to\epsilon$ transition.
Their data were subsequently used by \cite{Boettger:1997} as the basis for their theoretical study of the shock-induced transition.
A notable result of their study was their conclusion that the relaxation time of the transition decreases with increasing shock pressure and varies from 60 ns to 12 ns for shock pressures of 17 GPa to 30 GPa.

In the following we focus on ramp loading of iron and employ our PT kinetics model to obtain quantitative predictions for the dynamics of the $\alpha\to\epsilon$ transition.
Specifically, we calculate the $\epsilon$ volume fraction as a function of the ramp pressure and loading rate, and also the relaxation time versus loading rate.
Ideally our predictions will be checked by ramp compression wave experiments.
Such experiments are difficult because ramp wave transmission through phase-changing materials generally leads to a multi-wave structure that is similar to those seen in shock wave experiments \citep{Hayes:2007}.
%
We shall obtain results for constant loading rates $\dot{P}$ (as indicated in our figures) 
that include
those that are experimentally accessible.
For example, in a series of experiments on the $\alpha$(hcp)$\to\omega$(hex-3)
transition in zirconium the ramp loading rates were $\sim$250 GPa$\mu$s$^{-1}$ \citep{Rigg:2009}.

A key ingredient for our calculations of the time dependence of the $\epsilon$ volume fraction is $\Delta G(P,T)$, the difference of the specific (energy per unit volume) Gibbs free energies of $\alpha$ and $\epsilon$, which in part controls the interphase interface speed and the activation energies for homogeneous nucleation, nucleation at grain sites, and nucleation on dislocations.
We obtain $\Delta G(P,T)$ from the Helmholtz free energies $F^{(\alpha)}(V,T)$ and $F^{(\epsilon)}(V,T)$ calculated by \cite{Boettger:1997} (BW).
The BW Helmholtz free energies have dimensions of J/mol and $V$ is the volume per mole.
The Gibbs free energy is $\mathcal{G}=F+PV$ and $P=-\left(\partial F/\partial V\right)_T$.
Since $P=P(V,T)$ it is necessary to solve for $V(P,T)$ and then $\mathcal{G}(P,T)=F(V(P,T),T) + PV(P,T)$.
The Gibbs free energy, $G$, defined in subsection \ref{sec:Gibbs} 
is the specific (per unit volume) Gibbs free energy, therefore $G(P,T)=\mathcal{G}(P,T)/V(P,T)$.

The Helmholtz free energies of $\alpha$-Fe and $\epsilon$-Fe are comprised of three contributions \citep{Boettger:1997}:
\begin{equation}
F(V,T) = \Phi_0(V) + F_H(V,T) + F_E(V,T)
\,.
\end{equation}
Here $\Phi_0$ is the static lattice potential, $F_H$ is the quasiharmonic phonon free energy, and $F_E$ is due to thermal excitation of electrons from the ground state.
We have dropped a negligible anharmonic contribution in accord with \cite{Boettger:1997}.
In the following we present each of the contributions to $F(V,T)$ for both phases and obtain approximations for $V_\alpha(P,T)$ and $V_\epsilon(P,T)$.
Substitution of those approximations in $F^{(\alpha)}(V_\alpha,T)+PV_\alpha$ and $F^{(\epsilon)}(V_\epsilon,T)+PV_\epsilon$ yields $\mathcal{G}_\alpha(P,T)$ and $\mathcal{G}_\epsilon(P,T)$.

The static lattice potential is
\begin{align}
\Phi_0(V) &= \Phi^*+\frac{4V^*B^*}{\left(B^*_1-1\right)^2}\left[1-(1+\eta)e^{-\eta}\right]
\,, &
\eta &= \frac32\left(B_1^*-1\right)\left[\sqrt[3]{\frac{V}{V^*}}-1\right]
\end{align}
with
\begin{align}
V^*_\alpha &= 7.0047\, \text{cm$^3$/mol} \,,& 
B^*_\alpha &= 176.64\, \text{GPa}\,, & 
B^*_{1,\alpha} &= 4.7041\,, & 
\Phi^*_\alpha &= 0 \,,
\nonumber\\
V^*_\epsilon &= 6.5984 \,\text{cm$^3$/mol} \,, & B^*_\epsilon &=  181.5\,\text{GPa} \,, & B^*_{1,\epsilon} &=  5.74\,, & \Phi^*_\epsilon &=  5533\,\text{J/mol} 
\,, \label{eq:iron-eq4}
\end{align}
for $\alpha$-Fe and $\epsilon$-Fe.
Volume, $V$, is in units of cm$^3$/mol and $\Phi_0$ is thus in units of J/mol.
The contribution of $\Phi_0$ to the total pressure in each phase is
\begin{align}
P^\alpha_\phi(V) &= -d\Phi_0^{(\alpha)}/dV=
 -95.375\eta_\alpha e^{-\eta_\alpha}\left(\frac{7.0047}{V}\right)^{2/3}\text{ GPa}
\,,\nonumber\\
P^\epsilon_\phi(V) &= -d\Phi_0^{(\epsilon)}/dV=
-76.582\eta_\epsilon e^{-\eta_\epsilon}\left(\frac{6.5984}{V}\right)^{2/3}\text{ GPa}
\,. 
\end{align}

The quasiharmonic free energy is given by
\begin{equation}
F_H(V,T) = 3 N_A k_B T\left\{-\ln\left[\frac{T}{\theta_0(V)}\right] + \frac1{40}\left[\frac{\theta_2(V)}{T}\right]^2\right\}
\,, \label{eq:iron-eq7}
\end{equation}
where
\begin{align}
\theta_0^{(\alpha)}(V) &=301\exp\left[1.82\left(1-V/7.093\right)\right]\text{K}
\,,&
\theta_0^{(\epsilon)}(V) &=261\exp\left[2.8\left(1-V/6.73\right)\right]\text{K}
\,,\nonumber\\
\theta_2^{(\alpha)}(V) &=420\exp\left[1.82\left(1-V/7.093\right)\right]\text{K}
\,,&
\theta_2^{(\epsilon)}(V) &=364\exp\left[2.8\left(1-V/6.73\right)\right]\text{K}
\,; \label{eq:DebyeTempIron}
\end{align}
$3N_A k_B=24.94$ J/mol\,K $=0.02494$ GPa\,cm$^3$/mol\,K.

The contribution of $F_H$ to the total pressure is
\begin{align}
P_H^{(\alpha)}(V,T) &=-\partial F_H^{(\alpha)}/\partial V= 0.0064\frac{T}{\text{K}}\left[1+\frac{1}{20}\left(\frac{\theta_2^{(\alpha)}(V)}{T}\right)^2\right]\text{ GPa}
\,,\nonumber\\
P_H^{(\epsilon)}(V,T) &=-\partial F_H^{(\epsilon)}/\partial V= 0.0104\frac{T}{\text{K}}\left[1+\frac{1}{20}\left(\frac{\theta_2^{(\epsilon)}(V)}{T}\right)^2\right]\text{ GPa}
\,.
\end{align}

The electronic free energy, $F_E(V,T)$, is the sum of magnetic, $F_\text{mag}(T)$, and conduction-electron, $F_\text{cond}(V,T)$, free energies.
The magnetic free energy for $\alpha$-Fe (ferromagnetic ground states) is given by
\begin{equation}
F_\text{mag}^{(\alpha)}(T) = 4680\left[\left(1-\frac{T}{T_m}\right)\ln\left(\frac{1+\sqrt{T/T_m}}{1-\sqrt{T/T_m}}\right)-2\sqrt{\frac{T}{T_m}}+\frac43\left(\frac{T}{T_m}\right)^{3/2}\right]\text{ J/mol}
\,, \label{eq:iron-eq16}
\end{equation}
where $T_m=1135$ K.
Since $F_\text{mag}$ is a function of only $T$ it does not contribute to the pressure.
The $\epsilon$ phase is non-ferromagnetic, hence $F_\text{mag}^{(\epsilon)}=0$.
The conduction electron free energies are
\begin{align}
F^{(\alpha)}_\text{cond}(V,T)&=-0.00125\left(\frac{V}{7.093}\right)^{1.3}T^2\ \text{ J/mol}
\,, \nonumber\\
F^{(\epsilon)}_\text{cond}(V,T)&=-0.00125\left(\frac{V}{6.73}\right)^{1.3}T^2\ \text{ J/mol}
\,, \label{eq:iron-eq18}
\end{align}
and their contributions to the total pressure are
\begin{align}
P^{(\alpha)}_\text{cond}(V,T) &= 2.291\times10^{-7}\left(\frac{V}{7.093}\right)^{0.3}T^2 \text{ GPa}
\,,\nonumber\\
P^{(\epsilon)}_\text{cond}(V,T) &= 2.415\times10^{-7}\left(\frac{V}{6.73}\right)^{0.3}T^2 \text{ GPa}
\,.
\end{align}
Comparison of $P_\text{cond}$ with a coefficient $\mathcal{O}\left(10^{-7}\right)$ to $P_{\phi}(V)$ and $P_{H}(T)$ shows that $P_\text{cond}$ is negligible, hence the total pressure in each phase is accurately approximated by $P_{\phi}(V)+P_{H}(T)$.
Solving $P_{\phi}(V)=P-P_{H}(T)$ for $V$ gives
$V(P,T)$.
Given $P$, this can be done numerically for each phase.
The Helmholtz free energy is given by the sum of Equation \eqref{eq:iron-eq4} for $\Phi_0(V)$, Equations \eqref{eq:iron-eq7} and \eqref{eq:DebyeTempIron} for $F_H(V,T)$, Equation \eqref{eq:iron-eq16} for $F^{(\alpha)}_\text{mag}(T)$, and \eqref{eq:iron-eq18} for $F_\text{cond}(V,T)$.
Substitution of 
$V_\alpha(P,T)$ and $V_\epsilon(P,T)$ in $F(V,T)+PV$ then yields $\mathcal{G}_\alpha(P,T)$ and $\mathcal{G}_\epsilon(P,T)$.

The dimensions of $F$ and $\mathcal{G}$ are J/mol whereas G as defined in subsection \ref{sec:Gibbs} 
is the specific (per unit volume) Gibbs free energy.
Clearly we have $G_i(P,T)=\mathcal{G}_i(P,T)/V_i(P,T)$ where $i=\alpha,\epsilon$, hence
\begin{equation}
\Delta G(P,T) = \mathcal{G}_\alpha (P,T)/V_\alpha (P,T)-\mathcal{G}_\epsilon(P,T)/V_\epsilon(P,T)
\,.
\end{equation}
It is implicit in the LP phase field model that the densities of A and M in the stress-independent parts of the Gibbs free energies, $f_{0,1}(P,T)$, are the same.
(Changes in material density due to the transition are accounted for in the transformation strain tensor.)
We find that the change in density due to the $\alpha\to\epsilon$ transition at any $P$ and $T$ is in fact negligible.
Consider the ratios $\left[V_{\alpha,\epsilon}(P,T)+\bar{V}(P,T)\right]/\bar{V}(P,T)$ where $\bar{V}(P,T)=\left[V_\alpha(P,T)+V_\epsilon(P,T)\right]/2$.
At $P=0$ and 300K we find that $\left(V_\alpha-\bar{V}\right)/\bar{V}=0.025$ and $\left(V_\epsilon-\bar{V}\right)/\bar{V}=-0.025$.
At higher pressures and/or temperatures the magnitudes of these rates decrease from 0.025, therefore to a good approximation we have $V_\alpha(P,T)\approx V_\epsilon(P,T)\approx\bar{V}(P,T)$ and
\begin{equation}
\Delta G(P,T) \approx \left[\mathcal{G}_\alpha(P,T) - \mathcal{G}_\epsilon(P,T)\right]/\bar{V}(P,T)\equiv \Delta\mathcal{G}(P,T)/\bar{V}(P,T)
\,.
\end{equation}
The $\alpha$-$\epsilon$ coexistence curve is given by $\Delta\mathcal{G}(P,T)=0$, and so in this approximation it is given by $\Delta G(P,T)=0$.
In accordance with Equation \eqref{eq:G9} this implies the additional approximation $W_{t,\lambda}\approx0$.

We now expand $\Delta G(P,T)$ around the coexistence curve $P_e(T)$
\begin{equation}
\Delta G(P,T) = \Delta\mathcal{G}'_P(T)\left[P-P_e(T)\right]/\bar{V}(P_e(T),T)
\,,
\end{equation}
where
\begin{equation}
\Delta\mathcal{G}'_P(T) = \frac{\partial \mathcal{G}(P,T)}{\partial P}\Bigg|_{P_e(T)}
\,.
\end{equation}
For the $\alpha$-$\epsilon$ transition at 13 GPa and 300K we have $\bar{V}=6.46$ cm$^3$/mol.
Calculations give $\Delta\mathcal{G}'_P(300$K$)=335$ J/mol\,GPa;
the value of $\Delta\mathcal{G}'_P$ is insensitive to variation of $T$, e.g. $\Delta\mathcal{G}'_P(200\textrm{K})=326$ J/mol\,GPa and $\Delta\mathcal{G}'_P(400\textrm{K})=343$ J/mol\,GPa.
It follows that
\begin{equation}
\Delta G'_P(300\textrm{K}) = \frac{\Delta \mathcal{G}'_P(300\textrm{K})}{\bar{V}(13\textrm{GPa},300\textrm{K})} = 51.9\,\textrm{J\,cm$^{-3}$\,GPa}^{-1}
\,.
\end{equation}

In the absence of an athermal threshold, the $\alpha$-$\epsilon$ interface speed is given by Equation \eqref{eq:struct11}.
Assuming $\xi=1/2$, the interface speed in the $\epsilon$ phase $(x\ge0)$ is
\begin{equation}
c = 2\kappa\sqrt{3\beta\Delta G'_P\Delta P}\,x
\,. \label{eq:interfacespeediron}
\end{equation}
Equations (19) and (27) in \cite{Levitas:2010} imply $\kappa\approx 1300$ m$^2$/N\,s for NiAl, and Equation (26) gives $\beta=2.59\times10^{-10}$N for NiAl.
Here we use these values to obtain order-of-magnitude estimates for $\alpha$-$\epsilon$ interfaces:
\begin{align}
\kappa &\sim 10^3\, \textrm{m}^2\textrm{N}^{-1}\textrm{s}^{-1} = 10^9\, \textrm{cm}^3\textrm{J}^{-1}\textrm{s}^{-1}
\,,\nonumber\\
\beta &\sim 10^{-10}\,\textrm{N} = 10^{-12}\,J\,\textrm{cm}^{-1}
\,. \label{eq:kappabetairon}
\end{align}
The value of $\Delta P$, the pressure difference between the coexistence curve, $P_e(T)$, and the $\alpha\to\epsilon$ spinodal, $\bar{P}_c(T)$, is not known for iron.
We choose $\Delta P=10$ GPa as a very rough estimate and account for variations around this value.
The parameter $x$ in Equation \eqref{eq:interfacespeediron} is given by
\begin{equation}
x = (P-13\,\textrm{GPa})/\Delta P
\,; \label{eq:xofinterfacespeediron}
\end{equation}
$x=1$ on the spinodal.
Substitution of \eqref{eq:xofinterfacespeediron} and the above values for $\Delta G'_P$, $\kappa$, $\beta$, and $\Delta P$ in \eqref{eq:interfacespeediron} results in
\begin{equation}
c = 78.9\left(P-13\right)\,\textrm{m/s}
\,,
\end{equation}
where $P$ is in GPa.

\paragraph{Homogeneous nucleation.}\ 
\\
Equation \eqref{eq:epsilonchom} for $\epsilon_c^\textrm{hom}$ with $\Delta G=51.9 (P-13)$ J/cm$^3$ (with $P$ in GPa) and $\gamma_{\alpha\epsilon}=50$ mJ/m$^2$ (an approximation)
gives $\epsilon_c^\textrm{hom}=4.86 (P-13)^{-2}$ eV.
The homogeneous nucleation rate is
\begin{equation}
\dot{N}^\textrm{hom} = \nu_D n \exp\left[-\frac{188}{\left(\frac{P}{\textrm{GPa}}-13\right)^2\frac{T}{300\textrm{K}}}\right]
\,, \label{eq:Ndothomeiron}
\end{equation}
where $\nu_D=10^{13}$s$^{-1}$, $n=N_A/\bar{V}(13\textrm{GPa},300\textrm{K})=9.3\times10^{22}$cm$^{-3}$, and we used $k_\textrm{B}T=0.02585\left(T/300\textrm{K}\right)$\,eV.
The ramp pressure at time $t'$ is
\begin{equation}
P(t') = 13 + \dot{P}t'\ \textrm{GPa}
\,, \label{eq:Ptprimeiron}
\end{equation}
where $[\dot{P}]=$GPa/$\mu$s.
Equations \eqref{eq:Ndothomeiron} and \eqref{eq:Ptprimeiron} give
\begin{equation}
\dot{N}^\textrm{hom}(t') = 9.3\times10^{29}\exp\left[-\frac{188}{\left(\frac{\dot{P}\mu\textrm{s}}{\textrm{GPa}}\right)^2\left(\frac{t'}{\mu\textrm{s}}\right)^2\frac{T}{300\textrm{K}}}\right]\textrm{cm}^{-3}\mu\textrm{s}^{-1}
\,.
\end{equation}
In addition to the nucleation rate, we require $r(t,t')$ as given in Equation \eqref{eq:extendedlambda}
\begin{equation}
r(t,t') = \int_{t'}^t c(t'') dt'' = 0.004 \frac{\dot{P}\mu\textrm{s}}{\textrm{GPa}}\left[\left(\frac{t}{\mu\textrm{s}}\right)^2 - \left(\frac{t'}{\mu\textrm{s}}\right)^2 \right] \textrm{cm}
\,.
\end{equation}
The extended volume fraction \eqref{eq:extendedlambda} is
\begin{equation}
\lambda^\text{hom}_E(t) = \dot{P}^3\int_0^t dt'\exp\left(54 - \frac{188}{\dot{P}^2t'^2\frac{T}{300\textrm{K}}}\right)\left(t^2-t'^2\right)^3
\,,\label{eq:lambdaE_hom}
\end{equation}
where $[\dot{P}]=$GPa/$\mu$s and $t'$ is in $\mu$s.
We also include the effects of nucleation on dislocations as well as nucleation on grain boundaries.
The nucleation on grain edges and corners is sub-leading and can be neglected\footnote{Additionally, the nucleation on grain edges requires a 2-dimensional numerical integration---see Equation \eqref{eq:volfracevolution}, and is hence at least an order of magnitude slower in our Python implementation.}.

\paragraph{Nucleation on dislocations.}\ 
\\
The contribution from dislocations to the extended volume fraction was computed from
\begin{align}
\lambda_E^\text{dis} &= \dot{P}^3 \int_0^t dt' \left(t^2-t'^2\right)^3\exp\left(a-\frac{b}{(\dot{P} t')^2}\right)
\,,\nonumber\\
a &= \ln\left[\varrho_\text{dis}b_B^2\frac{4\pi \nu_D N_A}{3\bar{V}}\left(\frac{2\kappa\sqrt{3\beta\Delta G'_P\Delta P}}{2\Delta P}\right)^3\right]\approx \ln(14732.15\varrho_\text{dis}[\text{1/m}^2])
\,,\nonumber\\
b &= f_\text{dis}(\alpha(\dot{P}, t')) \left(\frac{16\pi}3\right)\frac{\gamma_{AM}^3}{\left(\Delta G_{300}'\right)^2k_\text{B}T}
 \approx 188.29 f_\text{dis}(\alpha(\dot{P}, t')) \,\text{GPa}^{-2}
  \,,\nonumber\\
  f_\text{dis}(\alpha) &\approx (1-\alpha)\left(1-\frac45\alpha\right)\theta(1-\alpha)
   \,,\nonumber\\
  \alpha (\dot{P},t') &= \frac{\mu b_B^2 \tilde\kappa \Delta G}{2\pi^2\gamma^2_\text{AM}}\approx  \frac{\mu b_B^2  \Delta G'_{300}\dot{P} t'}{2\pi^2\gamma^2_\text{AM}}\frac{(1-\nu/2)}{(1-\nu)}\approx 6.3750 \frac{\dot{P} t'}{\text{GPa}}
  \,,\label{eq:lambdaE_dis}
\end{align}
with (average) shear modulus $\mu=81.6$ GPa for $\alpha$-iron and Poisson's ratio $\nu=0.293$ and $\tilde\kappa\approx(1-\nu/2)/(1-\nu)$ (average over edge and screw dislocations in the isotropic limit) as discussed earlier.
The Burgers vector magnitude in $\alpha$ iron is $b_B=a\sqrt{3}/2\approx2.4825$\,{\AA}.
Note that $f_\text{dis}(\alpha)$ (and consequently the constant $b$ in \eqref{eq:lambdaE_dis} above) is zero unless $\dot{P}t'<0.1569$ GPa, a limit we frequently encounter in our calculations.
But $b=0$ implies that the volume fraction of $\epsilon$ iron, given by $\lambda=1-\exp(-\lambda_E)$ jumps to 1 almost instantaneously.
Thus, the relaxation time can in these cases shrink to almost zero leading to unphysically steep rises in volume fraction (resp. an almost instantaneous phase transformation).
We therefore include a cutoff parameter, which we have calibrated to $\varepsilon\sim10^{-2}$, and replace $f_\text{dis}(\alpha)\to \max\left(f_\text{dis}(\alpha),\varepsilon\right)$.
Future experimental data on the time dependence of the phase transformation as a function of dislocation density would allow a more accurate calibration of the cutoff.

\paragraph{Nucleation on grain boundaries.}\ 
\\
We use Equation \eqref{eq:volfracevolution_grains} for $\lambda^\text{grain}_{2,E}$with
\begin{align}
r(t,0)[\text{cm}] &= \left(\frac{\kappa\sqrt{3\beta\Delta G'_P\Delta P}}{\Delta P}\right){\dot{P}t^2}\approx 0.00197145 {\dot{P}[\text{GPa}/\mu\text{s}]\left(t[\mu\text{s}]\right)^2}
\,,\nonumber\\
A_2[\mu\text{s}^2] &= \left(\frac{16\pi}3\right)\frac{\gamma_{AM}^3}{\left(\Delta G_{P}'\right)^2k_\text{B}T}\frac{f_2^\text{grain}(k) }{\dot{P}^2}\approx \frac{25.3143}{\left(\dot{P}[\text{GPa/$\mu$s}]\right)^2}
\,.
\end{align}

The physical volume fraction is approximated as
\begin{align}
\lambda = 1-e^{-\lambda_E} \, ,
\end{align}
where $\lambda_E$ is the sum of contributions due to the different nucleation effects discussed above.

Figure \ref{fig:iron_fig1} is a plot of the $\epsilon$ volume fraction versus ramp pressure for constant loading rates of 1, 10, 100, and 1000 GPa/$\mu$s at 300K.
If only homogeneous nucleation is considered, the volume fraction rapidly increases to unity at pressures $\sim15$--16 GPa for the chosen loading rates.
Including nucleation on dislocations and grain boundaries reduces the overshoot pressure and accelerates the phase transformation such that the volume fraction rapidly increases to unity at pressures $\sim13$--14 GPa for the chosen loading rates.
In both cases, the transitions to $\epsilon$ are complete <3 GPa above the equilibrium pressure.
We define $\tau$, the relaxation time of the phase transition, as the time interval between $\lambda=0.05$ and $\lambda=0.95$.
Figure \ref{fig:iron_fig2} shows that $\ln\tau$ is a decreasing linear function of $\ln\dot{P}$.
The solid line in the left pane (homogeneous nucleation only), which was computed for $\kappa$ and $\beta$ in Equation \eqref{eq:kappabetairon} and $\Delta P=10$ GPa, can be approximated by the line $\tau\approx71\dot{P}^{-0.86}$ns.
The inclusion of nucleation on dislocations and grain boundaries leads to deviations from this linear relationship and speeds up the phase transformation.
Because of the significant uncertainties in the values of $\kappa$, $\beta$, and $\Delta P$, which fix the interface speed, we considered order-of-magnitude variations in $c$.
The lower (upper) dashed (dotted) line in the left pane of Figure \ref{fig:iron_fig2} shows the effect of increasing (decreasing) $c$ by a factor of ten.

\begin{figure}[pos=!ht]
\centering
\includegraphics[width=0.5\textwidth]{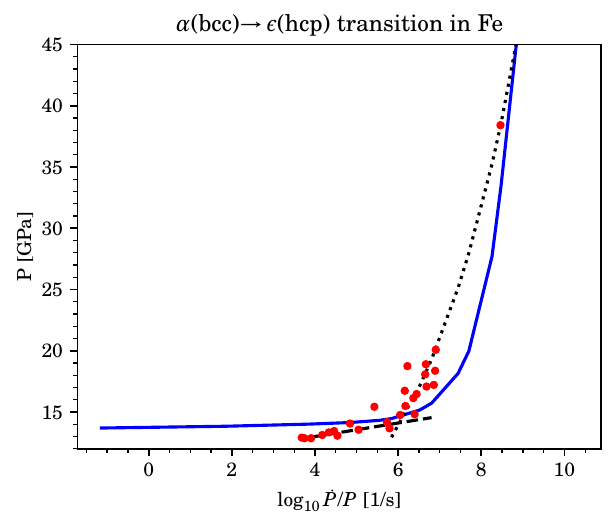}%
\includegraphics[width=0.5\textwidth]{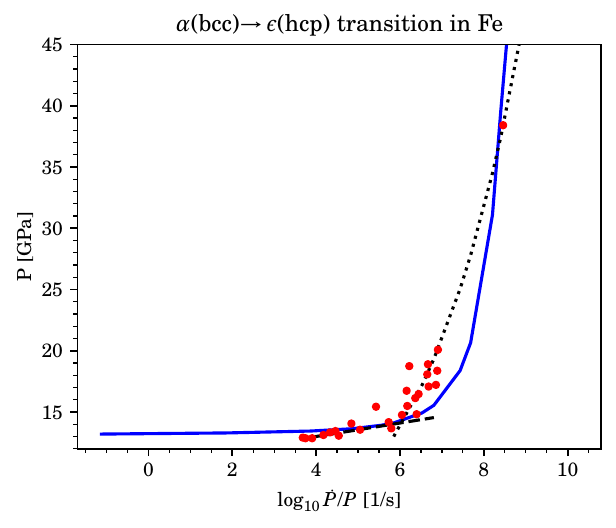}
\caption{We compare our simulations to experimental results of \cite[fig. 6b]{Smith:2013}.
The dashed / dotted lines represent the linear / non-linear fits to the latter (red dots), where the non-linear fit is valid for strain rates above 10$^6$/s and the linear fit is valid between $\sim5\!\times\!10^3$--10$^6$/s.
Our own simulations depend on several model and material parameters that can be tuned to get a match.
Here we show two examples: in the left pane we include only grain boundary nucleation, whereas in the right pane we add also nucleation on a low dislocation density of 10$^{10}$m$^{-2}$ and at the same time reduce model parameter $\kappa$ to 1.}
\label{fig:iron_fig3}
\end{figure}

We now compare to some data in the literature.
In particular, \cite{Smith:2013} measured the $\alpha\to\epsilon$ iron phase transition under ramp loading up to very high loading rates (i.e. compressive strain rates up to $\frac1\rho\frac{d\rho}{dt}\leq5\!\times\!10^8$s$^{-1}$) and found the onset pressure (i.e. the pressure under which the $\epsilon$ phase first appears in the measurement) to follow a roughly linear trend for compressive strain rates in the range 10$^3$--10$^6$/s and a non-linear one for strain rates exceeding 10$^6$/s.
In order to convert pressure rates within our own simulations to compressive strain rates, we divide $\dot{P}(t)$ by the current pressure to get a rough estimate sufficient for a log-plot.

Our simulations agree with these data only if we neglect homogeneous nucleation, indicating that homogeneous nucleation is not the driving mechanism in these data.
This is consistent with other authors' conclusions on martensitic transformation kinetics in other materials, see e.g. \cite{Gornostyrev:1999}.
We get good agreement if either only grain boundary nucleation is included with fairly large grains (as before in Figure \ref{fig:iron_fig1}), or if we include a low dislocation density in addition to grain boundary nucleation supplemented by a reduced interface speed $c$ (a freedom we have since we do not know the exact values for $\kappa$ and $\beta$).
Specifically, we get good agreement if we set the dislocation density to $10^{10}$m$^{-2}$ and $\kappa=1$ instead of 100 in our simulations.
Figure \ref{fig:iron_fig3} shows the comparison in both cases.
The dashed lines represent the linear/non-linear fits to the experimental data of \cite{Smith:2013}.
Note, however, that the equilibrium pressure is 13\,GPa and hence our simulations are valid to lower strain rates whereas the linear fit of \cite{Smith:2013} (dashed line) is valid only between 10$^3$--10$^6$/s.

The Python code we developed to create Figures \ref{fig:iron_fig1}--\ref{fig:iron_fig3} is available as open-source, see \cite{pyptkinetics}.

\section{Conclusion}
\label{sec:conclusion}

We developed a new model for solid-solid phase transformation kinetics in metals where we accounted for the role of microstructure in nucleating said phase transformation.
In particular, we determined the nucleation rates at grain sites and dislocations, as well as homogeneous nucleation.
Our model was implemented in an in-house Python research code which is publicly available, see \cite{pyptkinetics}.
We then calculated the change in volume fraction of the new phase as a function of time and a combination of microstructure parameters, such as dislocation density and average grain sizes and identified those model parameters to which the results are most sensitive.

We tested our new model for the $\alpha\to\epsilon$ iron transition under ramp loading conditions by comparing to experimental data published in \cite{Smith:2013}.
More than one combination of material and model parameters (such as dislocation density and interface speed) led to good agreement of our simulations to the experimental data,
thus highlighting the importance of more detailed knowledge of the microstructure in the metal sample for which one wishes to make phase transformation kinetics predictions.
However, agreement with the experimental data could only be achieved when homogeneous nucleation was neglected indicating that this is not the driving mechanism in the experiments to which we compared.
More accurate predictions using our model would be possible if characterization data were available for the sample in question.
In particular, the kinetics of the phase transition are sensitive to the average grain size and dislocation density.

Additionally, some of our model parameters are currently known only to an order of magnitude.
In particular, more accurate knowledge of the interface speed $c$, the difference in pressure $\Delta P$ between the coexistence curve and the spinodals, and the interfacial energies $\gamma_\text{AM}$, $\gamma_\text{AA}$ would facilitate more accurate predictions if the microstructure was measured and/or would allow us to back out said microstructure from multiple ramp loading phase transformation experiments.

To summarize, our model depends (as input parameters) on the accurate knowledge of the prefactor of the interface speed $c$ (via $\kappa$ and $\beta$), the pressure difference $\Delta P$ between coexistence curve and spinodals, interfacial energies $\gamma_\text{AM}$ and $\gamma_\text{AA}$, equations of state for both phases, as well as the microstructure characterization, i.e. dislocation density, average grain size, and grain boundary thickness.
We envision that some of the required model parameters could be determined from molecular dynamics (MD) simulations (e.g. the interface speed at a given pressure, see \cite{Daphalapurkar:2024}), others by future experiments (e.g. $\Delta P$, microstructure characterization), or both (interfacial energies).
Additionally, future sensitivity studies can be used to identify those model parameters whose values need to be determined most accurately by experiments or MD simulations.

The kinetics model also must be coupled to a suitable flow stress (or strength) model, such as \cite{PTW:2003}.
The future generalization to shock problems is straightforward in principle, but requires additional numeric integrations and hence significant computational overhead compared to the ramp loading example presented here.
Other potential generalizations are mentioned in Section \ref{sec:summarizeapprox}.

\section*{Acknowledgements}
\noindent
DNB would like to thank A. E. Mattsson, I. Chesser, and J. P. Allen for related discussions.
DNB, AH and DLP gratefully acknowledge support from the Materials project within the Physics and Engineering Models (PEM) Subprogram element of the Advanced Simulation and Computing (ASC) Program at Los Alamos National Laboratory (LANL) in the early stages of this work.
DNB also gratefully acknowledges support from the Equations of State (EOS) project within the ASC/PEM program.
LANL, an affirmative action/equal opportunity employer, is operated by Triad National Security, LLC, for the National Nuclear Security Administration of the U.S. Department of Energy under contract 89233218NCA000001.

\appendix

\section{Critical Nuclei in 1D}
\label{sec:criticalnuclei1D}

In this section we discuss A and M critical nuclei in one dimension.
We first obtain the profiles of the nuclei and then calculate the critical energies.

\subsection*{Profiles}

With the definition
\begin{equation}
	\widetilde{G}(s_{ij},P,T,\eta) = \frac{s_1}{\beta}\eta^2 \left[1-(6-\mathcal{P})\eta/3 + (4-\mathcal{P})\eta^2/4\right]
	\,,
\end{equation}
the time-independent Ginzburg-Landau (GL) equation reads
\begin{equation}
	2\frac{d^2\eta}{d x^2} = \frac{\partial \widetilde{G}}{\partial\eta}
	\,.
\end{equation}
Once again we convert to a non-autonomous equation via $d\eta/dx=f(\eta)$
\begin{equation}
	2\frac{d^2\eta}{dx^2} = 2f(\eta)f'(\eta) = \frac{d}{d\eta}f^2(\eta) = \frac{\partial \widetilde{G}}{\partial\eta}
	\,.
\end{equation}
A trivial integration gives
\begin{equation}
	f(\eta)=\frac{d\eta}{dx}=\sqrt{\widetilde{G}(\eta) - \widetilde{G}_0}
	\,, \label{eq:fofetasol}
\end{equation}
where $\widetilde{G}_0$ is an integration constant.

\subsection*{Martensitic nuclei profiles}

We now calculate the profiles of M nuclei in A (M stable, A metastable) for which $\eta\sim0$ as $x\to\pm\infty$.
Since $\widetilde{G}(\eta=0)=0$ we set $\widetilde{G}_0=0$ and it follows that $d\eta/dx\sim0$ as $\eta\to0$.
The derivative $d\eta/dx$ also vanishes at the roots of $\widetilde{G}(\eta)=0:$
\begin{equation}
	\eta_\pm = \frac{6}{6-\mathcal{P}\mp\sqrt{\mathcal{P}^2-3\mathcal{P}}}\,. \label{eq:etapm}
\end{equation}
For $\mathcal{P}<0$ ($\eta_\pm$ real) we have $\eta_+>1$, which is unphysical, and $0<\eta_-<1$.
We conclude that $\eta_-$ is the value of the order parameter at the center of a M nucleus.

The time-independent GL equation may be written
\begin{equation}
	\frac{d\eta}{dx} = \frac12\sqrt{\frac{s_1}{\beta}(4-\mathcal{P})} \,\,\eta \sqrt{(\eta_+-\eta)(\eta_--\eta)} \,,
\end{equation}
therefore
\begin{align}
	\sqrt{\frac{s_1}{\beta}} \int\limits_0^{x(\eta)}dx' &= \frac{2}{\sqrt{4-\mathcal{P}}} \int\limits_{\eta_-}^{\eta} \frac{d\eta'}{\eta'\sqrt{(\eta_+-\eta')(\eta_--\eta')}}
	\nonumber\\
	&=\ln\left\{\frac{(\eta_+-\eta_-)\eta}{2\eta_+\eta_--(\eta_++\eta_-)\eta+2\sqrt{\eta_+\eta_-(\eta_+-\eta)(\eta_--\eta)}}\right\}
	\,; \label{eq:xprimeintegral}
\end{align}
see \cite[p.97, 2.266]{Gradshteyn:2007}.
The solution of Equation \eqref{eq:xprimeintegral} for $\eta$ is
\begin{equation}
	\eta^\text{M}(x) = \frac{6}{6-\mathcal{P}+\sqrt{\mathcal{P}^2-3\mathcal{P}}\,\cosh\left(\sqrt{\frac{s_1}{\beta}}\, x\right)} \,,
	\label{eq:etaM}
\end{equation}
in agreement with (59) in \cite{Levitas:2003}; $\eta^\text{M}(0)=\eta_-$.

\subsection*{Austenitic nuclei profiles}

For A nuclei in M, $\eta\sim1$ as $x\to\pm\infty$.
In this case $\widetilde{G}(\eta=1)=s_1\mathcal{P}/12\beta$, which is the value of the integration constant $\widetilde{G}_0$.
Equation \eqref{eq:fofetasol} becomes
\begin{equation}
	\frac{d\eta}{dx} = \sqrt{\frac{s_1}{\beta}} \left[-\mathcal{P}/12 + \eta^2 - \left(6-\mathcal{P}\right)\eta^3/3 + \left(4-\mathcal{P}\right)\eta^4/4\right]^{1/2}
	\,.
\end{equation}
This quartic polynomial can be factored into $(1-\eta)^2$ and a quadratic.
With the definition $\zeta=1-\eta$ we obtain
\begin{equation}
	\frac{d\zeta}{dx} = \sqrt{\frac{s_1}{\beta}} \,\zeta \left[1-\frac{\mathcal{P}}{2} +2\left(\frac{\mathcal{P}}{3}-1\right)\zeta +\left(1-\frac{\mathcal{P}}{4}\right)\zeta^2\right]^{1/2}
	\,.
\end{equation}
The roots of the quadratic are
\begin{equation}
	\zeta_\pm = \frac{6(2-\mathcal{P})}{4(3-\mathcal{P}) \mp \sqrt{2\mathcal{P}(6-\mathcal{P})}}
	\,. \label{eq:zetapm}
\end{equation}
Like M nuclei, $\eta_-=1-\zeta_-$ is the value of the order parameter at the center of the A nucleus.
Following the same approach as for M nuclei we obtain
\begin{equation}
	\eta^\text{A}(x) = 1 - \frac{6(2-\mathcal{P})}{4(3-\mathcal{P}) + \sqrt{2\mathcal{P}(6-\mathcal{P})} \cosh\!\left(\sqrt{\frac{s_1}{2\beta}(2-\mathcal{P})}\,x\right)}
	\,; \label{eq:etaA}
\end{equation}
this result was presented previously in \cite{Levitas:2003}, Equation (77).
If $\mathcal{P}<0$ then $\eta^\text{A}$ is a complex number and if $\mathcal{P}>2$ then $\eta^\text{A}>1$, hence $0\leq\mathcal{P}\leq2$.

\subsection*{Critical nuclei energies}

The GL energy is
\begin{align}
	G_\text{GL} &= s_1\eta^2\left[1 - (6-\mathcal{P})\eta/3 + (4-\mathcal{P})\eta^2/4\right] + \beta\left(\frac{d\eta}{dx}\right)^2
	\nonumber\\
	&= \beta \widetilde{G}(\eta) + \beta \left(\frac{d\eta}{dx}\right)^2
	\,.
\end{align}
The energy of a critical nucleus is
\begin{align}
	E &= \beta\int_{-\infty}^\infty\left[\widetilde{G}(\eta) - \widetilde{G}_0 + \left(\frac{d\eta}{dx}\right)^2\right]dx = 2\beta\int_{-\infty}^\infty\left(\frac{d\eta}{dx}\right)^2dx
	\nonumber\\
	&= 4\beta\int d\eta \frac{d\eta}{dx} = 4\beta\int d\eta\sqrt{\widetilde{G}(\eta) - \widetilde{G}_0}
\end{align}
where use was made of Equation \eqref{eq:fofetasol}.
The limits of the $\eta$ integration and the value of $\widetilde{G}_0$ depend on the type of critical nucleus.

\subsubsection*{Martensitic nuclei energies}
For M nuclei we have $\widetilde{G}_0=0$ and $0\leq\eta\leq\eta_-$
\begin{align}
	E^\text{M} &= 4\sqrt{\beta s_1} \int\limits_0^{\eta_-}d\eta\, \eta\sqrt{1 - (6-\mathcal{P})\eta/3 + (4-\mathcal{P})\eta^2/4}
	\nonumber\\
	&= 2\sqrt{\beta s_1(4-\mathcal{P})} \int\limits_0^{\eta_-}d\eta\, \eta\sqrt{(\eta_+-\eta)(\eta_--\eta)}
\end{align}
where $\eta_\pm$ are given in Equation \eqref{eq:etapm}.
Using \cite[2.262]{Gradshteyn:2007} we get
\begin{equation}
	E^\text{M} = \frac89\sqrt{\beta s_1}\frac{1}{(4-\mathcal{P})^2} \left[12 - 6\mathcal{P} + \mathcal{P}^2 + \frac{\mathcal{P}(3-\mathcal{P})(6-\mathcal{P})}{3\sqrt{4-\mathcal{P}}}\ln\!\left(\frac{6-\mathcal{P} + 3\sqrt{4-\mathcal{P}}}{\sqrt{\mathcal{P}^2-3\mathcal{P}}}\right)\right]
	\,. \label{eq:critnucenergyM}
\end{equation}

\subsubsection*{Austenitic nuclei energies}
For A nuclei $\widetilde{G}_0=s_1\mathcal{P}/12\beta$ and $0\leq\zeta\leq\zeta_-$ where $\zeta=1-\eta$ and $\zeta_-$ is given in Equation \eqref{eq:zetapm}.
The energy of an A nucleus is
\begin{align}
	E^\text{A} &= \frac89\sqrt{\beta s_1}\frac{1}{(4-\mathcal{P})^2} \Bigg\{\frac12(12-6\mathcal{P}+\mathcal{P}^2)\sqrt{2(2-\mathcal{P})}
	\nonumber\\
	&\qquad\qquad - \frac{\mathcal{P}(3-\mathcal{P})(6-\mathcal{P})}{3\sqrt{4-\mathcal{P}}}\ln\!\left[\frac{4(3-\mathcal{P}) + 3\sqrt{2(2-\mathcal{P})(4-\mathcal{P})}}{\sqrt{2\mathcal{P}(6-\mathcal{P})}}\right]\Bigg\}
	\,. \label{eq:critnucenergyA}
\end{align}
$E^\text{M}/\sqrt{\beta s_1}$ equals $2/3$ at $\mathcal{P}=0$, goes to zero as $\mathcal{P}\to-\infty$, and is complex for $\mathcal{P}>0$.
$E^\text{A}/\sqrt{\beta s_1}$ also equals $2/3$ at $\mathcal{P}=0$, drops to zero as $\mathcal{P}\to2$, and is complex for $\mathcal{P}<0$ and $\mathcal{P}>2$.




\bibliographystyle{elsarticle-harv}
\bibliography{paper_bib}

\begin{thebibliography}{78}
\expandafter\ifx\csname natexlab\endcsname\relax\def\natexlab#1{#1}\fi
\providecommand{\url}[1]{\texttt{#1}}
\providecommand{\href}[2]{#2}
\providecommand{\path}[1]{#1}
\providecommand{\DOIprefix}{doi:}
\providecommand{\ArXivprefix}{arXiv:}
\providecommand{\URLprefix}{URL: }
\providecommand{\Pubmedprefix}{pmid:}
\providecommand{\doi}[1]{\href{http://dx.doi.org/#1}{\path{#1}}}
\providecommand{\Pubmed}[1]{\href{pmid:#1}{\path{#1}}}
\providecommand{\bibinfo}[2]{#2}
\ifx\xfnm\relax \def\xfnm[#1]{\unskip,\space#1}\fi
\bibitem[{Amadou and {de Ress\'eguier}(2023)}]{PhysRevB.108.174109}
\bibinfo{author}{Amadou, N.}, \bibinfo{author}{{de Ress\'eguier}, T.},
  \bibinfo{year}{2023}.
\newblock \bibinfo{title}{Phase transformations and plasticity in
  single-crystal iron from shock compression to spall fracture}.
\newblock \bibinfo{journal}{Physical Review B} \bibinfo{volume}{108},
  \bibinfo{pages}{174109}.
\newblock \DOIprefix\doi{10.1103/PhysRevB.108.174109}.
\bibitem[{Amadou and {de Ress{\'e}guier}(2025)}]{Amadou:2025}
\bibinfo{author}{Amadou, N.}, \bibinfo{author}{{de Ress{\'e}guier}, T.},
  \bibinfo{year}{2025}.
\newblock \bibinfo{title}{Pressure waves induced by the bcc-hcp phase
  transition in dynamically loaded single crystal iron}.
\newblock \bibinfo{journal}{Computational Materials Science}
  \bibinfo{volume}{247}, \bibinfo{pages}{113559}.
\newblock \DOIprefix\doi{10.1016/j.commatsci.2024.113559}.
\bibitem[{Amadou et~al.(2016)Amadou, de~Resseguier, Brambrink, Vinci,
  Benuzzi-Mounaix, Huser, Morard, Guyot, Miyanishi, Ozaki, Kodama and
  Koenig}]{PhysRevB.93.214108}
\bibinfo{author}{Amadou, N.}, \bibinfo{author}{de~Resseguier, T.},
  \bibinfo{author}{Brambrink, E.}, \bibinfo{author}{Vinci, T.},
  \bibinfo{author}{Benuzzi-Mounaix, A.}, \bibinfo{author}{Huser, G.},
  \bibinfo{author}{Morard, G.}, \bibinfo{author}{Guyot, F.},
  \bibinfo{author}{Miyanishi, K.}, \bibinfo{author}{Ozaki, N.},
  \bibinfo{author}{Kodama, R.}, \bibinfo{author}{Koenig, M.},
  \bibinfo{year}{2016}.
\newblock \bibinfo{title}{Kinetics of the iron
  {$\ensuremath{\alpha}\ensuremath{-}\ensuremath{\varepsilon}$} phase
  transition at high-strain rates: Experiment and model}.
\newblock \bibinfo{journal}{Physical Review B} \bibinfo{volume}{93},
  \bibinfo{pages}{214108}.
\newblock \DOIprefix\doi{10.1103/PhysRevB.93.214108}.
\bibitem[{Ansari et~al.(2021)Ansari, Huang and Shi}]{Ansari:2021}
\bibinfo{author}{Ansari, T.Q.}, \bibinfo{author}{Huang, H.},
  \bibinfo{author}{Shi, S.Q.}, \bibinfo{year}{2021}.
\newblock \bibinfo{title}{Phase field modeling for the morphological and
  microstructural evolution of metallic materials under environmental attack}.
\newblock \bibinfo{journal}{npj Computational Materials} \bibinfo{volume}{7},
  \bibinfo{pages}{143}.
\newblock \DOIprefix\doi{10.1038/s41524-021-00612-7}.
\bibitem[{Asta et~al.(2009)Asta, Beckermann, Karma, Kurz, Napolitano, Plapp,
  Purdy, Rappaz and Trivedi}]{Asta:2009}
\bibinfo{author}{Asta, M.}, \bibinfo{author}{Beckermann, C.},
  \bibinfo{author}{Karma, A.}, \bibinfo{author}{Kurz, W.},
  \bibinfo{author}{Napolitano, R.}, \bibinfo{author}{Plapp, M.},
  \bibinfo{author}{Purdy, G.}, \bibinfo{author}{Rappaz, M.},
  \bibinfo{author}{Trivedi, R.}, \bibinfo{year}{2009}.
\newblock \bibinfo{title}{Solidification microstructures and solid-state
  parallels: Recent developments, future directions}.
\newblock \bibinfo{journal}{Acta Materialia} \bibinfo{volume}{57},
  \bibinfo{pages}{941--971}.
\newblock \DOIprefix\doi{10.1016/j.actamat.2008.10.020}.
\bibitem[{Avrami(1939)}]{Avrami:1939}
\bibinfo{author}{Avrami, M.}, \bibinfo{year}{1939}.
\newblock \bibinfo{title}{Kinetics of phase change. {I General} theory}.
\newblock \bibinfo{journal}{Journal of Chemical Physics} \bibinfo{volume}{7},
  \bibinfo{pages}{1103}.
\newblock \DOIprefix\doi{10.1063/1.1750380}.
\bibitem[{Avrami(1940)}]{Avrami:1940}
\bibinfo{author}{Avrami, M.}, \bibinfo{year}{1940}.
\newblock \bibinfo{title}{Kinetics of phase change. {II Transformation-time}
  relations for random distributions of nuclei}.
\newblock \bibinfo{journal}{Journal of Chemical Physics} \bibinfo{volume}{8},
  \bibinfo{pages}{212}.
\newblock \DOIprefix\doi{10.1063/1.1750631}.
\bibitem[{Avrami(1941)}]{Avrami:1941}
\bibinfo{author}{Avrami, M.}, \bibinfo{year}{1941}.
\newblock \bibinfo{title}{Granulation, phase change, and microstructure
  kinetics of phase change. {III}}.
\newblock \bibinfo{journal}{Journal of Chemical Physics} \bibinfo{volume}{9},
  \bibinfo{pages}{177--184}.
\newblock \DOIprefix\doi{10.1063/1.1750872}.
\bibitem[{Barker and Hollenbach(1974)}]{Barker:1974}
\bibinfo{author}{Barker, L.M.}, \bibinfo{author}{Hollenbach, R.E.},
  \bibinfo{year}{1974}.
\newblock \bibinfo{title}{Shock wave study of the
  {$\alpha\leftrightarrow\varepsilon$} phase transition in iron}.
\newblock \bibinfo{journal}{Journal of Applied Physics} \bibinfo{volume}{45},
  \bibinfo{pages}{4872--4887}.
\newblock \DOIprefix\doi{10.1063/1.1663148}.
\bibitem[{Barton et~al.(2022)Barton, Luscher, Battaile, Brown, Buechler,
  Burakovsky, Crockett, Greeff, Mattsson, Prime and Schill}]{Barton:2022}
\bibinfo{author}{Barton, N.R.}, \bibinfo{author}{Luscher, D.J.},
  \bibinfo{author}{Battaile, C.}, \bibinfo{author}{Brown, J.L.},
  \bibinfo{author}{Buechler, M.}, \bibinfo{author}{Burakovsky, L.},
  \bibinfo{author}{Crockett, S.}, \bibinfo{author}{Greeff, C.},
  \bibinfo{author}{Mattsson, A.E.}, \bibinfo{author}{Prime, M.B.},
  \bibinfo{author}{Schill, W.J.}, \bibinfo{year}{2022}.
\newblock \bibinfo{title}{A multi-phase modeling framework suitable for dynamic
  applications}.
\newblock \bibinfo{journal}{Metals} \bibinfo{volume}{12},
  \bibinfo{pages}{1844}.
\newblock \DOIprefix\doi{10.3390/met12111844}.
\bibitem[{Bastea et~al.(2009)Bastea, Bastea and Becker}]{Bastea:2009}
\bibinfo{author}{Bastea, M.}, \bibinfo{author}{Bastea, S.},
  \bibinfo{author}{Becker, R.}, \bibinfo{year}{2009}.
\newblock \bibinfo{title}{High pressure phase transformation in iron under fast
  compression}.
\newblock \bibinfo{journal}{Applied Physics Letters} \bibinfo{volume}{95},
  \bibinfo{pages}{241911}.
\newblock \DOIprefix\doi{10.1063/1.3275797}.
\bibitem[{Bender and Orszag(1978)}]{Bender:1978}
\bibinfo{author}{Bender, C.M.}, \bibinfo{author}{Orszag, S.A.},
  \bibinfo{year}{1978}.
\newblock \bibinfo{title}{Advanced Mathematical Methods for Scientists and
  Engineers}.
\newblock \bibinfo{publisher}{McGraw-Hill Inc.}, \bibinfo{address}{New York}.
\bibitem[{Blaschke(2025)}]{pyptkinetics}
\bibinfo{author}{Blaschke, D.N.}, \bibinfo{year}{2025}.
\newblock \bibinfo{title}{{PyPTkinetics}}.
\newblock \URLprefix \url{https://github.com/lanl/PyPTkinetics},
  \DOIprefix\doi{10.11578/dc.20250822.5}. \bibinfo{note}{{O4940}}.
\bibitem[{Boettger and Wallace(1997)}]{Boettger:1997}
\bibinfo{author}{Boettger, J.C.}, \bibinfo{author}{Wallace, D.C.},
  \bibinfo{year}{1997}.
\newblock \bibinfo{title}{Metastability and dynamics of the shock-induced phase
  transition in iron}.
\newblock \bibinfo{journal}{Physical Review B} \bibinfo{volume}{55},
  \bibinfo{pages}{2840--2849}.
\newblock \DOIprefix\doi{10.1103/PhysRevB.55.2840}.
\bibitem[{Boettinger et~al.(2000)Boettinger, Coriell, Greer, Karma, Kurz,
  Rappaz and Trivedi}]{Boettinger:2000}
\bibinfo{author}{Boettinger, W.J.}, \bibinfo{author}{Coriell, S.R.},
  \bibinfo{author}{Greer, A.L.}, \bibinfo{author}{Karma, A.},
  \bibinfo{author}{Kurz, W.}, \bibinfo{author}{Rappaz, M.},
  \bibinfo{author}{Trivedi, R.}, \bibinfo{year}{2000}.
\newblock \bibinfo{title}{Solidification microstructures: recent developments,
  future directions}.
\newblock \bibinfo{journal}{Acta Materialia} \bibinfo{volume}{48},
  \bibinfo{pages}{43--70}.
\newblock \DOIprefix\doi{10.1016/S1359-6454(99)00287-6}.
\bibitem[{Bruna et~al.(2006)Bruna, Crespo and Gonz\'{a}lez-Cinca}]{Bruna:2006}
\bibinfo{author}{Bruna, P.}, \bibinfo{author}{Crespo, D.},
  \bibinfo{author}{Gonz\'{a}lez-Cinca, R.}, \bibinfo{year}{2006}.
\newblock \bibinfo{title}{On the validity of {Avrami} formalism in primary
  crystallization}.
\newblock \bibinfo{journal}{Journal of Applied Physics} \bibinfo{volume}{100},
  \bibinfo{pages}{054907}.
\newblock \DOIprefix\doi{10.1063/1.2337407}.
\bibitem[{Burgers(1934)}]{Burgers:1934}
\bibinfo{author}{Burgers, W.}, \bibinfo{year}{1934}.
\newblock \bibinfo{title}{On the process of transition of the
  cubic-body-centered modification into the hexagonal-close-packed modification
  of zirconium}.
\newblock \bibinfo{journal}{Physica} \bibinfo{volume}{1},
  \bibinfo{pages}{561--586}.
\newblock \DOIprefix\doi{10.1016/S0031-8914(34)80244-3}.
\bibitem[{Cahn(1956)}]{Cahn:1956}
\bibinfo{author}{Cahn, J.W.}, \bibinfo{year}{1956}.
\newblock \bibinfo{title}{The kinetics of grain boundary nucleated reactions}.
\newblock \bibinfo{journal}{Acta Metallurgica} \bibinfo{volume}{4},
  \bibinfo{pages}{449--459}.
\newblock \DOIprefix\doi{10.1016/0001-6160(56)90041-4}.
\bibitem[{Cahn(1957)}]{Cahn:1957}
\bibinfo{author}{Cahn, J.W.}, \bibinfo{year}{1957}.
\newblock \bibinfo{title}{Nucleation on dislocations}.
\newblock \bibinfo{journal}{Acta Metallurgica} \bibinfo{volume}{5},
  \bibinfo{pages}{169--172}.
\newblock \DOIprefix\doi{10.1016/0001-6160(57)90021-4}.
\bibitem[{Clayton(2024)}]{Clayton:2024}
\bibinfo{author}{Clayton, J.D.}, \bibinfo{year}{2024}.
\newblock \bibinfo{title}{Analysis of shear localization in viscoplastic solids
  with pressure-sensitive structural transformations}.
\newblock \bibinfo{journal}{Journal of the Mechanics and Physics of Solids}
  \bibinfo{volume}{193}, \bibinfo{pages}{105880}.
\newblock \DOIprefix\doi{10.1016/j.jmps.2024.105880}.
\bibitem[{Clemm and Fisher(1955)}]{Clemm:1955}
\bibinfo{author}{Clemm, P.J.}, \bibinfo{author}{Fisher, J.C.},
  \bibinfo{year}{1955}.
\newblock \bibinfo{title}{The influence of grain boundaries on the nucleation
  of secondary phases}.
\newblock \bibinfo{journal}{Acta Metallurgica} \bibinfo{volume}{3},
  \bibinfo{pages}{70--73}.
\newblock \DOIprefix\doi{10.1016/0001-6160(55)90014-6}.
  \bibinfo{note}{\href{https://doi.org/10.1016/0001-6160(55)90053-5}{Erratum-ibid.
  p. 315}}.
\bibitem[{Daphalapurkar(2024)}]{Daphalapurkar:2024}
\bibinfo{author}{Daphalapurkar, N.P.}, \bibinfo{year}{2024}.
\newblock \bibinfo{title}{Kinetics of {HCP-BCC} phase transition boundary in
  magnesium at high pressure}.
\newblock \bibinfo{journal}{Metals} \bibinfo{volume}{14}, \bibinfo{pages}{609}.
\newblock \DOIprefix\doi{10.3390/met14060609}.
\bibitem[{Davis and Hayes(2007)}]{Davis:2007}
\bibinfo{author}{Davis, J.P.}, \bibinfo{author}{Hayes, D.B.},
  \bibinfo{year}{2007}.
\newblock \bibinfo{title}{{Measurement of the dynamic {$\beta$-$\gamma$} phase
  boundary in tin}}.
\newblock \bibinfo{journal}{AIP Conference Proceedings} \bibinfo{volume}{955},
  \bibinfo{pages}{159--162}.
\newblock \DOIprefix\doi{10.1063/1.2832998}.
\bibitem[{Dup{\'e} et~al.(2013)Dup{\'e}, Amadon, Pellegrini and
  Denoual}]{Bertrand:2013}
\bibinfo{author}{Dup{\'e}, B.}, \bibinfo{author}{Amadon, B.},
  \bibinfo{author}{Pellegrini, Y.P.}, \bibinfo{author}{Denoual, C.},
  \bibinfo{year}{2013}.
\newblock \bibinfo{title}{Mechanism for the
  $\ensuremath{\alpha}\ensuremath{\rightarrow}\ensuremath{\epsilon}$ phase
  transition in iron}.
\newblock \bibinfo{journal}{Physical Review B} \bibinfo{volume}{87},
  \bibinfo{pages}{024103}.
\newblock \DOIprefix\doi{10.1103/PhysRevB.87.024103}.
\bibitem[{{Ehsan Esfahani} et~al.(2020){Ehsan Esfahani}, Ghamarian and
  Levitas}]{EhsanEsfahani:2020}
\bibinfo{author}{{Ehsan Esfahani}, S.}, \bibinfo{author}{Ghamarian, I.},
  \bibinfo{author}{Levitas, V.I.}, \bibinfo{year}{2020}.
\newblock \bibinfo{title}{Strain-induced multivariant martensitic
  transformations: {A} scale-independent simulation of interaction between
  localized shear bands and microstructure}.
\newblock \bibinfo{journal}{Acta Materialia} \bibinfo{volume}{196},
  \bibinfo{pages}{430--443}.
\newblock \DOIprefix\doi{10.1016/j.actamat.2020.06.059}.
\bibitem[{Elder et~al.(2007)Elder, Provatas, Berry, Stefanovic and
  Grant}]{Elder:2007}
\bibinfo{author}{Elder, K.R.}, \bibinfo{author}{Provatas, N.},
  \bibinfo{author}{Berry, J.}, \bibinfo{author}{Stefanovic, P.},
  \bibinfo{author}{Grant, M.}, \bibinfo{year}{2007}.
\newblock \bibinfo{title}{Phase-field crystal modeling and classical density
  functional theory of freezing}.
\newblock \bibinfo{journal}{Physical Review B} \bibinfo{volume}{75},
  \bibinfo{pages}{064107}.
\newblock \DOIprefix\doi{10.1103/PhysRevB.75.064107}.
\bibitem[{Giles et~al.(1971)Giles, Longenbach and Marder}]{Giles:1971}
\bibinfo{author}{Giles, P.M.}, \bibinfo{author}{Longenbach, M.H.},
  \bibinfo{author}{Marder, A.R.}, \bibinfo{year}{1971}.
\newblock \bibinfo{title}{High-pressure {$\alpha\leftrightarrow\epsilon$}
  martensitic transformation in iron}.
\newblock \bibinfo{journal}{Journal of Applied Physics} \bibinfo{volume}{42},
  \bibinfo{pages}{4290--4295}.
\newblock \DOIprefix\doi{10.1063/1.1659768}.
\bibitem[{Gomez et~al.(2019)Gomez, Bures and Moure}]{Gomez:2018}
\bibinfo{author}{Gomez, H.}, \bibinfo{author}{Bures, M.},
  \bibinfo{author}{Moure, A.}, \bibinfo{year}{2019}.
\newblock \bibinfo{title}{A review on computational modelling of
  phase-transition problems}.
\newblock \bibinfo{journal}{Philosophical Transactions of the Royal Society A}
  \bibinfo{volume}{377}, \bibinfo{pages}{20180203}.
\newblock \DOIprefix\doi{10.1098/rsta.2018.0203}.
\bibitem[{Gornostyrev et~al.(1999)Gornostyrev, Katsnel'son, Kuznetsov and
  Trefilov}]{Gornostyrev:1999}
\bibinfo{author}{Gornostyrev, {\relax{Yu}}.N.}, \bibinfo{author}{Katsnel'son,
  M.I.}, \bibinfo{author}{Kuznetsov, A.R.}, \bibinfo{author}{Trefilov, A.V.},
  \bibinfo{year}{1999}.
\newblock \bibinfo{title}{Microscopic description of the kinetics of a
  martensitic transition in real crystals: bcc-hcp transition in {Zr}}.
\newblock \bibinfo{journal}{JETP Letters} \bibinfo{volume}{70},
  \bibinfo{pages}{380--384}.
\newblock \DOIprefix\doi{10.1134/1.568184}.
\bibitem[{Gradshteyn and Ryzhik(2007)}]{Gradshteyn:2007}
\bibinfo{author}{Gradshteyn, I.S.}, \bibinfo{author}{Ryzhik, I.M.},
  \bibinfo{year}{2007}.
\newblock \bibinfo{title}{Table of Integrals, Series and Products}.
\newblock \bibinfo{edition}{seventh} ed., \bibinfo{publisher}{Academic Press}.
\bibitem[{Gunkelmann et~al.(2015)Gunkelmann, Tramontina, Bringa and
  Urbassek}]{Gunkelmann:2015}
\bibinfo{author}{Gunkelmann, N.}, \bibinfo{author}{Tramontina, D.R.},
  \bibinfo{author}{Bringa, E.M.}, \bibinfo{author}{Urbassek, H.M.},
  \bibinfo{year}{2015}.
\newblock \bibinfo{title}{Morphological changes in polycrystalline {Fe} after
  compression and release}.
\newblock \bibinfo{journal}{Journal of Applied Physics} \bibinfo{volume}{117},
  \bibinfo{pages}{085901}.
\newblock \DOIprefix\doi{10.1063/1.4913622}.
\bibitem[{Guo et~al.(2021)Guo, Shao and Lu}]{Guo:2021}
\bibinfo{author}{Guo, X.X.}, \bibinfo{author}{Shao, J.L.}, \bibinfo{author}{Lu,
  G.}, \bibinfo{year}{2021}.
\newblock \bibinfo{title}{Reversibility of the structural transition in single
  crystal iron driven by uniaxial and triaxial strains: {Atomistic} study}.
\newblock \bibinfo{journal}{International Journal of Mechanical Sciences}
  \bibinfo{volume}{191}, \bibinfo{pages}{106064}.
\newblock \DOIprefix\doi{10.1016/j.ijmecsci.2020.106064}.
\bibitem[{Hawreliak et~al.(2011)Hawreliak, El-Dasher, Lorenzana, Kimminau,
  Higginbotham, Nagler, Vinko, Murphy, Whitcher, Wark, Rothman and
  Park}]{Hawreliak:2011}
\bibinfo{author}{Hawreliak, J.A.}, \bibinfo{author}{El-Dasher, B.},
  \bibinfo{author}{Lorenzana, H.}, \bibinfo{author}{Kimminau, G.},
  \bibinfo{author}{Higginbotham, A.}, \bibinfo{author}{Nagler, B.},
  \bibinfo{author}{Vinko, S.M.}, \bibinfo{author}{Murphy, W.J.},
  \bibinfo{author}{Whitcher, T.}, \bibinfo{author}{Wark, J.S.},
  \bibinfo{author}{Rothman, S.}, \bibinfo{author}{Park, N.},
  \bibinfo{year}{2011}.
\newblock \bibinfo{title}{In situ x-ray diffraction measurements of the $c/a$
  ratio in the high-pressure $\ensuremath{\epsilon}$ phase of shock-compressed
  polycrystalline iron}.
\newblock \bibinfo{journal}{Physical Review B} \bibinfo{volume}{83},
  \bibinfo{pages}{144114}.
\newblock \DOIprefix\doi{10.1103/PhysRevB.83.144114}.
\bibitem[{Hayes(2007)}]{Hayes:2007}
\bibinfo{author}{Hayes, D.B.}, \bibinfo{year}{2007}.
\newblock \bibinfo{title}{Analyzing ramp compression wave experiments}.
\newblock \bibinfo{journal}{AIP Conference Proceedings} \bibinfo{volume}{955},
  \bibinfo{pages}{1181--1185}.
\newblock \DOIprefix\doi{10.1063/1.2832931}.
\bibitem[{Hunter and Preston(2022)}]{Hunter:2022}
\bibinfo{author}{Hunter, A.}, \bibinfo{author}{Preston, D.L.},
  \bibinfo{year}{2022}.
\newblock \bibinfo{title}{Analytic model of dislocation density evolution in
  fcc polycrystals accounting for dislocation generation, storage, and dynamic
  recovery mechanisms}.
\newblock \bibinfo{journal}{International Journal of Plasticity}
  \bibinfo{volume}{151}, \bibinfo{pages}{103178}.
\newblock \DOIprefix\doi{10.1016/j.ijplas.2021.103178}.
\bibitem[{Javanbakht and Levitas(2016)}]{Javanbakht:2016}
\bibinfo{author}{Javanbakht, M.}, \bibinfo{author}{Levitas, V.I.},
  \bibinfo{year}{2016}.
\newblock \bibinfo{title}{Phase field simulations of plastic strain-induced
  phase transformations under high pressure and large shear}.
\newblock \bibinfo{journal}{Physical Review B} \bibinfo{volume}{94},
  \bibinfo{pages}{214104}.
\newblock \DOIprefix\doi{10.1103/PhysRevB.94.214104}.
\bibitem[{Jensen et~al.(2009)Jensen, Gray and Hixson}]{Jensen:2009}
\bibinfo{author}{Jensen, B.J.}, \bibinfo{author}{Gray, {III}, G.T.},
  \bibinfo{author}{Hixson, R.S.}, \bibinfo{year}{2009}.
\newblock \bibinfo{title}{{Direct measurements of the {$\alpha\to\epsilon$}
  transition stress and kinetics for shocked iron}}.
\newblock \bibinfo{journal}{Journal of Applied Physics} \bibinfo{volume}{105},
  \bibinfo{pages}{103502}.
\newblock \DOIprefix\doi{10.1063/1.3110188}.
\bibitem[{Jiang et~al.(2024)Jiang, Li and Suo}]{JIANG2024103968}
\bibinfo{author}{Jiang, K.}, \bibinfo{author}{Li, J.}, \bibinfo{author}{Suo,
  T.}, \bibinfo{year}{2024}.
\newblock \bibinfo{title}{Extensive phase transformation in an equiatomic
  {CrCoNi} medium entropy alloy under extreme uniaxial tension}.
\newblock \bibinfo{journal}{International Journal of Plasticity}
  \bibinfo{volume}{176}, \bibinfo{pages}{103968}.
\newblock \DOIprefix\doi{10.1016/j.ijplas.2024.103968}.
\bibitem[{Johnson and Mehl(1939)}]{Johnson:1939}
\bibinfo{author}{Johnson, W.A.}, \bibinfo{author}{Mehl, P.A.},
  \bibinfo{year}{1939}.
\newblock \bibinfo{title}{Reaction kinetics in processes of nucleation and
  growth}.
\newblock \bibinfo{journal}{Transactions of the Metallurgical Society of AIME}
  \bibinfo{volume}{135}, \bibinfo{pages}{416}.
\bibitem[{Kadau et~al.(2005)Kadau, Germann, Lomdahl and Holian}]{Kadau:2005}
\bibinfo{author}{Kadau, K.}, \bibinfo{author}{Germann, T.C.},
  \bibinfo{author}{Lomdahl, P.S.}, \bibinfo{author}{Holian, B.L.},
  \bibinfo{year}{2005}.
\newblock \bibinfo{title}{Atomistic simulations of shock-induced
  transformations and their orientation dependence in bcc {Fe} single
  crystals}.
\newblock \bibinfo{journal}{Physical Review B} \bibinfo{volume}{72},
  \bibinfo{pages}{064120}.
\newblock \DOIprefix\doi{10.1103/PhysRevB.72.064120}.
\bibitem[{Kalantar et~al.(2005)Kalantar, Belak, Collins, Colvin, Davies,
  Eggert, Germann, Hawreliak, Holian, Kadau, Lomdahl, Lorenzana, Meyers,
  Rosolankova, Schneider, Sheppard, St{\"o}lken and Wark}]{Kalantar:2005}
\bibinfo{author}{Kalantar, D.H.}, \bibinfo{author}{Belak, J.F.},
  \bibinfo{author}{Collins, G.W.}, \bibinfo{author}{Colvin, J.D.},
  \bibinfo{author}{Davies, H.M.}, \bibinfo{author}{Eggert, J.H.},
  \bibinfo{author}{Germann, T.C.}, \bibinfo{author}{Hawreliak, J.},
  \bibinfo{author}{Holian, B.L.}, \bibinfo{author}{Kadau, K.},
  \bibinfo{author}{Lomdahl, P.S.}, \bibinfo{author}{Lorenzana, H.E.},
  \bibinfo{author}{Meyers, M.A.}, \bibinfo{author}{Rosolankova, K.},
  \bibinfo{author}{Schneider, M.S.}, \bibinfo{author}{Sheppard, J.},
  \bibinfo{author}{St{\"o}lken, J.S.}, \bibinfo{author}{Wark, J.S.},
  \bibinfo{year}{2005}.
\newblock \bibinfo{title}{Direct observation of the {$\alpha$--$\varepsilon$}
  transition in shock-compressed iron via nanosecond x-ray diffraction}.
\newblock \bibinfo{journal}{Physical Review Letters} \bibinfo{volume}{95},
  \bibinfo{pages}{075502}.
\newblock \DOIprefix\doi{10.1103/PhysRevLett.95.075502}.
\bibitem[{Kolmogorov(1937)}]{Kolmogorov:1937}
\bibinfo{author}{Kolmogorov, A.N.}, \bibinfo{year}{1937}.
\newblock \bibinfo{title}{On the statistical theory of crystallization of
  metals}.
\newblock \bibinfo{journal}{Bulletin of the Academy of Sciences of the USSR.
  Mathematics Series} \bibinfo{volume}{3}, \bibinfo{pages}{355--359}.
\bibitem[{Lahiri(2022)}]{Lahiri:2022}
\bibinfo{author}{Lahiri, A.}, \bibinfo{year}{2022}.
\newblock \bibinfo{title}{Phase-field modeling of phase transformations in
  multicomponent alloys: {A} review}.
\newblock \bibinfo{journal}{Journal of the Indian Institute of Science}
  \bibinfo{volume}{102}, \bibinfo{pages}{39--57}.
\newblock \DOIprefix\doi{10.1007/s41745-022-00288-y}.
\bibitem[{Lazicki et~al.(2015)Lazicki, Rygg, Coppari, Smith, Fratanduono,
  Kraus, Collins, Briggs, Braun, Swift and Eggert}]{Lazicki:2015}
\bibinfo{author}{Lazicki, A.}, \bibinfo{author}{Rygg, J.R.},
  \bibinfo{author}{Coppari, F.}, \bibinfo{author}{Smith, R.},
  \bibinfo{author}{Fratanduono, D.}, \bibinfo{author}{Kraus, R.G.},
  \bibinfo{author}{Collins, G.W.}, \bibinfo{author}{Briggs, R.},
  \bibinfo{author}{Braun, D.G.}, \bibinfo{author}{Swift, D.C.},
  \bibinfo{author}{Eggert, J.H.}, \bibinfo{year}{2015}.
\newblock \bibinfo{title}{X-ray diffraction of solid tin to 1.2 {TPa}}.
\newblock \bibinfo{journal}{Physical Review Letters} \bibinfo{volume}{115},
  \bibinfo{pages}{075502}.
\newblock \DOIprefix\doi{10.1103/PhysRevLett.115.075502}.
\bibitem[{Levitas(2021)}]{Levitas:2021}
\bibinfo{author}{Levitas, V.I.}, \bibinfo{year}{2021}.
\newblock \bibinfo{title}{Phase transformations, fracture, and other structural
  changes in inelastic materials}.
\newblock \bibinfo{journal}{International Journal of Plasticity}
  \bibinfo{volume}{140}, \bibinfo{pages}{102914}.
\newblock \DOIprefix\doi{10.1016/j.ijplas.2020.102914}.
\bibitem[{Levitas et~al.(2010)Levitas, Lee and Preston}]{Levitas:2010}
\bibinfo{author}{Levitas, V.I.}, \bibinfo{author}{Lee, D.W.},
  \bibinfo{author}{Preston, D.L.}, \bibinfo{year}{2010}.
\newblock \bibinfo{title}{Interface propagation and microstructure evolution in
  phase field models of stress-induced martensitic phase transformations}.
\newblock \bibinfo{journal}{International Journal of Plasticity}
  \bibinfo{volume}{26}, \bibinfo{pages}{395--422}.
\newblock \DOIprefix\doi{10.1016/j.ijplas.2009.08.003}.
\bibitem[{Levitas and Preston(2002a)}]{Levitas:2002}
\bibinfo{author}{Levitas, V.I.}, \bibinfo{author}{Preston, D.L.},
  \bibinfo{year}{2002}a.
\newblock \bibinfo{title}{Three-dimensional {Landau} theory for multivariant
  stress-induced martensitic phase transformations. {I.
  Austenite$\leftrightarrow$martensite}}.
\newblock \bibinfo{journal}{Physical Review B} \bibinfo{volume}{66},
  \bibinfo{pages}{134206}.
\newblock \DOIprefix\doi{10.1103/PhysRevB.66.134206}.
\bibitem[{Levitas and Preston(2002b)}]{Levitas2:2002}
\bibinfo{author}{Levitas, V.I.}, \bibinfo{author}{Preston, D.L.},
  \bibinfo{year}{2002}b.
\newblock \bibinfo{title}{Three-dimensional {Landau} theory for multivariant
  stress-induced martensitic phase transformations. {II. Multivariant} phase
  transformations and stress space analysis}.
\newblock \bibinfo{journal}{Physical Review B} \bibinfo{volume}{66},
  \bibinfo{pages}{134207}.
\newblock \DOIprefix\doi{10.1103/PhysRevB.66.134207}.
\bibitem[{Levitas et~al.(2003)Levitas, Preston and Lee}]{Levitas:2003}
\bibinfo{author}{Levitas, V.I.}, \bibinfo{author}{Preston, D.L.},
  \bibinfo{author}{Lee, D.W.}, \bibinfo{year}{2003}.
\newblock \bibinfo{title}{Three-dimensional {Landau} theory for multivariant
  stress-induced martensitic phase transformations. {III. Alternative}
  potentials, critical nuclei, kink solutions, and dislocation theory}.
\newblock \bibinfo{journal}{Physical Review B} \bibinfo{volume}{68},
  \bibinfo{pages}{134201}.
\newblock \DOIprefix\doi{10.1103/PhysRevB.68.134201}.
\bibitem[{Li et~al.(2025)Li, Xu, Takagi, Wang, Yang, Zhou, Chen, Miyanishi,
  Sueda, Yabuuchi, Ichiyanagi, Nakamura, Amouretti, Ozaki, Wu and Hu}]{Li:2025}
\bibinfo{author}{Li, J.}, \bibinfo{author}{Xu, L.}, \bibinfo{author}{Takagi,
  S.}, \bibinfo{author}{Wang, Q.}, \bibinfo{author}{Yang, J.},
  \bibinfo{author}{Zhou, M.}, \bibinfo{author}{Chen, S.},
  \bibinfo{author}{Miyanishi, K.}, \bibinfo{author}{Sueda, K.},
  \bibinfo{author}{Yabuuchi, T.}, \bibinfo{author}{Ichiyanagi, K.},
  \bibinfo{author}{Nakamura, H.}, \bibinfo{author}{Amouretti, A.},
  \bibinfo{author}{Ozaki, N.}, \bibinfo{author}{Wu, Q.}, \bibinfo{author}{Hu,
  J.}, \bibinfo{year}{2025}.
\newblock \bibinfo{title}{Kinetics of the bct-bcc phase transformation in tin
  revealed by ultrafast x-ray diffraction}.
\newblock \bibinfo{journal}{Applied Physics Letters} \bibinfo{volume}{126},
  \bibinfo{pages}{264102}.
\newblock \DOIprefix\doi{10.1063/5.0274580}.
\bibitem[{Liu et~al.(2023)Liu, Jing, Geng, Li, Zhang, Li, Li, Chen, Gao and
  Wu}]{Liu:2023}
\bibinfo{author}{Liu, L.}, \bibinfo{author}{Jing, Q.}, \bibinfo{author}{Geng,
  H.Y.}, \bibinfo{author}{Li, Y.}, \bibinfo{author}{Zhang, Y.},
  \bibinfo{author}{Li, J.}, \bibinfo{author}{Li, S.}, \bibinfo{author}{Chen,
  X.}, \bibinfo{author}{Gao, J.}, \bibinfo{author}{Wu, Q.},
  \bibinfo{year}{2023}.
\newblock \bibinfo{title}{Revisiting the high-pressure behaviors of zirconium:
  {Nonhydrostaticity} promoting the phase transitions and absence of the
  isostructural phase transition in {$\beta$}-zirconium}.
\newblock \bibinfo{journal}{Materials} \bibinfo{volume}{16},
  \bibinfo{pages}{5157}.
\newblock \DOIprefix\doi{10.3390/ma16145157}.
\bibitem[{Luu et~al.(2019)Luu, Veiga and Gunkelmann}]{Luu:2019}
\bibinfo{author}{Luu, H.T.}, \bibinfo{author}{Veiga, R.G.A.},
  \bibinfo{author}{Gunkelmann, N.}, \bibinfo{year}{2019}.
\newblock \bibinfo{title}{Atomistic study of the role of defects on
  {$\alpha\to\epsilon$} phase transformations in iron under hydrostatic
  compression}.
\newblock \bibinfo{journal}{Metals} \bibinfo{volume}{9}, \bibinfo{pages}{1040}.
\newblock \DOIprefix\doi{10.3390/met9101040}.
\bibitem[{Ma and Dongare(2022)}]{Ma:2022}
\bibinfo{author}{Ma, K.}, \bibinfo{author}{Dongare, A.M.},
  \bibinfo{year}{2022}.
\newblock \bibinfo{title}{Role of {$\alpha\to\epsilon\to\alpha$} phase
  transformation on the spall behavior of iron at atomic scales}.
\newblock \bibinfo{journal}{Journal of Materials Science} \bibinfo{volume}{57},
  \bibinfo{pages}{12556--12571}.
\newblock \DOIprefix\doi{10.1007/s10853-022-07381-8}.
\bibitem[{Merkel et~al.(2020)Merkel, Lincot and Petitgirard}]{Merkel:2020}
\bibinfo{author}{Merkel, S.}, \bibinfo{author}{Lincot, A.},
  \bibinfo{author}{Petitgirard, S.}, \bibinfo{year}{2020}.
\newblock \bibinfo{title}{Microstructural effects and mechanism of bcc-hcp-bcc
  transformations in polycrystalline iron}.
\newblock \bibinfo{journal}{Physical Review B} \bibinfo{volume}{102},
  \bibinfo{pages}{104103}.
\newblock \DOIprefix\doi{10.1103/PhysRevB.102.104103}.
\bibitem[{Pang et~al.(2014)Pang, Zhang, Zhang, Xu and Zhao}]{Pang2014SR}
\bibinfo{author}{Pang, W.W.}, \bibinfo{author}{Zhang, P.},
  \bibinfo{author}{Zhang, G.C.}, \bibinfo{author}{Xu, A.G.},
  \bibinfo{author}{Zhao, X.G.}, \bibinfo{year}{2014}.
\newblock \bibinfo{title}{Morphology and growth speed of hcp domains during
  shock-induced phase transition in iron}.
\newblock \bibinfo{journal}{Scientific Reports} \bibinfo{volume}{4},
  \bibinfo{pages}{3628}.
\newblock \DOIprefix\doi{10.1038/srep03628}.
\bibitem[{Pei et~al.(2024)Pei, Pan, Zhong, Xiang and Lv}]{Pei:2024}
\bibinfo{author}{Pei, G.}, \bibinfo{author}{Pan, C.}, \bibinfo{author}{Zhong,
  D.}, \bibinfo{author}{Xiang, J.}, \bibinfo{author}{Lv, X.},
  \bibinfo{year}{2024}.
\newblock \bibinfo{title}{Crystal structure, phase transitions, and
  thermodynamic properties of magnesium metavanadate ({MgV2O6})}.
\newblock \bibinfo{journal}{Journal of Magnesium and Alloys}
  \bibinfo{volume}{12}, \bibinfo{pages}{1449--1460}.
\newblock \DOIprefix\doi{10.1016/j.jma.2022.05.011}.
\bibitem[{Pitsch and Schrader(1958)}]{PitschSchrader}
\bibinfo{author}{Pitsch, W.}, \bibinfo{author}{Schrader, A.},
  \bibinfo{year}{1958}.
\newblock \bibinfo{title}{{Die Ausscheidungsform des $\varepsilon$-Karbids im
  Ferrit und im Martensit beim Anlassen}}.
\newblock \bibinfo{journal}{Archiv f{\"u}r das Eisenh{\"u}ttenwesen}
  \bibinfo{volume}{29}, \bibinfo{pages}{715--721}.
\newblock \DOIprefix\doi{10.1002/srin.195803018}.
\bibitem[{Preston et~al.(2003)Preston, Tonks and Wallace}]{PTW:2003}
\bibinfo{author}{Preston, D.L.}, \bibinfo{author}{Tonks, D.L.},
  \bibinfo{author}{Wallace, D.C.}, \bibinfo{year}{2003}.
\newblock \bibinfo{title}{Model of plastic deformation for extreme loading
  conditions}.
\newblock \bibinfo{journal}{Journal of Applied Physics} \bibinfo{volume}{93},
  \bibinfo{pages}{211--220}.
\newblock \DOIprefix\doi{10.1063/1.1524706}.
\bibitem[{Rawat and Rastogi(2025)}]{Rawat:2025}
\bibinfo{author}{Rawat, S.}, \bibinfo{author}{Rastogi, V.},
  \bibinfo{year}{2025}.
\newblock \bibinfo{title}{Role of crystal orientations and loading conditions
  on the microstructure evolution and void evolution dynamics in single crystal
  iron: {An} atomistic investigation}.
\newblock \bibinfo{journal}{Mechanics of Materials} \bibinfo{volume}{207},
  \bibinfo{pages}{105387}.
\newblock \DOIprefix\doi{10.1016/j.mechmat.2025.105387}.
\bibitem[{de~Ress{\'e}guier and Hallouin(2008)}]{Resseguier:2008}
\bibinfo{author}{de~Ress{\'e}guier, T.}, \bibinfo{author}{Hallouin, M.},
  \bibinfo{year}{2008}.
\newblock \bibinfo{title}{Effects of the
  {$\ensuremath{\alpha}\text{\ensuremath{-}}\ensuremath{\epsilon}$} phase
  transition on wave propagation and spallation in laser shock-loaded iron}.
\newblock \bibinfo{journal}{Physical Review B} \bibinfo{volume}{77},
  \bibinfo{pages}{174107}.
\newblock \DOIprefix\doi{10.1103/PhysRevB.77.174107}.
\bibitem[{Ricks et~al.(2024)Ricks, Gleason, Miozzi, Yang, Chariton, Prakapenka,
  Sinogeikin, Sandberg, Mao and Pandolfi}]{PhysRevResearch.6.013316}
\bibinfo{author}{Ricks, M.}, \bibinfo{author}{Gleason, A.E.},
  \bibinfo{author}{Miozzi, F.}, \bibinfo{author}{Yang, H.},
  \bibinfo{author}{Chariton, S.}, \bibinfo{author}{Prakapenka, V.B.},
  \bibinfo{author}{Sinogeikin, S.V.}, \bibinfo{author}{Sandberg, R.L.},
  \bibinfo{author}{Mao, W.L.}, \bibinfo{author}{Pandolfi, S.},
  \bibinfo{year}{2024}.
\newblock \bibinfo{title}{Phase transition kinetics revealed by in situ x-ray
  diffraction in laser-heated dynamic diamond anvil cells}.
\newblock \bibinfo{journal}{Physical Review Research} \bibinfo{volume}{6},
  \bibinfo{pages}{013316}.
\newblock \DOIprefix\doi{10.1103/PhysRevResearch.6.013316}.
\bibitem[{Rigg et~al.(2009)Rigg, Greeff, Knudson, Gray and Hixson}]{Rigg:2009}
\bibinfo{author}{Rigg, P.A.}, \bibinfo{author}{Greeff, C.W.},
  \bibinfo{author}{Knudson, M.D.}, \bibinfo{author}{Gray, G.T.},
  \bibinfo{author}{Hixson, R.S.}, \bibinfo{year}{2009}.
\newblock \bibinfo{title}{Influence of impurities on the {$\alpha$ to $\omega$}
  phase transition in zirconium under dynamic loading conditions}.
\newblock \bibinfo{journal}{Journal of Applied Physics} \bibinfo{volume}{106},
  \bibinfo{pages}{123532}.
\newblock \DOIprefix\doi{10.1063/1.3267325}.
\bibitem[{Righi et~al.(2023)Righi, Briggs, Deluigi, Stan, Singh, Clarke,
  Bringa, Smith, Rudd, Park and Meyers}]{Righi:2023}
\bibinfo{author}{Righi, G.}, \bibinfo{author}{Briggs, R.},
  \bibinfo{author}{Deluigi, O.R.}, \bibinfo{author}{Stan, C.V.},
  \bibinfo{author}{Singh, S.}, \bibinfo{author}{Clarke, S.M.},
  \bibinfo{author}{Bringa, E.M.}, \bibinfo{author}{Smith, R.F.},
  \bibinfo{author}{Rudd, R.E.}, \bibinfo{author}{Park, H.S.},
  \bibinfo{author}{Meyers, M.A.}, \bibinfo{year}{2023}.
\newblock \bibinfo{title}{A spall and diffraction study of nanosecond pressure
  release across the iron {$\epsilon$-$\alpha$} phase boundary}.
\newblock \bibinfo{journal}{Acta Materialia} \bibinfo{volume}{257},
  \bibinfo{pages}{119148}.
\newblock \DOIprefix\doi{10.1016/j.actamat.2023.119148}.
\bibitem[{Sikka et~al.(1982)Sikka, Vohra and Chidambaram}]{Sikka:1982}
\bibinfo{author}{Sikka, S.K.}, \bibinfo{author}{Vohra, Y.K.},
  \bibinfo{author}{Chidambaram, R.}, \bibinfo{year}{1982}.
\newblock \bibinfo{title}{Omega phase in materials}.
\newblock \bibinfo{journal}{Progress in Materials Science}
  \bibinfo{volume}{27}, \bibinfo{pages}{245--310}.
\newblock \DOIprefix\doi{10.1016/0079-6425(82)90002-0}.
\bibitem[{Smith et~al.(2013)Smith, Eggert, Swift, Wang, Duffy, Braun, Rudd,
  Reisman, Davis, Knudson and Collins}]{Smith:2013}
\bibinfo{author}{Smith, R.F.}, \bibinfo{author}{Eggert, J.H.},
  \bibinfo{author}{Swift, D.C.}, \bibinfo{author}{Wang, J.},
  \bibinfo{author}{Duffy, T.S.}, \bibinfo{author}{Braun, D.G.},
  \bibinfo{author}{Rudd, R.E.}, \bibinfo{author}{Reisman, D.B.},
  \bibinfo{author}{Davis, J.P.}, \bibinfo{author}{Knudson, M.D.},
  \bibinfo{author}{Collins, G.W.}, \bibinfo{year}{2013}.
\newblock \bibinfo{title}{Time-dependence of the alpha to epsilon phase
  transformation in iron}.
\newblock \bibinfo{journal}{Journal of Applied Physics} \bibinfo{volume}{114},
  \bibinfo{pages}{223507}.
\newblock \DOIprefix\doi{10.1063/1.4839655}.
\bibitem[{Sohrabi et~al.(2024)Sohrabi, Mehranpour, Heydarinia, Kalhor, Lee,
  Mirzadeh, Mahmudi, Parsa, Rodak and Kim}]{Sohrabi:2024}
\bibinfo{author}{Sohrabi, M.J.}, \bibinfo{author}{Mehranpour, M.S.},
  \bibinfo{author}{Heydarinia, A.}, \bibinfo{author}{Kalhor, A.},
  \bibinfo{author}{Lee, J.H.}, \bibinfo{author}{Mirzadeh, H.},
  \bibinfo{author}{Mahmudi, R.}, \bibinfo{author}{Parsa, M.H.},
  \bibinfo{author}{Rodak, K.}, \bibinfo{author}{Kim, H.S.},
  \bibinfo{year}{2024}.
\newblock \bibinfo{title}{Deformation-induced martensitic transformation
  kinetics in {TRIP}-assisted steels and high-entropy alloys}.
\newblock \bibinfo{journal}{Acta Materialia} \bibinfo{volume}{280},
  \bibinfo{pages}{120354}.
\newblock \DOIprefix\doi{10.1016/j.actamat.2024.120354}.
\bibitem[{Sosso et~al.(2016)Sosso, Chen, Cox, Fitzner, Pedevilla, Zen and
  Michaelides}]{Sosso:2016}
\bibinfo{author}{Sosso, G.C.}, \bibinfo{author}{Chen, J.},
  \bibinfo{author}{Cox, S.J.}, \bibinfo{author}{Fitzner, M.},
  \bibinfo{author}{Pedevilla, P.}, \bibinfo{author}{Zen, A.},
  \bibinfo{author}{Michaelides, A.}, \bibinfo{year}{2016}.
\newblock \bibinfo{title}{Crystal nucleation in liquids: Open questions and
  future challenges in molecular dynamics simulations}.
\newblock \bibinfo{journal}{Chemical Reviews} \bibinfo{volume}{116},
  \bibinfo{pages}{7078--7116}.
\newblock \DOIprefix\doi{10.1021/acs.chemrev.5b00744}.
\bibitem[{Takahashi and Bassett(1964)}]{Takahashi:1964}
\bibinfo{author}{Takahashi, T.}, \bibinfo{author}{Bassett, W.A.},
  \bibinfo{year}{1964}.
\newblock \bibinfo{title}{High-pressure polymorph of iron}.
\newblock \bibinfo{journal}{Science} \bibinfo{volume}{145},
  \bibinfo{pages}{483--486}.
\newblock \DOIprefix\doi{10.1126/science.145.3631.483}.
\bibitem[{Taylor et~al.(1991)Taylor, Pasternak and Jeanloz}]{Taylor:1991}
\bibinfo{author}{Taylor, R.D.}, \bibinfo{author}{Pasternak, M.P.},
  \bibinfo{author}{Jeanloz, R.}, \bibinfo{year}{1991}.
\newblock \bibinfo{title}{Hysteresis in the high pressure transformation of
  bcc- to hcp-iron}.
\newblock \bibinfo{journal}{Journal of Applied Physics} \bibinfo{volume}{69},
  \bibinfo{pages}{6126--6128}.
\newblock \DOIprefix\doi{10.1063/1.348779}.
\bibitem[{Tol{\'e}dano and Dmitriev(1996)}]{Toledano:1996}
\bibinfo{author}{Tol{\'e}dano, P.}, \bibinfo{author}{Dmitriev, V.},
  \bibinfo{year}{1996}.
\newblock \bibinfo{title}{Reconstructive Phase Transitions}.
\newblock \bibinfo{publisher}{World Scientific}, \bibinfo{address}{Singapore}.
\newblock \DOIprefix\doi{10.1142/2848}.
\bibitem[{Tourret et~al.(2022)Tourret, Liu and LLorca}]{Tourret:2022}
\bibinfo{author}{Tourret, D.}, \bibinfo{author}{Liu, H.},
  \bibinfo{author}{LLorca, J.}, \bibinfo{year}{2022}.
\newblock \bibinfo{title}{Phase-field modeling of microstructure evolution:
  {Recent} applications, perspectives and challenges}.
\newblock \bibinfo{journal}{Progress in Materials Science}
  \bibinfo{volume}{123}, \bibinfo{pages}{100810}.
\newblock \DOIprefix\doi{10.1016/j.pmatsci.2021.100810}. \bibinfo{note}{{A
  Festschrift in Honor of Brian Cantor}}.
\bibitem[{Wallace(1972)}]{Wallace:1972}
\bibinfo{author}{Wallace, D.C.}, \bibinfo{year}{1972}.
\newblock \bibinfo{title}{Thermodynamics of Crystals}.
\newblock \bibinfo{publisher}{J. Wiley \& Sons Inc.}, \bibinfo{address}{New
  York}.
\bibitem[{Weinstock(1952)}]{Weinstock:1952}
\bibinfo{author}{Weinstock, R.}, \bibinfo{year}{1952}.
\newblock \bibinfo{title}{Calculus of Variations}.
\newblock \bibinfo{publisher}{McGraw-Hill}, \bibinfo{address}{New York}.
\bibitem[{Wu and Shao(2023)}]{WU2023103730}
\bibinfo{author}{Wu, Y.C.}, \bibinfo{author}{Shao, J.L.}, \bibinfo{year}{2023}.
\newblock \bibinfo{title}{{FCC-BCC} phase transformation induced simultaneous
  enhancement of tensile strength and ductility at high strain rate in
  high-entropy alloy}.
\newblock \bibinfo{journal}{International Journal of Plasticity}
  \bibinfo{volume}{169}, \bibinfo{pages}{103730}.
\newblock \DOIprefix\doi{10.1016/j.ijplas.2023.103730}.
\bibitem[{Xu et~al.(2025)Xu, Yu, Xiong, Hu, Kan, Wang, Wang and Kang}]{Xu:2025}
\bibinfo{author}{Xu, B.}, \bibinfo{author}{Yu, C.}, \bibinfo{author}{Xiong,
  J.}, \bibinfo{author}{Hu, J.}, \bibinfo{author}{Kan, Q.},
  \bibinfo{author}{Wang, C.}, \bibinfo{author}{Wang, Q.},
  \bibinfo{author}{Kang, G.}, \bibinfo{year}{2025}.
\newblock \bibinfo{title}{Progress in phase field modeling of functional
  properties and fracture behavior of shape memory alloys}.
\newblock \bibinfo{journal}{Progress in Materials Science}
  \bibinfo{volume}{148}, \bibinfo{pages}{101364}.
\newblock \DOIprefix\doi{10.1016/j.pmatsci.2024.101364}.
\bibitem[{Yang et~al.(2025)Yang, Qin, Feng, Ren, Chen, Wang and
  Chen}]{YANG2025104403}
\bibinfo{author}{Yang, X.}, \bibinfo{author}{Qin, G.}, \bibinfo{author}{Feng,
  L.}, \bibinfo{author}{Ren, H.}, \bibinfo{author}{Chen, Y.},
  \bibinfo{author}{Wang, Q.}, \bibinfo{author}{Chen, R.}, \bibinfo{year}{2025}.
\newblock \bibinfo{title}{Development of high-strength and high-ductility
  high-entropy alloys via directional solidification: {A} multifaceted
  strengthening approach}.
\newblock \bibinfo{journal}{International Journal of Plasticity}
  \bibinfo{volume}{191}, \bibinfo{pages}{104403}.
\newblock \DOIprefix\doi{10.1016/j.ijplas.2025.104403}.
\bibitem[{Yao et~al.(2024)Yao, Yu, Pei, Cui, Zhang, Peng, Li and Wu}]{Yao:2024}
\bibinfo{author}{Yao, S.}, \bibinfo{author}{Yu, J.}, \bibinfo{author}{Pei, X.},
  \bibinfo{author}{Cui, Y.}, \bibinfo{author}{Zhang, H.},
  \bibinfo{author}{Peng, H.}, \bibinfo{author}{Li, Y.}, \bibinfo{author}{Wu,
  Q.}, \bibinfo{year}{2024}.
\newblock \bibinfo{title}{A coupled phase-field and crystal plasticity model
  for understanding shock-induced phase transition of iron}.
\newblock \bibinfo{journal}{International Journal of Plasticity}
  \bibinfo{volume}{173}, \bibinfo{pages}{103860}.
\newblock \DOIprefix\doi{10.1016/j.ijplas.2023.103860}.
\bibitem[{Zong et~al.(2014)Zong, Lookman, Ding, Luo and Sun}]{Zong:2014}
\bibinfo{author}{Zong, H.}, \bibinfo{author}{Lookman, T.},
  \bibinfo{author}{Ding, X.}, \bibinfo{author}{Luo, S.N.},
  \bibinfo{author}{Sun, J.}, \bibinfo{year}{2014}.
\newblock \bibinfo{title}{Anisotropic shock response of titanium:
  {Reorientation} and transformation mechanisms}.
\newblock \bibinfo{journal}{Acta Materialia} \bibinfo{volume}{65},
  \bibinfo{pages}{10--18}.
\newblock \DOIprefix\doi{10.1016/j.actamat.2013.11.047}.

\end{thebibliography}

\end{document}